\def\Lya{Ly$\alpha$~}

\def\degsq{$\,{\rm deg}^2$}

\documentclass[useAMS,usenatbib]{mn2e}
\usepackage{upgreek}
\usepackage{graphicx}
\usepackage{url}
\usepackage{amsmath}
\usepackage{amssymb}
\usepackage{subfig}
\usepackage{caption}
\usepackage{verbatim}
\title[The bright end of the $z \simeq 6$ galaxy LF]{The galaxy luminosity function at ${\bf z \simeq 6}$ and evidence for rapid evolution in the bright end from ${\bf z \simeq 7}$ to ${\bf 5}$}  
\author[R. A .A. Bowler et al.]{R. A. A. Bowler$^{1}$\thanks{E-mail:
raab@roe.ac.uk}, J. S. Dunlop$^{1}$, R. J. McLure$^{1}$, H. J. McCracken$^2$, \and B. Milvang-Jensen$^{3}$, H. Furusawa$^{4}$, Y. Taniguchi$^{5}$, O. Le F\`{e}vre$^{6}$, \and J. P. U. Fynbo$^{3}$, M. J. Jarvis$^{7,8}$, B. H{\"a}u{\ss}ler$^{7,9}$ \\ 
$^{1}$SUPA\thanks{Scottish Universities Physics Alliance}, Institute for Astronomy, University of Edinburgh, Royal Observatory, Edinburgh, EH9 3HJ \\
$^{2}$Institut d'Astrophysique de Paris, UMR 7095 CNRS, Universit\'{e} Pierre et Marie Curie, 98 bis Boulevard Arago, 75014 Paris, France \\
$^{3}$Dark Cosmology Centre, Niels Bohr Institute, University of Copenhagen, Juliane Maries Vej 30, 2100 Copenhagen, Denmark \\
$^{4}$Astronomical Data Center, National Astronomical Observatory of Japan, Mitaka, Tokyo 181-8588, Japan \\
$^{5}$Research Institute for Space and Cosmic Evolution, Ehime University, 2-5 Bunkyo-cho, Matsuyama 790-8577, Japan \\
$^{6}$Laboratoire d'Astrophysique de Marseille, CNRS and Aix-Marseille Universit\'{e}, 38 rue Fr\'{e}d\'{e}ric Joliot-Curie, 13388 Marseille Cedex 13, France \\
$^{7}$Sub-department of Astrophysics, University of Oxford, Denys Wilkinson Building, Keble Road, Oxford OX1 2DL, UK \\
$^{8}$Department of Physics, University of the Western Cape, Bellville 7535, South Africa \\
$^{9}$Centre for Astrophysics, Science \& Technology Research Institute, University of Hertfordshire, Hatfield, Herts AL10 9AB, UK\\
}

\begin{document}
\date{}

\pagerange{\pageref{firstpage}--\pageref{lastpage}} \pubyear{2014}

\maketitle
\label{firstpage}

\begin{abstract}
We present the results of a search for bright ($-22.7 \leq M_{\rm UV} \leq -20.5$) Lyman-break galaxies at $z \simeq 6$ within a total of 1.65 square degrees of imaging in the UltraVISTA/Cosmological Evolution Survey (COSMOS) and UKIRT Deep Sky Survey (UKIDSS) Ultra Deep Survey (UDS) fields.
The deep near-infrared imaging available in the two independent fields, in addition to deep optical (including $z'$-band) data, enables the sample of $z \simeq 6$ star-forming galaxies to be securely detected long-ward of the break (in contrast to several previous studies).
We show that the expected contamination rate of our initial sample by cool galactic brown dwarfs is $\lesssim 3$ per cent and demonstrate that they can be effectively removed by fitting brown dwarf spectral templates to the photometry.
At $z \simeq 6$ the galaxy surface density in the UltraVISTA field exceeds that in the UDS by a factor of $\simeq 1.8$, indicating strong cosmic variance even between degree-scale fields at $z > 5$.
We calculate the bright end of the rest-frame Ultra-Violet (UV) luminosity function (LF) at $z \simeq 6$.
The galaxy number counts are a factor of $\sim 1.7$ lower than predicted by the recent LF determination by Bouwens et al..
In comparison to other smaller area studies, we find an evolution in the characteristic magnitude between $z \simeq 5$ and $z \simeq 7$ of $\Delta M^* \sim 0.4$, and show that a double power-law or a Schechter function can equally well describe the LF at $z = 6$.
Furthermore, the bright-end of the LF appears to steepen from $z \simeq 7$ to $z \simeq 5$, which could indicate the onset of mass quenching or the rise of dust obscuration, a conclusion supported by comparing the observed LFs to a range of theoretical model predictions.

\end{abstract}

\begin{keywords}galaxies: evolution - galaxies: formation - galaxies: high-redshift
\end{keywords}

\section{Introduction}\label{sect:intro}

The luminosity and mass functions of galaxies (i.e. the comoving number density as a function of intrinsic luminosity or stellar mass) are key observables in astronomy, as they trace the build-up and evolution of galaxies through cosmic time (\citealp{Madau2014}).
Through the comparison between the observed functions and the predictions of theoretical models and simulations,  it is possible to gain an insight into the dominant processes that control the formation and evolution of galaxies (e.g.~\citealp{Peng2010, Bower2012}).
When observing high-redshift galaxies ($z \gtrsim 3$), the rest-frame UV emission is redshifted into the optical/near-infrared and galaxies can be efficiently selected via their strong Lyman-break (at $\lambda_{\rm rest} = 1216\,$\AA), as pioneered by~\citet{Guhathakurta1990} and~\citet{Steidel1992}.
When characterising the number densities of these Lyman-break galaxies (LBGs) at the highest redshifts, the determination of the luminosity function of galaxies at rest-frame UV wavelengths has therefore become a standard practice~\citep{Steidel1999}.
More recent progress in determining the rest-frame UV LF at high-redshift has been primarily driven by data from the~\emph{Hubble Space Telescope} \emph{(HST)}, where the superior near-infrared sensitivity provided by the Wide Field Camera 3 (WFC3) has enabled the detection of hundreds of galaxies at $z > 6$ since it was installed in 2009 (e.g.~\citealp{McLure2010,Oesch2010, Bouwens2010}), with samples of objects now extending up to $z \simeq 9$~\citep{Ellis2013,Oesch2013}.
The tightest constraints on the LF have come primarily from a ``wedding-cake'' like combination of these \emph{HST} surveys (\citealp{McLure2013, Oesch2013, Bouwens2015, Finkelstein2014}), with extremely deep, small-area surveys such as the \emph{Hubble} Ultra Deep Field (HUDF, area = 4.5 arcmin$^2$, typical limiting magnitude in the near-infrared $m_{\rm AB} \sim 29.5$) and parallel fields detecting the faintest objects, being combined with samples of brighter galaxies from the wider-area Cosmic Assembly Near-infrared Deep Extragalactic Legacy Survey (CANDELS;~\citealp{Grogin2011, Koekemoer2011}, area $\simeq$ 0.2 deg$^2$, $m_{\rm AB} \sim 26$--$27$).
Furthermore, specifically at $z \simeq 8$, the Brightest of the Reionizing Galaxies (BoRG,~\citet{Trenti2011}, area $\gtrsim 350$ armin$^2$, $m_{\rm AB} \sim 27$) pure-parallel survey has enhanced the samples of $L \gtrsim L^*$ galaxies.
Overall, there has been a strong consensus between different analyses, using both the classical `colour-colour' selection~\citep{Bouwens2011, Schenker2013, Oesch2013} and photometric redshift fitting approaches~\citep{McLure2009, McLure2010, McLure2013, Finkelstein2014}. 

At the bright end, determining the number of $L \gg L^*$ galaxies with \emph{HST} surveys alone becomes challenging due to the declining number counts of objects brighter than the characteristic luminosity ($L^*$), and the relatively small area provided by the HUDF and CANDELS imaging.
Consequently, the very bright end of the LF at $z = 5-7$ has been successfully studied using ground-based surveys, which provide degree-scale imaging albeit to shallower depths (e.g. at $z = 5-6$: \citealp{McLure2006,McLure2009,Willott2013}, and at $z \simeq 7$:~\citealp{Ouchi2009, Castellano2010a,Castellano2010b,Bowler2012,Bowler2014}).
Previous comparisons between the ground- and space-based determinations have generally shown good agreement (e.g.~\citealp{McLure2009, Bouwens2007}), however, the recent, expanded analysis of $\sim 0.2\,{\rm deg}^2$ of~\emph{HST} imaging from the UDF and CANDELS surveys by \citet{Bouwens2015} has revealed new tensions between the different approaches.
In particular, the results of~\citet{Bouwens2015} at $z \simeq 6$ and $7$ are in excess of the previous results at the bright-end by a substantial factor ($\gtrsim 5 \times$).
Such a result is surprising as the inferred high density of bright objects should have been detected in the existing ground-based searches.
While~\citet{Bouwens2015} note that the previous under-estimation of the absolute magnitudes of galaxies, uncertain contamination fractions and over-estimated selection volumes can mostly explain the differences between previous~\emph{HST} based results (see Appendix F of~\citealp{Bouwens2015}), the disagreement with the ground-based results lacks a clear explanation.

The inferred high number density of bright galaxies found by~\citet{Bouwens2015} and similarly by~\citet{Finkelstein2014} (who used $\simeq 300\,{\rm arcmin}^2$ of data from the UDF and the Great Observatories Origins Deep Surveys North and South fields to select LBGs from $z = 4$--$8$) has changed the derived form of the evolution in the rest-frame UV LF from $z \simeq 8$ to $z \simeq 4$.
The luminosity function at high redshift is typically fitted using a Schechter function~\citep{Schechter1976}, where the observed number density, $\phi$, follows a power law with slope $\alpha$ to faint luminosities and an exponential cut-off brightward of the characteristic luminosity $L^*$, as $\phi (L) = \phi^* (L/L^*)^{\alpha}\,e^{-L/L^*}$.
The majority of previous studies of the form of the evolution of the LF at high-redshift have tended to favour a pure luminosity-evolution, with an approximately constant normalisation, $\phi^*$~\citep{Bouwens2007, Bouwens2008, Bouwens2011, McLure2009}.
Such an evolution is to be expected if the rest-frame UV luminosity of galaxies follows approximately the hierarchical assembly of the host dark matter halos at high redshift (e.g. see figure 10 of~\citealp{Bouwens2008}).
In contrast the new analyses by both~\citet{Bouwens2015} and~\citet{Finkelstein2014} find an approximately constant characteristic magnitude of the best-fitting Schechter functions of $M^* \simeq -21$ from $z \simeq 4$--$7$, and instead invoke a strong evolution in the faint-end slope and the overall normalisation to reproduce the observed evolution.
Whether the observed evolution in the rest-frame UV LF from $z \simeq 7$ to $z \simeq 4$ occurs primarily as density or luminosity evolution depends critically on the combination of astrophysical processes with the underlying dark matter halo mass function (HMF), for example the existence, origin and onset of any cut-off luminosity or quenching mass (e.g.~\citealp{Peng2010}).
Theoretical models tend to predict a more power-law type form for the LF at high-redshift (e.g. in the Illustris simulation;~\citealp{Genel2014}), and the implementation of astrophysical processes necessary to quench (e.g. feedback from accretion onto a black hole) or obscure (e.g. dust production) star formation in the most massive halos is required to bring the LFs into agreement (e.g.~\citealp{Bower2012, Cai2014}).
Hence the accurate determination of the evolution in the rest-frame UV LF at high redshifts is key for constraining the implementation of such cut-off mechanisms in these models.

Furthermore, at $z \simeq 7$,~\citet{Bowler2014} found evidence for a shallower drop-off in the number density of bright galaxies to that expected from the standard Schechter-function fit, and instead found that a double power law (DPL) form was preferred.
The observed $z \simeq 7$ LF from~\citet{Bowler2014} follows closely the form of the HMF, suggesting that the onset of significant quenching or dust obscuration occurs at $z < 7$.
A detailed analysis of the bright end of the $z \simeq 6$ LF is therefore essential to clarify the dominant form of the evolution of the rest-frame UV LF at high redshift, and to investigate how the functional form of the bright end changes as a result of the potential build-up of dust or the role of feedback in quenching the most massive galaxies.

In this work we follow the methodology of~\citet{Bowler2012} and~\citet{Bowler2014} to perform a search for bright $z \simeq 6$ star-forming galaxies within the UltraVISTA and UDS fields.
These fields contain a wealth of multiwavelength imaging (including deep $z'$-band and near-infrared data essential for the secure selection of $z > 5$ galaxies) covering an area over eight times greater than that analysed by~\citet{Bouwens2015}, and almost $20$ times that utilised by~\citet{Finkelstein2014}.
Recent improvements in the depth of imaging in both fields (e.g. $\gtrsim 1 \,{\rm mag}$ deeper in the $z'$ and/or $Y, J, H$ and $K_s$ bands) also allows us to directly compare to, and reassess, previous results at $z \simeq 6$ determined using these survey fields by~\citet{McLure2009} and~\citet{Willott2013}.

The paper is structured as follows.
We start with a description of the UltraVISTA and UDS datasets in Section~\ref{sect:data}, followed by the details of our $z \simeq 6$ galaxy selection in Section~\ref{sect:selection}.
The potential contamination of our sample by dwarf stars is quantified in Section~\ref{sect:dwarfstars}, and we detail our methodology to determine the rest-frame UV LF in Section~\ref{sect:lf}.
In Section~\ref{sect:sample} we present the basic properties of our sample of $z \sim 6$ LBGs, and compare our objects to previous $z = 6$ samples extracted from these survey fields by~\citet{Willott2013} and~\citet{McLure2009}.
We present our rest-frame UV LF results (including a comparison to previous studies) in Section~\ref{sect:discussion}.
Finally, we investigate the form of the $z \simeq 6 $ LF and the observed evolution from $z \simeq 5$ to $z \simeq 7$ in Section~\ref{sect:evolution}, and compare the observed evolution in the LF to the predictions from a range of theoretical models.
We end with our conclusions in Section~\ref{sect:conclusions}.
Throughout we assume a standard $\Lambda$CDM cosmology with $\Omega_{\rm m} = 0.3$, $\Omega_{\rm \Lambda} = 0.7$ and $H_{0} = 70\,{\rm km}\,{\rm s}^{-1}$.
All magnitudes are quoted in the AB system~\citep{Oke1974,Oke1983} where $m_{\rm AB} = -2.5\,{\rm log}_{10}[F_{\nu}({\rm Jy})/3631\,{\rm Jy}]$, and redshifts are photometric unless otherwise specified.
The functional forms for the Schechter, DPL and Saunders functions in magnitudes are presented in Appendix A for reference.

\section{Data}\label{sect:data}

\begin{figure*}
\begin{center}

\includegraphics[width = \textwidth]{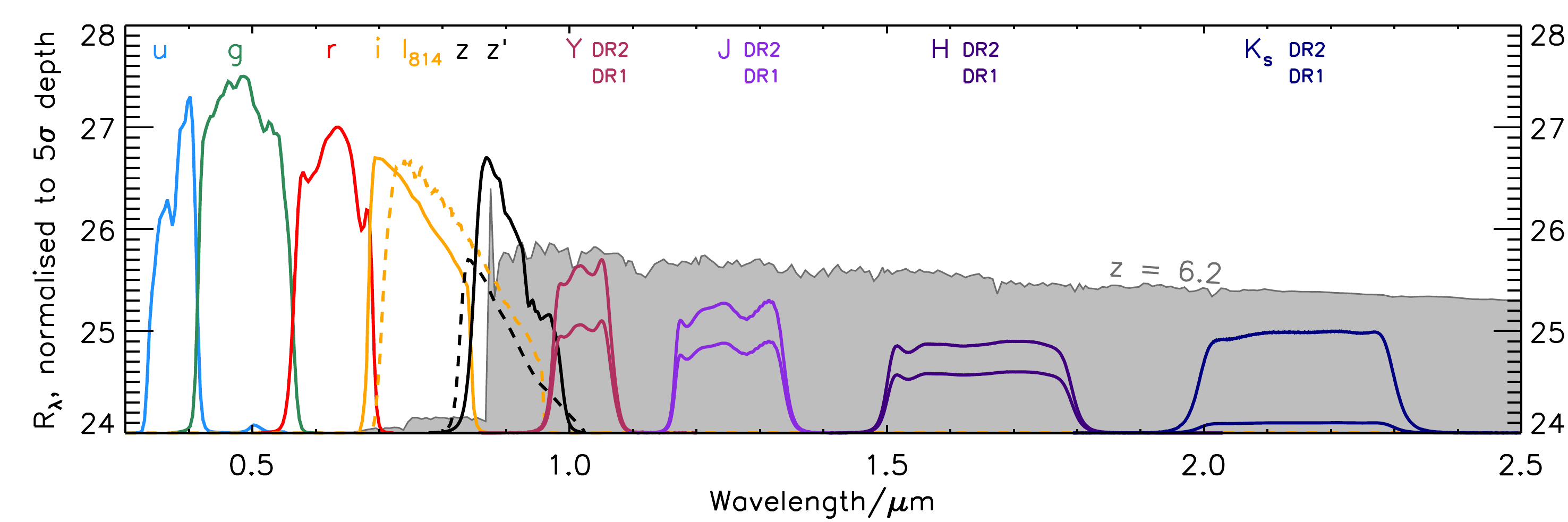}
\includegraphics[width = \textwidth]{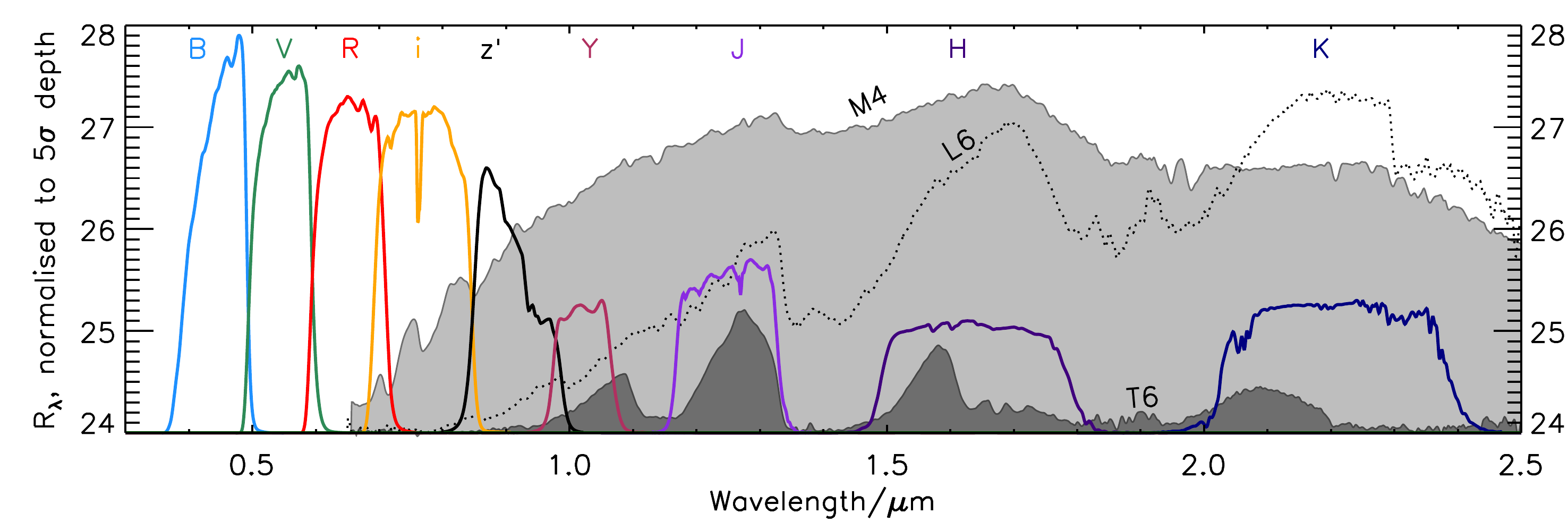}

\caption{The transmission curves for the filters used in this study in the UltraVISTA/COSMOS and UDS/SXDS fields are shown in the upper and low plots respectively.
Each filter curve has been peak normalised to the median $5\sigma$ depth (calculated in $1.8$-arcsec diameter circular apertures) and is shown as the response per unit wavelength.
The two curves shown for the $Y$, $J$, $H$ and $K_s$ filters in the UltraVISTA/COSMOS field illustrate the depths from the two epochs of imaging.
The deeper DR2 imaging exists over $\sim 70$ per cent of the UltraVISTA field with overlapping imaging from CFHTLS (for schematics see figures 1 and 2 of~\citealp{Bowler2014}).
In the upper plot an example redshifted galaxy SED from the~\citet{Bruzual2003} library is shown (this SED assumes an exponentially declining star-formation history with $\tau = 50\,{\rm Myr}$, an age of $100\,{\rm Myr}$ and $A_{\rm V} = 0.1$), in flux per unit frequency, where the mean IGM absorption from~\citet{Madau1995} has been applied.
The Lyman-break can clearly be seen at $\sim 0.9 \umu m$ in the SED, along with an example Lyman-$\alpha$ emission line of $EW_{0} = 10$\AA.
The lower plot shows example M, L and T-dwarf spectra taken from the SpeX library.}
\label{fig:filters}
\end{center}
\end{figure*}

The available optical/near-infrared imaging in the two extragalactic survey fields utilised here is summarised below, and further details of the data processing can be found in~\citet{Bowler2012, Bowler2014}.
Fig.~\ref{fig:filters} summarises the photometric bands used and illustrates the relative depths of the imaging (calculated in $1.8$-arcsec diameter circular apertures).
The search for galaxies at $z \simeq 6$ requires deep $z'$-band imaging for selection, and the UltraVISTA and UDS fields both benefit from deep Subaru $z'$-band imaging extending faint-ward of $m_{\rm AB} = 26$.
The presence of imaging in filters long-ward of the Lyman-break, in the near-infrared at $z > 6$, allows the robust removal of contaminant populations such as Galactic brown dwarfs and dusty low-redshift galaxies, while deep optical imaging is necessary to accurately determine the redshift defined by the strong Lyman-break (which moves through the $i$-band filter at these redshifts).

\subsection{UltraVISTA/COSMOS}

The UltraVISTA near-infrared imaging essential for this work lies within the COSMOS field~\citep{Scoville2007a}, which contains an abundance of multiwavelength imaging over 1--2 square degrees on the sky\footnote{\url{http://cosmos.astro.caltech.edu/}}.
We obtained deep Subaru/SuprimeCam $z'$-band imaging of the central square degree of the field, which reaches a $5\sigma$ limit of ${\rm m}_{\rm AB} = 26.6-26.8$ (1.8 arcsec diameter circular aperture) over four separate SuprimeCam pointings.
The COSMOS field also contains one of four `deep' fields (D2) imaged as part of the Canada-France-Hawaii Legacy Survey (CFHTLS), each corresponding to a single $1\,{\rm deg}^2$ pointing of CFHT/MegaCam.
We use optical imaging in the $u^{*},  g, r$ and $i$ filters from the T0007 release of the CFHTLS.
  The near-infrared imaging from UltraVISTA provides data in the $Y$, $J$, $H$ and $K_s$ filters over a total of 1.5 square degrees of the COSMOS field.
 The UltraVISTA survey consists of a `deep' component with depths of ${\rm m}_{\rm AB} \sim 25$ over the full 1.5 square degrees~\citep{McCracken2012}, superseded by the `ultra-deep' imaging that covers $\sim 70$\% of the full field in the form of four strips.
 Here we utilise the second data release (DR2) of UltraVISTA~\citep{McCracken2013}, in which the `ultra-deep' part has reached depths of ${\rm m}_{\rm AB} = 25.8$ in the $Y$ band (see table 1 of B14 and Fig.~\ref{fig:filters}).
 The maximal area of overlapping optical/near-infrared imaging utilised here is defined by the region covered by the CFHTLS ($1\,{\rm deg}^2$).
 Of this area, after the masking of large diffraction haloes, $0.62\,{\rm deg}^2$ is covered by near-infrared imaging from the `ultra-deep' part of UltraVISTA, and the remaining $0.29\,{\rm deg}^2$ has shallower imaging from the `deep' component.
 Due to the difference in depth, we treat these two regions separately in the selection of LBGs and in the analysis of the LF.

\subsection{UKIDSS UDS/SXDS}

The UKIDSS UDS field includes deep optical imaging taken as part of the Subaru \emph{XMM-Newton} Deep Survey (SXDS) in the $B$, $V$, $R$, $i$ and $z'$ filters~\citep{Furusawa2008}.
Further to this public data, we obtained deep Subaru/SuprimeCam imaging in the $z'$ band (see~\citealp{Bowler2014} for details) which reaches depths of ${\rm m}_{\rm AB} = 26.5-26.7$ (1.8 arcsec diameter circular aperture) over four separate SuprimeCam pointings.
The DR10 of the UKIDSS UDS~\citep{Lawrence2007} provides the key near-infrared imaging over the field in the $J$, $H$ and $K$ filters.
Finally, we use the second data release of the VISTA Deep Extragalactic Observations survey (VIDEO;~\citealp{Jarvis2013}), which has observed the UDS field in the $Y$ band to a $5\sigma$ limiting depth of ${\rm m}_{\rm AB} = 25.3$, extending $\simeq0.5$ mag deeper than the previous imaging utilised in~\citet{Bowler2014}.
The total area of overlapping optical and near-infrared imaging in the UDS/SXDS field was $0.74\,{\rm deg}^2$.

\section{Candidate Selection}\label{sect:selection}

\subsection{Catalogue production and initial cuts}\label{cuts}
Candidate high-redshift galaxies were selected from the deep Subaru $z'$-band imaging in the UltraVISTA/COSMOS and UDS/SXDS fields using {\sc SExtractor} v2.8.6~\citep{Bertin1996}, with the fiducial photometry measured in 1.8 arcsec diameter circular apertures.
Multiwavelength catalogues were produced using the `dual-image' mode of {\sc SExtractor}, with the Subaru $z'$-band used as the detection image and the other available bands used as the measurement images (e.g. $u^{*}griYJHK_{s}$ for the UltraVISTA/COSMOS field and $BVRiYJHK$ for the UDS/SXDS field).

The full catalogues were first cut at a $z'$-band magnitude of 26.0 in the UltraVISTA DR2 (`ultra-deep') and UDS fields, and at 25.0 for the shallower UltraVISTA DR1 (`deep') regions, resulting in a minimum significance of $7\sigma$ in the shallowest Subaru/Suprime-Cam tiles for each field, extending to $\sim 10\sigma$ in the deepest tiles.
To ensure a non-detection in the bluest bands we removed any object with a detection at greater than $2\sigma$ significance in either the $u$ or $g$ band in the UltraVISTA field, or the $B$ or $V$ band in the UDS, using local depth estimates within the field.
Local depths were calculated at each point in the images from the median absolute deviation (MAD) estimator (using $\sigma = 1.48 \times {\rm MAD}$) of the counts within the closest 200, randomly-placed, blank apertures.
For the purpose of SED fitting, we corrected all magnitudes to total assuming a point-source correction.
Many of the candidate high-redshift galaxies are resolved in the ground-based imaging (see~\citealp{Bowler2014}), however the size is still dominated by the seeing of the images and therefore the point source correction dominates any colour difference due to intrinsic size variation between the bands.

\subsection{Photometric redshift fitting}\label{photoz}

The final sample of $z \simeq 6$ galaxies was defined by fitting~\citet{Bruzual2003} SED models to the photometry, coupled with careful visual checks to remove artefacts and objects with low-level flux in the two bluest optical bands.
Given the degeneracies between metallicity, age and dust reddening when using broad-band photometry, we fit our candidate galaxy photometry with a reduced set of model galaxy SEDs (numbering $\simeq 500$ models in total, before the application of dust attenuation).
Exponentially declining ($\tau$) models with characteristic timescales in the range $50\,{\rm Myr} \le \tau \le 10\,{\rm Gyr}$ were used, and the~\citet{Calzetti2000} attenuation law was assumed in all cases.
The `high-redshift' model set, designed to identify good high-redshift candidates, consisted of models with ages from 10 Myr to the age of the Universe at $z = 5$, $A_{\rm V} = 0.0$--$2.0$ and a single metallicity of $0.2\,{\rm Z}_{\sun}$ motivated by recent measurements of the metallicity in low-redshift LBG analogues~\citep{Stanway2014}.
The photometric redshifts for our sample of $z\simeq 6$ LBGs have typical errors of $\Delta z \simeq 0.1$--$0.2$, depending on observed magnitude.
The `contaminant' model set was designed to provide red and dusty galaxy SEDs in the range $z = 1$--$3$,  with strong Balmer or 4000\AA~breaks, and therefore we allow ages up to the present age of the Universe, $A_{\rm V} = 0.0$--$6.0$ and a single metallicity of $1.0\,{\rm Z}_{\sun}$.
In addition to the suite of galaxy SEDs, we also fit standard stellar templates with types M4 to T8 (taken from the SpeX library~\footnote{\url{http://pono.ucsd.edu/~adam/browndwarfs/spexprism/}}), as cool galactic brown dwarfs can mimic the colours of high-redshift LBGs with $z = 5-7$.
We discuss and quantify potential brown dwarf contamination of our sample in Section~\ref{sect:dwarfstars}.

For inclusion in our final $z \simeq 6$ galaxy sample, we required that the object had a best fitting SED at $5.5~<~z~<~6.5$ with an acceptable $\chi^2$ ($ \leq 11.3$; calculated as the $\chi^2$ value that corresponds to $2\sigma$ significance given the number of degrees of freedom in the fitting), and a $\Delta \chi^2 > 4$ between the high-redshift solution and the next best-fitting $z < 5.5$ model (a $2\sigma$ condition when marginalising over all parameters but the redshifts; see~\citealp{Press1992}).
This step reduced the sample of several thousand objects that passed the magnitude cuts described above to a total sample of  335 objects, with 205, 3 and 127 objects from the UltraVISTA DR2 strips, DR1 and UDS respectively.
The sample was then carefully visually checked to remove single-band detections (including $z'$-band CCD bleeds) and to identify false optical non-detections due to the negative haloes around stars in the CFHTLS and Subaru optical imaging, which resulted in the removal of a further 26 objects (15 in the UltraVISTA DR2 and 11 in the UDS).
At this point we also removed the extreme LAE `Himiko' from our UDS sample, because of the known spectroscopic redshift of $z = 6.595$, which places it outside our desired redshift range (see Section~\ref{sect:mclure}).
The result of the SED fitting and visual checks described was a sample of 309 objects (190, 3 and 116 in the UltraVISTA DR2, DR1 and UDS respectively) that are consistent with being $5.5 < z < 6.5$ LBGs.
However, one final source of contamination must be considered, namely cool galactic brown dwarfs.
The removal of candidates consistent with being a brown dwarf using the criterion described in detail in the next section, resulted in the removal of 37 and 10 objects in the UltraVISTA DR2 and UDS samples respectively, producing a final sample of 266 galaxies with photometric redshifts in the range $5.5 < z < 6.5$.

\section{Contamination by brown dwarfs}\label{sect:dwarfstars}

Cool galactic brown dwarfs (with spectral types M, L and T) have SEDs that peak in the near-infrared and drop steeply towards the optical bands (e.g. the M8 type dwarf shown in Fig.~\ref{fig:filters}), potentially mimicking the colours of $z > 5$ galaxies.
The number density of brown dwarfs begins to drop at $J > 25$~\citep{Ryan2011} as the number counts of galaxies rapidly rise (e.g see Fig.~\ref{fig:numdensities}), and therefore they are in practice a negligible contaminant for extremely deep, small-area imaging programs such as the HUDF.
In the search for the brightest high-redshift $z \simeq 6$ LBGs however, the number densities of brown dwarfs can begin to dominate (see Fig.~\ref{fig:numdensities}), and the high-redshift galaxy samples become increasingly susceptible to contamination by the relatively more numerous M- and L-type dwarfs.
The comparatively poor resolution of the ground-based imaging surveys utilised here compared to \emph{HST} imaging, coupled with the small measured sizes of LBGs~\citep{CurtisLake2014, Ono2012}, precludes any discrimination based on size, and hence we must carefully assess the available multiwavelength information to remove brown dwarfs from the $z \simeq 6 $ sample.
The UltraVISTA/COSMOS and UDS/SXDS fields utilised here contain the best deep optical to near-infrared photometric data available on the degree scale, and hence, as we show quantitatively in the following section, the removal of brown dwarfs using SED fitting of standard spectral templates can be cleanly performed.
The possibility of photometric scattering of brown dwarfs (which are considerably more numerous than $z > 5$ LBGs at $m_{\rm AB} \lesssim 25$)  into our sample must be carefully considered however, and we use a simple model of the Galactic stellar distribution to estimate the likely number of contaminant brown dwarfs in Section~\ref{numdensities}.

\subsection{Injection and recovery simulations}\label{starsims}

We quantify the potential brown dwarf contamination of our sample by injecting and recovering synthetic dwarf star photometry into the UltraVISTA/COSMOS and UDS/SXDS images, passing the fake stars through an identical selection procedure as described above for our $5.5 < z < 6.5$ LBGs.
Each standard spectral template for dwarf star types M4--T8 from the SpEX library was integrated through the appropriate filters for the UltraVISTA/COSMOS and UDS/SXDS fields, scaled to a total $z'$-band magnitude in the range $m_{\rm AB} = 23$--$26$ and injected into the images using the model PSF (determined using the method described in~\citealp{Bowler2014}) at a random position.
As described further in Section~\ref{sect:completeness}, we used representative sub-sections of the full UltraVISTA and UDS mosaics to reduce computing time.
If the object was recovered ($\sim 20$\% of injected objects are lost due to blending with other sources), and passed the $z'$-band magnitude cut and the optical non-detection conditions (Section~\ref{cuts}),  the photometry was corrected to a total magnitude in each band and fitted with the galaxy and stellar spectral templates used in the LBG selection.
The injected stellar photometry was then classified as an LBG contaminant if $\chi^2_{gal} < 11.3$ and $5.5 < z_{\rm phot} < 6.5$, following exactly our LBG selection procedure.

The resulting fraction of each stellar type classified as an LBG contaminant depended strongly on the assumed magnitude, with injected stars brighter than $z' \simeq 25$ rarely being classified as an LBG.
We find that up to $\sim 10$--$20$ per cent of the input dwarf stars with $z' = 25$--$26$ are classified as $5.5 < z < 6.5$ galaxies, depending on sub-type and field (where the differing relative depths of the imaging results in different vulnerabilities).
The predicted number and redshift distribution of dwarf stars is discussed in the next section.

Crucially, the simulations show that the overwhelming majority ($> 95$\%) of injected stellar templates that are recovered as LBG candidates remain good stellar fits.
We also find that we can exclude the majority of the contaminant brown dwarfs from our galaxy sample by requiring that the object has a poor stellar fit quantified as $\chi^2_{\star} > 10.0$.
A small number of genuine LBGs will also be excluded as a result of this criterion, however we account for the additional incompleteness of our sample when calculating the LF, by recreating this selection criterion in our galaxy injection and recovery simulations.
By applying such a selection criterion, we remove 37 objects from the UltraVISTA DR2 sample, 0 from the DR1 and 10 from the UDS sample.
In Appendix B we present the 47 potential high-redshift galaxies that were excluded from the original sample as possible dwarf star contaminants based on a good stellar fit with $\chi^2_{\star} < 10.0$.

\subsection{Number density model}\label{numdensities}

\begin{figure}
\includegraphics[width = 0.47\textwidth]{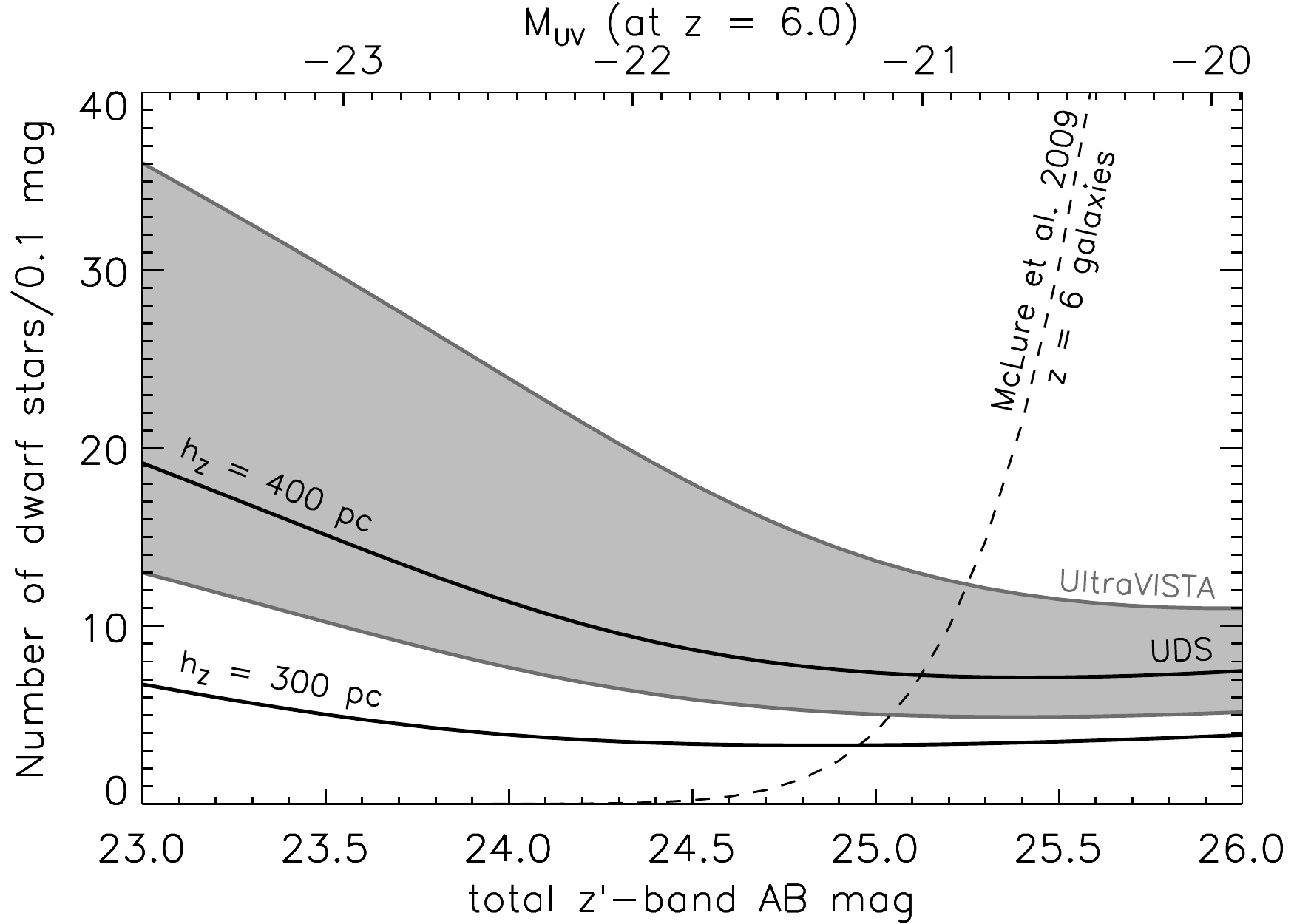}
\caption{The total number of dwarf stars (with spectral types M4--T8) predicted in the $0.62\,$deg$^2$ UltraVISTA/COSMOS DR2 field is shown as the grey shaded region, and the corresponding values for the $0.74\,$deg$^2$ UDS/SXDS field are shown as the black lines.  
The lower and upper curves for each field were calculated assuming a Galactic scale-height of $h_{\rm Z} = 300\,{\rm pc}$ and $h_{\rm Z} = 400\,{\rm pc}$ respectively, using the exponential disk model of~\citet{Caballero2008}.
The predicted number densities of $z = 6$ LBGs determined from the LF of~\citet{McLure2009} is shown as the dashed line, for an example survey area of $0.7$ deg$^2$.
The upper axis shows an estimate of the absolute UV magnitude corresponding to the continuum magnitude on the lower axis, assuming the object is at $z = 6$.
}
\label{fig:numdensities}
\end{figure}

Despite the apparent success of our stellar fitting method for removing dwarf stars masquerading as high-redshift LBGs, even a small contamination rate at the bright end could be significant for the determination of the $z \simeq 6$ LF (Fig.~\ref{fig:numdensities}).
Hence to constrain the likely contamination rate, an estimate of the number of each stellar type as a function of magnitude is required. 

\begin{figure}

\includegraphics[width = 0.48\textwidth]{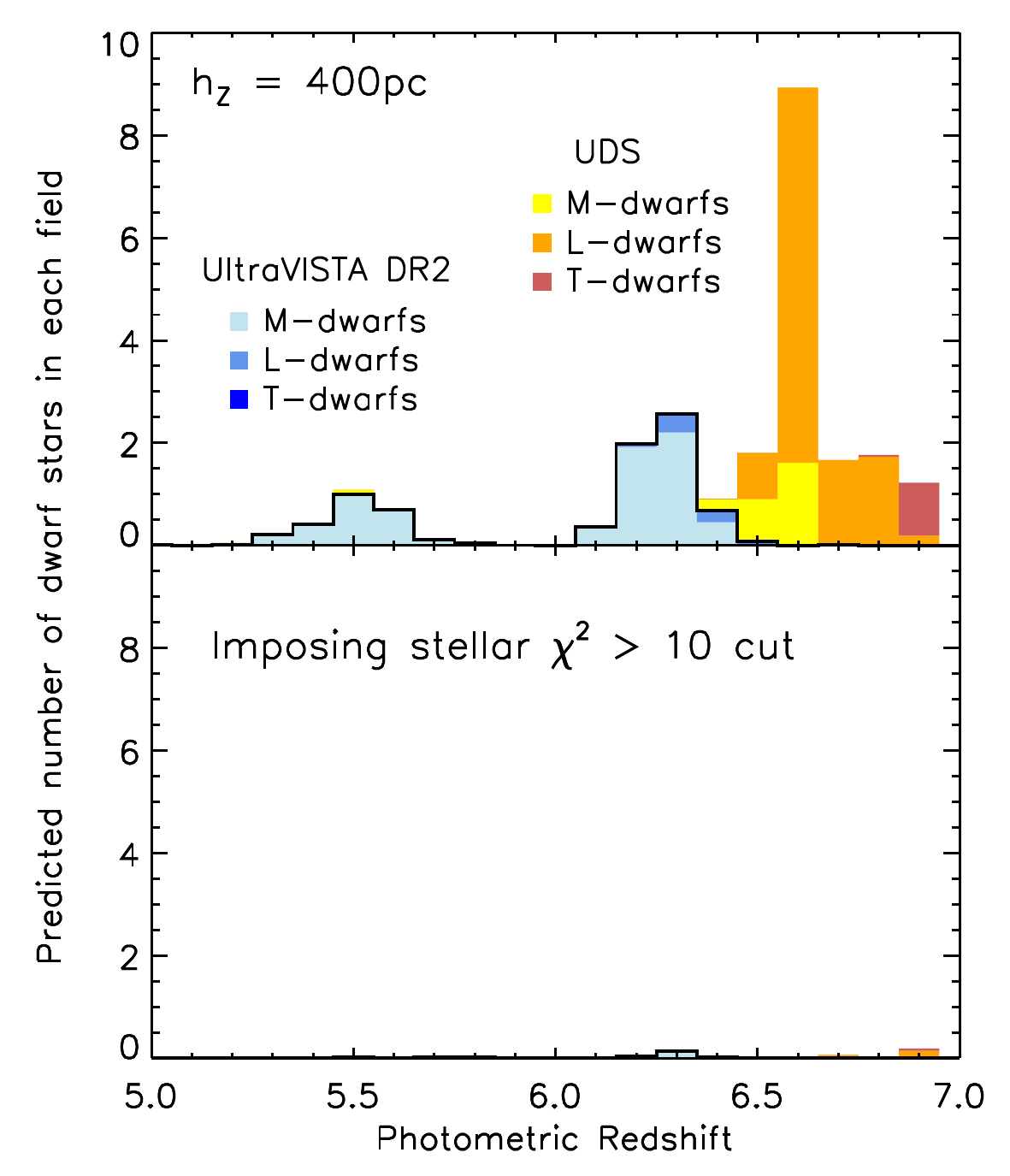}

\caption{The expected photometric redshift distribution of dwarf stars that are acceptable $z > 5$ LBG candidates in the UltraVISTA DR2 and UDS fields.
The UltraVISTA DR2 results are shown as the blue histograms with a thick black outline, where each stellar type is highlighted in a different shade of blue, whereas the distribution for the UDS is shown in yellow and orange.
A vertical Galactic scale height of $h_{\rm Z} = 400$ pc is assumed to illustrate an upper limit on the number of potential stellar contaminants.
The upper panel shows the predicted number of brown dwarfs in our sample when no attempt is made to remove objects with good stellar fits.
By requiring that robust high-redshift galaxies have a poor stellar fit, with $\chi_{\star}^2 > 10.0$, the expected number of contaminant brown dwarfs in the sample drops essentially to zero as shown in the lower panel.}
\label{zdist}
\end{figure}

The dwarf stars relevant for high-redshift galaxy studies are typically distant objects in Galactic terms, e.g. an M4 dwarf star with observed magnitude $z' = 26$--$24$ probes the galaxy at a radius of $1.5$--$4\,$kpc (or 250--600 pc for a T8 dwarf), and determining the scale-height of the disk components is challenging.
Previous searches for M, L and T-type stars have commonly assumed a single disk model (e.g.~\citealp{Holwerda2014, Ryan2011}) to describe the observed number of stars, as the small samples of objects preclude a more complicated analysis.
We follow such an approach using a single exponential disk model as described in~\citet{Caballero2008}.
Here the number density of each dwarf star type $s$, as a function of galactic longitude and latitude ($l,b$) and heliocentric distance ($d$), is given by:
\begin{equation}\label{equationn}
n_s = n_s(d = 0) \,{\rm e}^{-\frac{R(d,l,b) - R_{\odot}}{h_{\rm R}}}{\rm e}^{-\frac{|Z_{\odot} + d{\rm sin} b|}{h_{\rm Z}}}
\end{equation}
where $R_{\odot}$ and $Z_{\odot}$ denote the Galactocentric solar radius and height above the galactic disk, and $h_{\rm R}$ and  $h_{\rm Z}$ are the radial scale length and scale height for the model.
We use the approximations to Equation~\ref{equationn} outlined in~\citet{Caballero2008} relevant for deep extragalactic survey fields, and predict the number density of each spectral type as a function of magnitude by integrating along a line-of-sight through the galactic disk.
The local number densities and stellar absolute magnitudes ($M_{I}$) were taken from~\citet{Caballero2008}, converting the magnitudes from the Vega to the AB system according to~\citet{Frei1994}.
We then calculated the absolute $z'$-band magnitude from $M_{I}$ using the $i - J$ colours from~\citet{Caballero2008} and the standard spectral templates described in Section~\ref{photoz} to convert from $M_{J}$ to $M_{z}$.
We adopt the main parameters of the galactic thin disk model from~\citet{Chen2001}, with $Z_{\odot} = 27 \pm 4$pc, $R_{\odot} = 8600 \pm 200$pc and $h_{\rm R} = 2250 \pm 1000$ pc.
The most relevant quantity for this study is the galactic vertical scale height, where we use two values at $h_{\rm Z} = 300\,{\rm pc}$ and $h_{\rm Z} = 400\,{\rm pc}$ to reflect the uncertainty in this quantity~\citep{Holwerda2014, Ryan2011, Pirzkal2009}.
Fig.~\ref{fig:numdensities} shows the predicted number of dwarf stars as a function of magnitude in the two fields.
The analytic model shows that we expect more dwarf stars in the UltraVISTA/COSMOS field (galactic coordinates $b = 42.1$, $l = 236.8$) despite the smaller area covered (area of DR2 = 0.62 deg$^2$ vs. UDS = 0.74 deg$^2$), due to the line of sight intersecting with more of the galactic disk and being at a lower galactic latitude to the UDS/SXDS ($b = -60.0$, $l = 169.9$).
The intersection of a pencil-beam survey with this exponentially-declining distribution results in each dwarf star spectral type (which has a corresponding intrinsic absolute magnitude, $M_{J, {\rm AB}} \sim 9$--$17$ for spectral types M4--T8) having a peak in number density at an increasingly faint apparent magnitude, with M-dwarfs peaking at $z' < 23$, L-dwarfs at $z' \simeq 25$ and T-dwarfs at $z' > 26$.

\subsection{Predicted number of contaminant brown dwarfs}

\begin{figure*}
\includegraphics[width = \textwidth]{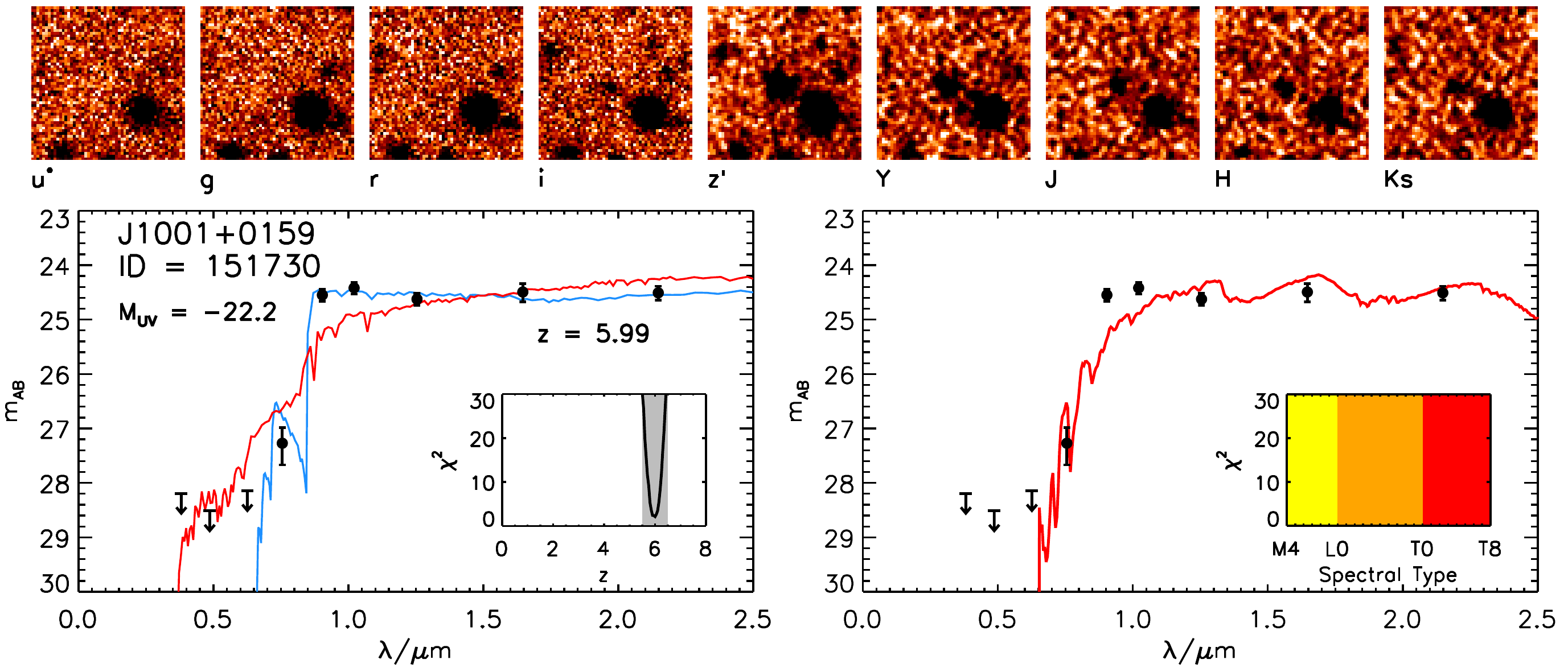}\\
\vspace{0.1cm}
\includegraphics[width = \textwidth]{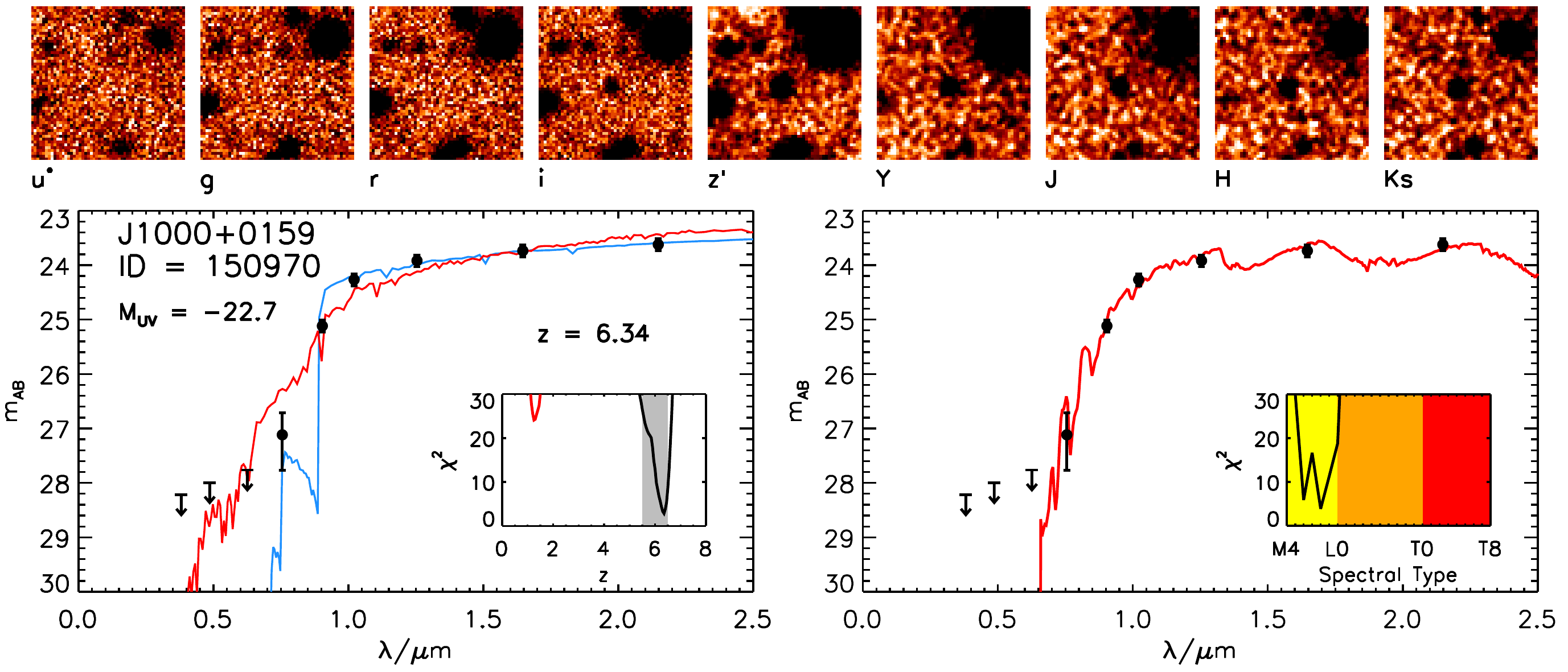}
\caption{SED fitting results and postage-stamp images of two objects from our penultimate sample of $z \simeq 6$ galaxies.
For each object cut-outs (10-arcsec on the side) from the images are shown in the upper panels.
The two plots show the observed photometry as the black points, with the best-fitting high and low-redshift galaxy templates shown as the blue and red lines in the left-hand panel, and the best-fitting stellar template shown in the right-hand panel.
The inset in each plot shows the variation of $\chi^2$ with either redshift or stellar spectral-type.
The upper object at $z \simeq 6$ shows a clear break between the $i$ and $z'$-bands that cannot be recreated by the low-redshift galaxy or stellar templates.
The lower object however, is both a good stellar and high-redshift galaxy fit, illustrating the difficulty in removing stars in particular at $z > 6.3$ and $z < 5.7$, where the optical to near-infrared break appears more gradual.
The lower object was removed from our final sample based on a good stellar fit (taken as $\chi_{\star}^2 < 10.0$).
}\label{fig:example}
\end{figure*}

By combining the predicted number densities of dwarf stars illustrated in Fig.~\ref{fig:numdensities} with the probability of a given dwarf stellar type of a given magnitude passing the LBG selection criterion (from the simulations described in Section~\ref{starsims}), we can predict the number of contaminant brown dwarfs expected in our $ z \simeq 6$ sample.
The expected pseudo-redshift distribution for these stars can also be calculated and is shown in Fig.~\ref{zdist}, assuming a vertical scale height of $h_{\rm Z} = 400$ pc to provide an upper limit on the number of objects.
Within the UltraVISTA DR2 area we would predict a total of 2--7 dwarf stars with $h_{\rm Z} = 300$--$400$ pc, and in the UDS this drops to 0.1--0.4.
Applying a $\chi_{\star}^2 > 10.0$ condition to exclude objects with good stellar fits, the number of contaminant stars drops dramatically to 0.1--0.4 in UltraVISTA DR2 and 0.02-0.06 in the UDS samples as shown in the lower panel of Fig.~\ref{zdist}.
The predicted redshift distribution for the contaminant stars also illustrates the differences between the selection functions of the two fields.
Most notably, the UDS field is more prone to contamination by late type dwarfs, such as L- and T-types, as a result of the deep optical imaging excluding early-type M-dwarfs, and the shallower $Y$-band imaging making the separation of a sharp break and the gentle rise of a dwarf star through the $z'$, $Y$ and $J$ filters difficult to distinguish.
These late-type dwarfs have higher photometric redshifts when identified as LBG candidates, and hence in any case are excluded as part of our selection procedure for $5.5 < z < 6.5$ objects.
The UltraVISTA/COSMOS field is more prone to contamination by M-dwarf stars, due to the shallower optical data available in the field.
The lack of dwarf stars showing a best-fitting redshift in the range $5.7 < z_{\rm phot} < 6.1$ arises due to stellar templates being unable to reproduce the resulting large $i - z'$ colour produced from an LBG in this redshift range, this is clearly seen in the example SEDs shown in Fig~\ref{fig:example}.
~\citet{McLure2009} restricted the redshift range of their sample of $z \sim 6$ LBGs to the range $5.7 < z < 6.3$, which our simulations show is also a relatively clean dwarf star region.

In summary, our simulations would suggest a $\sim 3$ per cent contamination of our initial sample, which can be reduced to $\ll 1$ object by ensuring all galaxy candidates have a bad stellar fit (with $\chi_{\star} > 10.0$).
Imposing a $\chi_{\star} > 10.0$ condition on our penultimate sample resulted in the removal of 37 and 10 objects from the UltraVISTA DR2 and UDS fields respectively, corresponding to $\sim 20$ per cent of the initial sample.
Inspection of the $\chi^2$ values for the stellar and galaxy fits for objects removed from the sample as potential stars (shown in Table~\ref{table:stars}), reveals that many of these objects are relatively poor stellar fits with $\chi_{\star}^2 \sim 5$--$9$.
Therefore we are likely excluding some genuine galaxies by imposing such a cut, a hypothesis supported by the redshift distribution of our sample shown in Fig.~\ref{fig:megafigure} which, after the imposition of a $\chi_{\star}^2 > 10.0$ condition, shows a deficit of objects at $5.5 < z < 5.7$ as compared to our expected distribution.
The deficit is particularly obvious in the UDS field, and inspection of the $\chi_{\star}^2$ values for the objects classified as stars reveals that they are all $\chi_{\star}^2> 5$.
As the key aim of this work is the accurate determination of the LF at $z \simeq 6$, free from contamination by brown dwarfs that can dominate the number counts of the brightest LBGs, we choose to apply such a cut with the acknowledgement that genuine galaxies will be excluded at this step.
This incompleteness is taken into account in our injection and recovery simulations.
In addition, to ensure that dwarf star contamination and our removal methodology has a minimum impact on our LF determination, we choose to restrict our redshift range to $5.7 < z < 6.3$ in the LF analysis.
Such a redshift restriction has the additional benefit of reducing the impact of the evolving LF on our analysis and makes the median redshift ($z_{\rm med} \simeq 5.9$) more in-line with previous determinations~\citep{McLure2009, Willott2013}.

\section{Determination of the LF}\label{sect:lf}

In the calculation of the rest-frame UV LF from an observed galaxy sample it is necessary to account for the impact of photometric scatter and the particular selection methodology implemented, as these effects, unless corrected for,  can strongly affect the derived intrinsic number density of galaxies.
The high signal-to-noise ratio of our galaxies ($> 7\sigma$ in the $z'$-band) strongly reduces the occurrence of spurious detections, and the application of our careful SED fitting procedure can remove low-redshift interlopers and Galactic brown dwarfs.
Our samples will, however, still suffer incompleteness from blending with foreground objects and from misidentification as dwarf stars or low-redshift galaxies at the faint end of our sample.
In the following section we describe our injection and recovery simulations that quantify our completeness and the methodology we use to calculate the binned LF including this correction.
Our rest-frame UV LF results are presented in Section~\ref{sect:discussion}. 
For both synthetic and observed galaxies in this work we calculate the rest-frame UV absolute magnitude at a central wavelength of $\lambda_{\rm rest} = 1500\,$\AA, in concordance with previous work~\citep{Bowler2014, McLure2013} by integrating the best-fitting SED in the rest-frame with a top hat filter of width $100\,$\AA.

\subsection{Completeness simulations}\label{sect:completeness}

We estimate the completeness of our sample following the methodology presented previously in~\citet{McLure2009},~\citet{McLure2013} and~\citet{Bowler2014}, by injecting synthetic $z = 5-7$ LBGs into the UltraVISTA/COSMOS and UDS/SXDS datasets, and attempting to recover them using an identical procedure as for the real $z = 6$ objects selected in this work.
Due to the differing depths of available optical/near-infrared imaging in the two fields, and the different selection criteria used for the `deep' and `ultra-deep' parts of the UltraVISTA survey, separate simulations for each region were performed to provide field dependent completeness correction factors.
In each case, the $z'$-band magnitude cut applied ($z' < 26.0$ or $z' <25.0$ in the `ultra-deep' and `deep' components respectively) was recreated in the simulations.

Photometric errors can scatter injected objects in magnitude and redshift, and therefore we inject galaxies from $z = 5$ to $z = 7$ and with magnitudes as faint as one magnitude below the $z'$-band cut we use for each field.
Whereas the number of objects scattered out of a bin is symmetric, the number scattered into a given bin depends on the underlying luminosity function (e.g. a Schechter function exponential decline will result in more objects up scattered into a given bin than a shallower function).
Since we do not know a priori the form of the bright end of the LF, we calculate the incompleteness using a range of functional forms for the injected galaxy population and compare the results.
In addition, the luminosity function of LBGs is evolving with time, which must be taken into account and could potentially affect the derived parameters at each redshift (\citealp{Munoz2008}).
In the Schechter function case, we assume the parameters from~\citet{McLure2009} at $z = 5$ with ($[M^*, \phi^*, \alpha] = [-20.73, \:0.00094/{\rm Mpc^3}, -1.66]$) and \citet{McLure2013} at $z = 7$ ($[M^*, \phi^*, \alpha] = [-19.90,\: 0.0011/{\rm Mpc^3}, -1.90]$), with a simple linear evolution in all the Schechter parameters between these redshifts.
We use an analogous approach for the DPL, using the fitted parameters from~\citet{Bowler2014} at $z = 5$ ($[M^*, \phi^*, \alpha] = [-21.0,\: 0.00039/{\rm Mpc^3}, -1.9]$) and $z = 7$ ($[M^*, \phi^*, \alpha] = [-20.4, \:0.00031/{\rm Mpc^3}, -2.2]$) with a fixed bright-end slope ($\beta_{} = -4.4$).
Using the evolving LFs, we then randomly populated an input $M_{\rm UV}-z$ plane, assigning each galaxy a rest-frame UV slope, $\beta_{\rm UV}$, drawn at random from a Gaussian distribution with mean $\beta_{\rm UV} = - 1.8$ and standard deviation $\sigma = 0.3$, to mimic the slightly red $\beta_{\rm UV}$ and intrinsic scatter found for bright galaxies at $z = 5$  by~\citet{Rogers2014}. 
The objects were then selected using~{\sc SExtractor} and the high-redshift candidates extracted using the magnitude cuts and SED fitting analysis as described in Section~\ref{sect:selection}.

We find that $\simeq 20$ per cent of injected objects are not recovered due to blending by foreground images in the crowded optical bands.
Of the objects that are recovered, we find a completeness of $\simeq 80$ per cent, where objects here are lost either because they do not pass the optical drop-out criterion or because they were misclassified as a dwarf star or low-redshift galaxy contaminant.

\subsection{The ${ \boldsymbol 1/\boldsymbol V_{\rm max}}$ estimator}

We used the $1/V_{\rm max}$ estimator~\citep{Schmidt1968} to derive the LF, where the number density of objects in a given magnitude bin, $\phi(M)$, depends on the maximum volume ($V_{\rm max}$) each galaxy could have been selected in, modulated by a completeness correction factor ($C_{ f}$) which accounts for the impact of both object blending and photometric scatter on the observed number of galaxies in a given bin:

\begin{equation}
\phi(M) = \sum_{i=1}^{N} \frac{C_{\rm f}(M_i,z_i)}{V_{{\rm max},i}}
\end{equation}
Here the sum is over the $N$ galaxies in the magnitude bin in question, where we chose magnitude bins of width $\Delta M = 0.25 $--$ 0.5\,$mag depending on the number of objects available.
The $V_{\rm max}$ was calculated by artificially redshifting the best-fitting SED of each galaxy to find the maximum redshift it could have been observed at and still be included in the sample (e.g. by passing the initial $z'$-band magnitude cut).
The comoving volume was then calculated from this $z_{\rm max}$ and the minimum allowed redshift of $z_{\rm min} = 6.7$, and the survey area.
The completeness correction factor depended on the field or region of the imaging in which the galaxy was selected, which has been denoted here as ${f}$.
The three regions in which objects were selected were the UltraVISTA/COSMOS DR2 (or `ultra-deep' strips) comprising $0.62\,{\rm deg}^2$ of imaging, the shallower UltraVISTA/COSMOS DR1 (or `deep' region) comprising $0.29\,{\rm deg}^2$ or the UDS/SXDS field which provided $0.74\,{\rm deg}^2$.
The errors on the number densities are assumed to be Poissonian, however there is also an additional error in the derived number density resulting from the error in the completeness value.
Hence, we include an estimate of this error by bootstrap resampling the galaxies within each bin.
The Poisson error dominates in the bright bins, however the error in the completeness becomes comparable for the faintest bins.

\subsection{The binned LF}

We restrict the redshift range to $5.7 < z < 6.3$ to enable direct comparison to the work of~\citet{McLure2009} and to ensure we are not influenced by any residual brown dwarf contamination or any incompleteness due to our dwarf star removal methodology.
Our results are robust to the underlying function assumed in our completeness simulations, due to the high signal-to-noise of our galaxies which reduces the effect of up-scattering.
We assume the DPL parameterisation as in~\citet{Bowler2014} in the final LFs presented.

We determine the LF in the range $-22.625 < M_{\rm UV} < -21.125$, using 0.25 or 0.5 mag bins depending on the number of objects available.
The faintest bin is defined by the point at which our observed absolute magnitude counts begin to drop, as shown in Fig.~\ref{fig:megafigure}, indicating that our sample is becoming increasingly incomplete.
The maximum volume available to our brightest objects is $7 \times 10^6\,$Mpc$^3$ in the UltraVISTA DR2 and UDS fields combined, with an additional $1 \times 10^6\,$Mpc$^3$ available for the brightest objects that could have been selected in the $0.29$\degsq~of UltraVISTA DR1 imaging.
Our LF results calculated following the described methodology are presented and discussed in Section~\ref{sect:discussion}.

\section{The Sample}\label{sect:sample}

The final sample of $5.5 < z < 6.5$ galaxies consists of 266 objects, with 156, 3 and 107 coming from the the UltraVISTA/COSMOS DR2, DR1 and UDS/SXDS fields respectively.
In the reduced redshift range $5.7 < z < 6.3$ used by~\citet{McLure2009}, we find 105, 2 and 70 objects in these fields.
We postpone a full discussion of the SED properties and sizes of these objects to a future paper (Bowler et al. in prep), however the basic sample properties are discussed below.

\subsection{Galaxy colours}

\begin{figure}
\includegraphics[width = 0.47\textwidth]{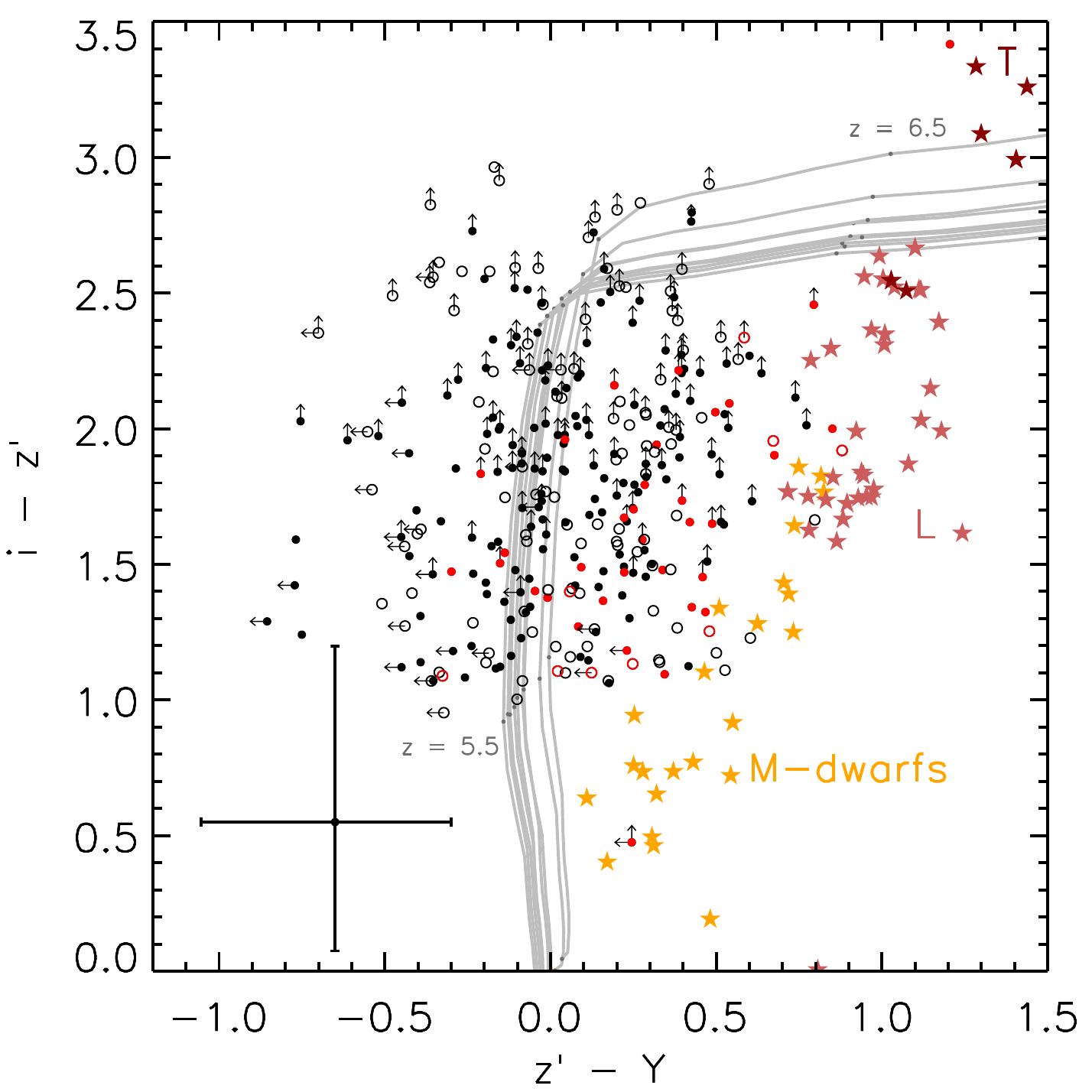}
\caption{The $z' - Y$ vs. $i-z'$ colours of the full $5.5 < z < 6.5$ sample.
The colours derived from high-redshift galaxy SEDs (shown as grey tracks) and the colours of galactic dwarf stars (star symbols) are shown for comparison.
The galaxy sample is shown as the black points, with the UltraVISTA/COSMOS and UDS/SXDS galaxies shown as the filled and open circles respectively.
Objects that were excluded from these samples as possible brown dwarfs are shown as the red points.
The typical error bar is shown in the lower left corner, and where an object is detected at less than the $2\sigma$-level in a given filter the magnitude here is set to the local $2\sigma$ depth, and we display the colour as a limit using an arrow.
The M-, L- and T-dwarf star colours were calculated from spectra taken from the compilation of stellar spectra described in~\citet{Findlay2012}.
High-redshift LBG colours taking from~\citet{Bruzual2003} models are shown as the grey lines (constant star-formation history, $Z = 0.2\,{\rm Z}_{\sun}$, $A_{V} = 0.0-0.5$, age = $50-500\,$Myr).
}
\label{fig:ccplot}
\end{figure}

Galaxies at $z \sim 6$ are typically selected as $i$-band dropout objects, and in Fig.~\ref{fig:ccplot} we show the $i-z'$ colour (which straddles the break at $z \simeq 5.8$) against the $z'-Y$ colour (which determines the rest-frame UV colour) of our sample of galaxies selected by their photometric redshift.
Both the UltraVISTA/COSMOS and UDS/SXDS samples occupy a similar region of colour space, indicating no strong biases in the galaxy colours due to the different relative imaging depths.
The colours of the objects can be reproduced by~\citet{Bruzual2003} models with $A_{V} = 0.0$--$0.5$ within the errors (\citealp{Calzetti2000} attenuation law), with no strong evidence for extremely red objects (to be compared with the stack of $z \sim 6$ galaxies found by~\citealp{Willott2013} which had a best-fitting $A_{V} = 0.75$, as discussed further in Section~\ref{beta}).
We also show the colours of the possible dwarf stars excluded from our sample, along with the stellar locus as derived from stellar spectra compiled by~\citet{Findlay2012}.
Several of these objects do lie on the stellar locus, however a considerable fraction are found with colours differing by up to $\simeq 0.5$ mag.
The results of the injection and recovery simulations described in Section~\ref{sect:dwarfstars} showed that the majority of dwarf stars enter our LBG sample as a result of scattering of the photometry, hence a wide range of colours for potential dwarf stars is to be expected.
The identification of potential stellar contaminants that would have identical colours to LBGs illustrates the power of using the full multiwavelength photometry,  as several of the likely dwarf stars would be indistinguishable from LBGs based on a simple colour selection.
Furthermore, Fig.~\ref{fig:ccplot} clearly shows that if a strict colour selection for $z \sim 6$ objects, designed to remove dwarf star candidates, was implemented, the true colour distribution of LBGs would have been biased to bluer objects.

\subsection{Redshift, ${\bf M_{\rm UV}}$ and ${\bf {m}_{\rm AB}}$ distributions}

\begin{figure*}
\includegraphics[width = \textwidth]{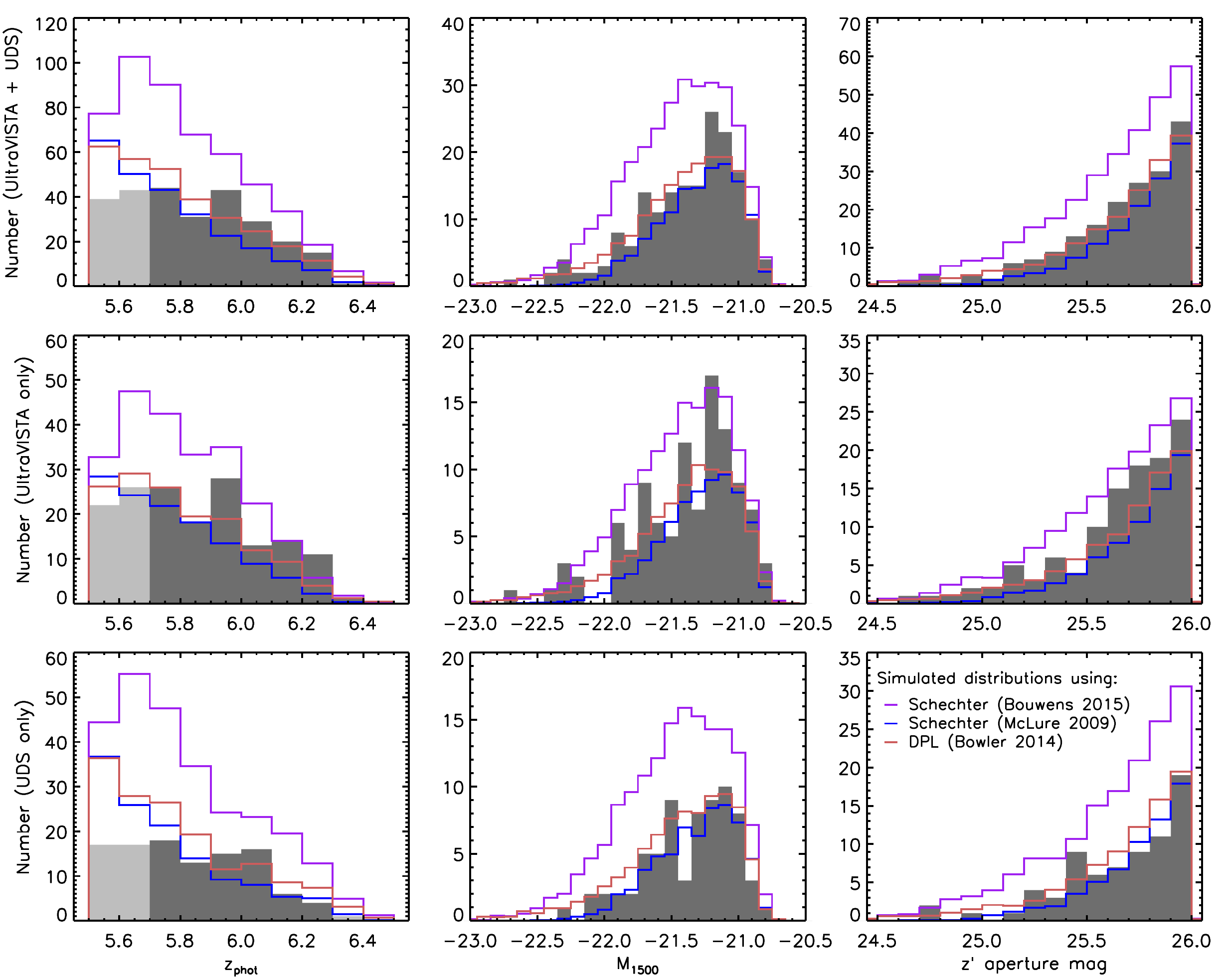}
\caption{The distributions of the $z \simeq 6$ galaxy sample with best-fitting photometric redshift, $M_{\rm 1500}$ and observed $z'$-band aperture magnitude.
The upper row of plots shows the full sample, with the middle and lower rows showing the UltraVISTA/COSMOS and UDS/SXDS samples separately.
The lines show the predicted distributions from the injection and recovery simulations described in Section~\ref{sect:completeness}, for evolving LF models derived from the Schechter function fits of~\citet{McLure2009} and~\citet{Bouwens2015} and the double power-law fits from~\citet{Bowler2014} in blue, purple and orange respectively.
In the $M_{\rm UV}$ and $m_{\rm AB}$ plots we show only the objects in the restricted redshift range $5.7 < z < 6.3$.}
\label{fig:megafigure}
\end{figure*}

In Fig.~\ref{fig:megafigure} we show the redshift distribution for our full sample of $5.5 < z < 6.5$ LBGs, and the $M_{\rm UV}$ and $m_{\rm AB}$ distributions in the restricted redshift range ($5.7 < z < 6.3$) to allow a direct  comparison with our LF determination.
In each panel, predicted distributions from the simulations described in Section~\ref{sect:completeness} for an evolving LF model are shown.
For each simulated distribution, the number of galaxies predicted by a linearly evolving model according to a Schechter function (derived from~\citealt{Bouwens2015} or~\citealt{McLure2009}) or a DPL function (derived from~\citealt{Bowler2014}) were injected into the images, and the resulting $z_{\rm phot}$, $M_{\rm UV}$ and $m_{\rm AB}$ histograms are displayed.
The LF determination of~\citet{Bouwens2015} over-predicts the number of galaxies we should find by around a factor of $\simeq 1.7$, whereas the LF determination of~\citet{McLure2009} is in better agreement, under-predicting the observed number by $\simeq 20$ per cent.
The results of the model using the DPL fits from~\citet{Bowler2014} agree well with the final sample of galaxies we find.
Splitting the sample by field (middle and lower panel for the UltraVISTA/COSMOS and UDS/SXDS samples respectively) reveals an excess of galaxies in the UltraVISTA/COSMOS field as opposed to the~\citet{McLure2009} and~\citet{Bowler2014} models, which is present over a range of redshifts and absolute magnitudes.
In both fields there exist more $M_{\rm UV} < -22.0$ galaxies than predicted by the Schechter function model of~\citet{McLure2009}.

The redshift distribution differs from the model prediction, with a flatter distribution than expected.
Some of the flattening could be a result of too strictly removing galaxy candidates that have good stellar fits, which would cause a drop in the number of objects in the range $5.5 < z < 5.7$.
However, the precise details of the form and evolution of the LF in the range $z = 5$--$7$ are not well constrained, and therefore the model predictions shown are rough estimates of the predicted distributions.

\subsection{Cosmic variance between the fields}\label{sect:cv}

The galaxies in our sample are not uniformly spread between the UltraVISTA/COSMOS and UDS/SXDS fields.
Given that the UltraVISTA DR2 region is only $\sim 84\,$\% of the area of the UDS data, the discrepancy becomes more significant, with the ratio of the surface density between the UltraVISTA DR2 and the UDS $= 1.7$ (or $1.8$ in the redshift range $5.7 < z < 6.3$).
Here we compare the smaller DR2 region of the UltraVISTA field to the UDS, because they have similar depths in the near-infrared and were cut at an identical $z'$-band magnitude ($z' = 26.0$).
The difference is present over a full range of redshifts and magnitudes as shown in Fig.~\ref{fig:megafigure}.
Both~\citet{Willott2013} and~\citet{Bouwens2015} also note an overdensity of galaxies in the UltraVISTA/COSMOS field.
Furthermore, when comparing the derived $\phi^*$ between the five CANDELS fields at $z \simeq 6$,~\citet{Bouwens2015} finds the largest discrepancy occurs between the COSMOS and UDS fields, with the UDS appearing marginally under-dense.

Correcting for the different areas of the surveys ($0.62\,{\rm deg}^2$ vs. $0.74\,{\rm deg}^2$), we find a surface density of $169 \pm 13$ and $95 \pm 10$ galaxies per \degsq~in the UltraVISTA/COSMOS and UDS/SXDS fields respectively (Poisson errors).
We calculated the predicted cosmic variance from the Cosmic Variance Calculator v1.02\footnote{\url{http://casa.colorado.edu/~trenti/CosmicVariance.html}}~\citep{Trenti2008} with a Sheth-Tormen halo MF, $\sigma_{8} = 0.9$ and a unity halo filling factor.
The result of the calculation was an expected error on the average density of $132 \pm 21$ galaxies per \degsq, where both cosmic variance and Poisson errors are included.
Hence the difference in number counts just exceeds that expected due to the large scale structure variations over fields of this size.

The predicted number counts of brown dwarfs are higher in the UltraVISTA/COSMOS field as a result of the galactic coordinates which, if dwarf star contamination was significant, could produce a higher number of objects in our sample in this field.
The simulations described in Section~\ref{sect:dwarfstars} show that dwarf stars can be effectively identified and removed using our stellar fitting procedure and furthermore, the expected pseudo-redshift distribution of dwarf stars (shown in Fig.~\ref{zdist}) does not match that observed in our sample.
The strongest evidence for a genuine discrepancy between the number of objects between the two fields comes from considering the reduced redshift range, which cannot be dominated by stellar contamination due to the large $i-z'$ colour required.
Restricting the redshift range of our sample to the dwarf star free region with $5.7 < z < 6.1$ results in 82, 2 and 61 objects in the UltraVISTA DR2, DR1 and UDS fields, producing a number density ratio of $ \sim 1.6$.

We consider the possibility of gravitational lensing and further large scale structure effects that could be responsible for the field-to-field variance in Section~\ref{sect:discussion}.

\subsection{Overlap with previous studies}\label{sect:previous}

\subsubsection{\citet{McLure2009} and \citet{CurtisLake2013}}\label{sect:mclure}

The first search for $z \gtrsim 5$ galaxies using the UKIDSS UDS near-infrared imaging was undertaken by~\citet{McLure2006}, who found 9 massive $z \simeq 5$ galaxy candidates by combining the SXDS optical imaging with the $J$, $H$ and $K$ images from the UDS early data release.
Using the subsequent release of the UKIDSS UDS imaging (DR1, which is $\sim 1.5$ mag shallower than the data utilised in this work),~\citet{McLure2009} were able to calculate the bright-end of the LF at $z = 5$ and $z = 6$.
In addition, the brightest 14 galaxies from the $z \simeq 6$ sample of~\citet{McLure2009} in the range $6.0 < z < 6.5$ were targeted spectroscopically and the results were presented by~\citet{CurtisLake2013} who detected Lyman-$\alpha$ emission in 11 of the objects. 
Comparing our sample of $z \simeq 6$ galaxies in the UDS, we find 8 of the 10 galaxies presented in~\citet{CurtisLake2013}.
Inspection of our initial catalogues reveals that the two excluded objects, UUDS\_J021922.01$-$045536.3 and UUDS\_J021701.44$-$050309.4, were both removed because they are fainter than our imposed $z'$-band magnitude limit, with 1.8 arcsec aperture magnitudes of $z' = 26.2$ and $z' = 26.4$ respectively.
The brightest object targeted by~\citet{CurtisLake2013} was identified as a faint AGN based on the broad \Lya line~\citep{Willott2009}.
This object was excluded from our galaxy sample based on a $\chi^2 = 12.6$, which slightly exceeds our acceptable criterion.
Closer inspection of the SED reveals that the poor galaxy fit is driven by an enhanced $z'$-band flux, as a result of strong \Lya emission at the very blueward edge of the $z'$-band filter (at $z = 6.01$), which contributes $\sim 70$\% of the $z'$-band flux~\citep{Willott2009}.
The presence of strong \Lya emission sufficient to significantly change the redshift of a galaxy candidate in our sample is unlikely, as the space density of quasars is extremely low (e.g. one object in the full UDS area) and the most luminous LBGs exhibit significantly lower EWs ($EW_{0} \ll 100$\AA,~\citealp{CurtisLake2013, Stark2011}).
We note that the one exception to this rule is the extreme LAE `Himiko', which, consistent with its discovery in a narrow-band survey, stands out as unusual in our SED fitting analysis.

\subsubsection{\citet{Willott2013}}\label{sect:willott}

Using the 4\degsq~of multiwavelength imaging from the CFHTLS `deep' component,~\citet{Willott2013} found 40 $i$-band dropout galaxies at $z \sim 6$.
The CFHTLS data consisted of $u^*, g, r, i$ and $z$-band imaging in 4 separate MegaCam pointings, including the COSMOS field (D2) which is utilised in this work.
When available, further near-infrared data from a variety of different observing programs in each field was utilised (e.g. from the UltraVISTA DR1 and VISTA VIDEO surveys).
\citet{Willott2013} were sensitive to the very brightest $z \simeq 6$ galaxies, with a $z$-band magnitude limit of $25.3$ (in a $2$ arcsec diameter circular aperture), and the imposition of a strict $i - z > 2$ criterion resulted in a higher median redshift of the candidates with $z \gtrsim 5.8$ (see Fig.~\ref{fig:ccplot}) .
In our final $z \simeq 6$ sample we find 7 of the 15 galaxies presented in~\citet{Willott2013} in the COSMOS/UltraVISTA field.
One object was removed only in the final stage of potential star removal, where it was classified as a possible brown dwarf with $\chi_{\star}^2 = 9.1$.
Another object in the sample, WMH18, was not selected in our initial $z'$-band catalogue and inspection of the imaging shows that it is heavily blended with a low-redshift galaxy 3-arcsec away.
The six further missing objects however, were initially selected in our sample but subsequently removed.

The removal of four of the objects (WMH11, WMH12, WMH19 and WMH21) is simply because they lie in the shallower DR1 region of UltraVISTA, where we applied a conservative magnitude limit of $z' < 25.0$.
We nevertheless extracted the photometry for these objects and performed SED fitting as for our $z \simeq 6$ objects, finding that all the objects excluding WHM11 would indeed have been selected as high-redshift galaxies.
WMH11 has low-level flux in the optical bands and was excluded based on our local depth cuts in the $u^*$ and $g$-bands.

Of the final two objects, WMH23 was also removed based on a marginal detection in the $u^*$-band, although the results of SED fitting of this object also show it to be a plausible high redshift candidate.
We note here that occasionally genuine $z \simeq 6$ objects will be lost during our selection process as a result of our optical drop-out criterion, however this incompleteness is carefully simulated and included in our LF analysis.
Finally, object WMH14 was excluded because it has a marginally unacceptable $\chi^2 = 12.6$, which exceeds our formal good fit criterion ($\chi^2 < 11.3$).
Inspection of the imaging reveals the object to be close to a low-redshift companion, which is likely contaminating the photometry for this galaxy.

Two of the 15 objects presented in the $z \simeq 6$ sample from~\citet{Willott2013} in the COSMOS field have spectroscopic confirmations, WMH13 at $z_{\rm spec} = 5.983$ and WMH15 at $z_{\rm spec} = 5.847$.
Reassuringly we find photometric redshifts of $z = 5.92_{-0.08}^{+0.13}$ and $z = 5.99_{-0.23}^{+0.09}$ respectively.

We therefore find good agreement with the bright sample of $z \simeq 6$ galaxies found in~\citet{Willott2013} with 12 out of the 15 objects present being consistently classified as high-redshift galaxies here.
However as presented in Section~\ref{sect:lf}, the derived rest-frame UV LF from~\citet{Willott2013} falls below our determination, with the difference in derived number densities suggesting that a large proportion ($\sim 60$ per cent at $M_{\rm UV} \simeq -22.0$) of objects selected in this work are not present in the~\citet{Willott2013} sample.
Using the new, deeper, optical and near-infrared photometry available for the UltraVISTA/COSMOS field in this study we applied the~\citet{Willott2013} selection criterion to the full $z \simeq 6$ sample we derive using photometric redshift fitting.
Of the 159 LBGs we find in the UltraVISTA/COSMOS field, 31 are sufficiently bright ($z < 25.3$, 2-arcsec diameter circular apertures) to have been included by~\citet{Willott2013}, and of these, 16 objects now also pass the $i - z > 2$ criterion imposed by~\citet{Willott2013}.
Although these 16 objects pass the required selection criterion, only 5 were previously found by~\citet{Willott2013}, suggesting that the origin of the discrepancy between the $z \simeq 6$ LF determinations by~\citet{Willott2013} and this work is a result of the selection procedure employed using the shallower CFHTLS $z$-band and near-infrared imaging.
In particular the strict $i - z > 2$ colour selection criterion applied by~\citet{Willott2013} was not robustly applicable given the relative depths of the $i$ and $z$-band imaging available, likely resulting in the exclusion of many genuine LBGs at $z \simeq 6$ as demonstrated above.

\subsubsection{Bowler et al. (2014)}

In~\citet{Bowler2014}, we presented a sample of $z \simeq 7$ LBGs  found within the UltraVISTA/COSMOS and UDS/SXDS datasets utilised in this work.
As part of the selection procedure, which was primarily aimed at finding $6.5 < z < 7.5$ galaxies, candidates with $z > 6$ were retained and presented if the presence of \Lya emission in the spectrum could shift the object into the primary redshift window.
We therefore expect some overlap with the sample presented in~\citet{Bowler2014}.
Taking the subset of the 34 objects presented in~\citet{Bowler2014} that have $z' < 26.0$, we find 6 galaxies in the UltraVISTA field and 1 in the UDS (the spectroscopically confirmed LAE `Himiko' at $z = 6.595$).
Comparing to the $z \simeq 6$ sample, we find all 7 objects (Himiko was identified and removed from the final sample), with photometric redshifts that agree within the errors.
The full recovery of these objects in the present work is a strong vindication of our selection methodology, as the samples presented in~\citet{Bowler2014} were selected in a different band ($Y$ and $J$-band selected) and refined using a slightly different SED fitting analysis.

\subsection{Rest-frame UV slope (${\boldsymbol \beta_{\rm UV}}$)}\label{beta}

\begin{figure}
\includegraphics[width = 0.47\textwidth]{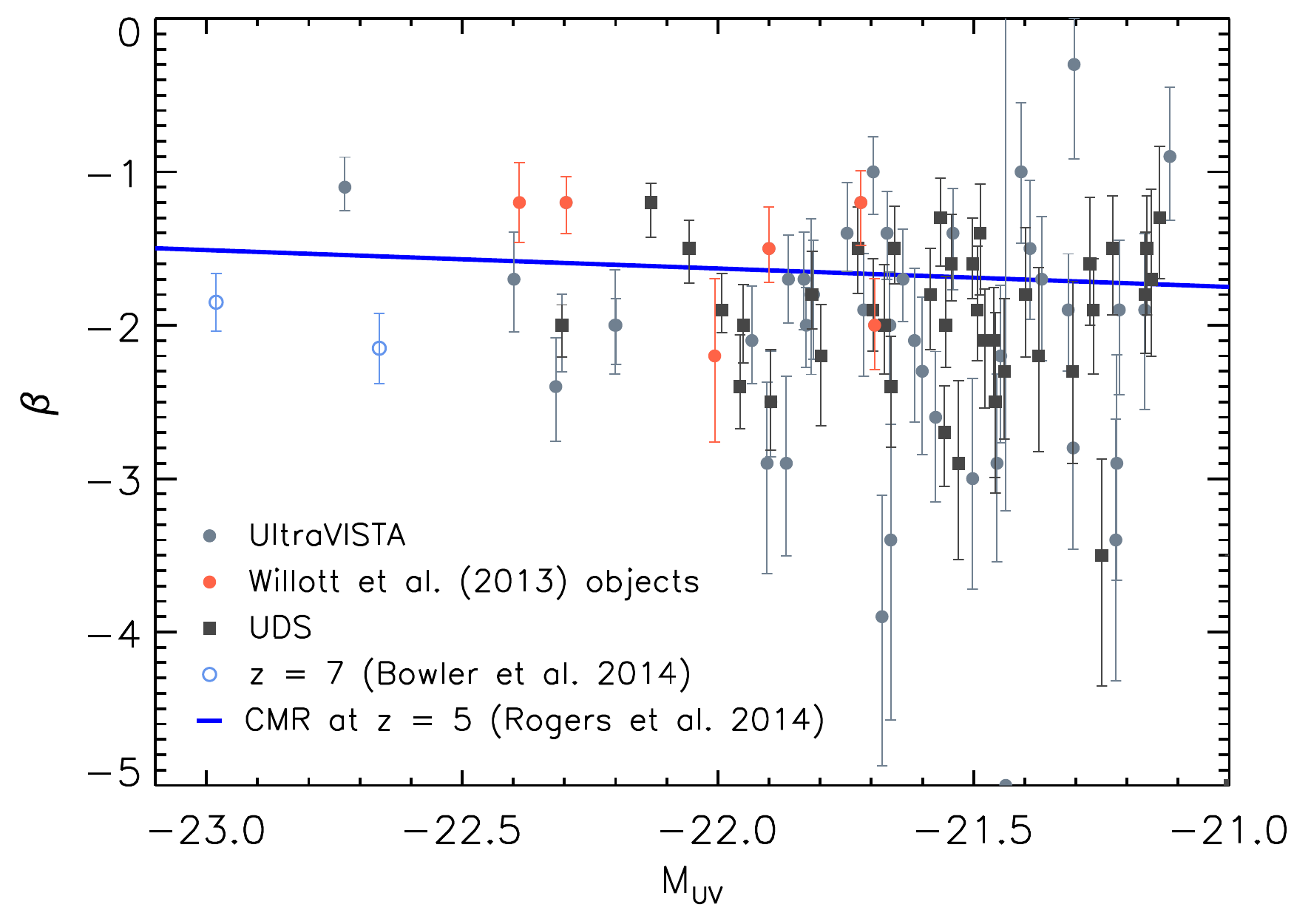}
\caption{The rest-frame UV slope ($\beta$) measured for our sample in the reduced redshift range $5.7 < z < 6.3$, plotted against absolute UV magnitude ($M_{\rm UV}$).
The galaxies shown have a detection in the $Y,J,H$ or $K_{s}$ bands at greater than $5\sigma$ significance.
Objects in the UltraVISTA/COSMOS field are shown as grey circles, with the galaxies previously detected by~\citet{Willott2013} highlighted in red, and the objects in the UDS/SXDS field are shown as dark grey squares.
The colour magnitude relation at $z \simeq 5$ determined by~\citet{Rogers2014} is shown as the blue line, and the derived $\beta_{\rm UV}$ values for the two brightest $z \simeq 7 $ galaxies from~\citet{Bowler2014} are shown as open blue circles.
}
\label{fig:beta}
\end{figure}

The rest-frame UV slope, $\beta_{\rm UV}$, of each galaxy was measured by fitting a power law (parameterised as $F_{\lambda} \propto \lambda^{\beta_{\rm UV}}$) to the $YJHK$ photometry for each object.
The $z'$-band was excluded from the fitting process, as at $z \gtrsim 5.8$ the Lyman-break is moving through the filter and furthermore there could be contamination by Lyman-$\alpha$ emission.
In Fig.~\ref{fig:beta} we show the derived $\beta_{\rm UV}$ values for the 87 LBGs in the $5.7 < z < 6.3$ range that have detections in one near-infrared band at greater than $5\sigma$ significance, where the reduced redshift range was chosen to allow more direct comparison with the results of~\citet{Willott2013}.
Using a sub-set of the 40 objects covered by sufficiently deep near-infrared data,~\citet{Willott2013} was able to measure $\beta_{\rm UV}$ using an identical method, finding a mean value of $\beta_{\rm UV} = -1.38 \pm 0.2$, which is redder than that found for fainter galaxies which tend to exhibit $\beta_{\rm UV} \simeq -2.0$~\citep{Dunlop2013, Rogers2013, Bouwens2014beta}.
We measured $\beta_{\rm UV}$ for the 7 objects from the~\citet{Willott2013} that are present in our sample, using the deeper near-infrared imaging now available, and highlight these objects as red points in Fig.~\ref{fig:beta}.
Excluding the faintest object (WHM22) that has a poorly constrained $\beta_{\rm UV}$ value (and is not included in Fig.~\ref{fig:beta} due to low significance near-infrared detections), we find a mean $\beta_{\rm UV} = -1.55 \pm 0.05$ for the sub-set of the~\citet{Willott2013} sample, where the error is the standard error on the mean.
For the galaxies in our sample with $M_{\rm UV} < -22.0$ we find on average slightly bluer values with a mean $\beta_{\rm UV} = -1.8 \pm 0.1$ (excluding the brightest object as discussed below).
For the brightest galaxies in our sample therefore, we do find redder rest-frame UV slopes than in similarly bright galaxies at $z \simeq 7$ (by $\Delta \beta_{\rm UV} \simeq 0.2$), however our full sample does not show the particularly red average $\beta_{\rm UV}$ found by~\citet{Willott2013}.
At $M_{\rm UV} \lesssim -22$ where our $\beta_{\rm UV}$ values are sufficiently accurate, our results follow the colour-magnitude relation derived at $z \simeq 5$ by~\citet{Rogers2014} well, and tentatively show an increased scatter as expected from their analysis, although careful modelling of potential biases in the selection process are required to show this quantitatively.

The very brightest object in our sample lies within the UltraVISTA/COSMOS field and shows a particularly red slope ($\beta_{\rm UV} = -1.1 \pm 0.2$), in contrast to the bluer values ($\beta_{\rm UV} \simeq -2.0$) found for the very brightest $z \simeq 7$ galaxies~\citep{Bowler2014}.
Larger samples are clearly needed, however an increase in dust obscuration for the brightest objects in our sample is one theoretical process by which the number density at the bright end of the rest-frame UV LF could be suppressed (e.g. see comparison with theoretical models in Section~\ref{sect:sims}).
The rest-frame UV slope can be linked to the predicted attenuation in the rest-frame UV according to the `IRX-$\beta_{\rm UV}$ relation'~\citep{Meurer1999}, where the IRX is the `infrared excess' and is given by $L_{\rm IR}/L_{\rm UV}$.
While a $\beta_{\rm UV} \simeq -2.0$ as observed at $z \simeq 7$ suggest little or no dust attenuation, if the updated relation of~\citet{Takeuchi2012} is assumed, we predict an attenuation in the range $A_{\rm UV} \simeq  1.5$--$0.7$ for rest-frame UV slopes of $\beta_{\rm UV} = -1.0$ to $ -1.5$ (as we observe in some of the brightest galaxies in our sample).
The predicted total far-infrared luminosity implied for objects in this $\beta_{\rm UV}$ range according to the IRX-$\beta_{\rm UV}$ relation is $L_{\rm TIR} \simeq 0.4$--$1.5\times 10^{11}\,{\rm L}_{\odot}$, similar to the observed luminosity of two $z \simeq 6$ galaxies found by~\citet{Willott2015}.

\section{The luminosity function}\label{sect:discussion}

\begin{table}
\caption{The binned rest-frame UV LF points at $z_{\rm med} \sim 5.9$ from this work, as shown in Fig.~\ref{fig:lf}.
The upper section of the table shows the results from the UltraVISTA/COSMOS and UDS/SXDS fields combined, with the middle and lower sections showing the results including only the UltraVISTA/COSMOS and UDS/SXDS fields respectively (e.g. the inset in Fig.~\ref{fig:lf}).
Columns 1 and 2 show the central $M_{\rm UV}$ of the bin and the width, where we calculate the $M_{\rm UV}$ by integrating the best-fitting SED through a top-hat filter centred on 1500\AA~with a width of 100\AA.
The weighted bin centre is shown in Column 3, given by the median completeness corrected $M_{\rm UV}$ of the galaxies in that bin.
The number density is shown in Column 4 and the number of galaxies in each bin is shown in Column 5.
}
\begin{tabular}{c c c c c}
\hline
 $M_{\rm UV}$  & $\Delta M_{\rm UV} $ &  $M_{\rm UV, {\rm w}}$ & $\phi$ & \# \\[3pt]
 /mag & /mag & /mag &  /mag/Mpc$^3$& \\

\hline
$-22.625$ & $  0.500$ & $ -22.52$ & $1.16\pm 0.67 \times 10^{-6}$ & $  3$\\
$-22.125$ & $  0.500$ & $ -22.08$ & $5.98\pm 1.64 \times 10^{-6}$ & $ 17$\\
$-21.750$ & $  0.250$ & $ -21.74$ & $1.90\pm 0.41 \times 10^{-5}$ & $ 23$\\
$-21.500$ & $  0.250$ & $ -21.49$ & $3.92\pm 0.70 \times 10^{-5}$ & $ 35$\\
$-21.250$ & $  0.250$ & $ -21.22$ & $9.14\pm 1.39 \times 10^{-5}$ & $ 49$\\
\hline
$-22.625$ & $  0.500$ & $ -22.52$ & $2.20\pm 1.27 \times 10^{-6}$ & $  3$\\
$-22.125$ & $  0.500$ & $ -22.11$ & $7.60\pm 2.92 \times 10^{-6}$ & $ 10$\\
$-21.750$ & $  0.250$ & $ -21.75$ & $2.92\pm 0.76 \times 10^{-5}$ & $ 16$\\
$-21.500$ & $  0.250$ & $ -21.48$ & $4.76\pm 1.17 \times 10^{-5}$ & $ 19$\\
$-21.250$ & $  0.250$ & $ -21.22$ & $1.34\pm 0.25 \times 10^{-4}$ & $ 31$\\
\hline
$-22.125$ & $  0.500$ & $ -22.04$ & $4.54\pm 1.74 \times 10^{-6}$ & $  7$\\
$-21.625$ & $  0.500$ & $ -21.57$ & $2.15\pm 0.47 \times 10^{-5}$ & $ 23$\\
$-21.250$ & $  0.250$ & $ -21.22$ & $5.54\pm 1.45 \times 10^{-5}$ & $ 18$\\
\hline
\end{tabular}
\label{table:lf}
\end{table}

\begin{figure*}

\includegraphics[width = 0.65\textwidth]{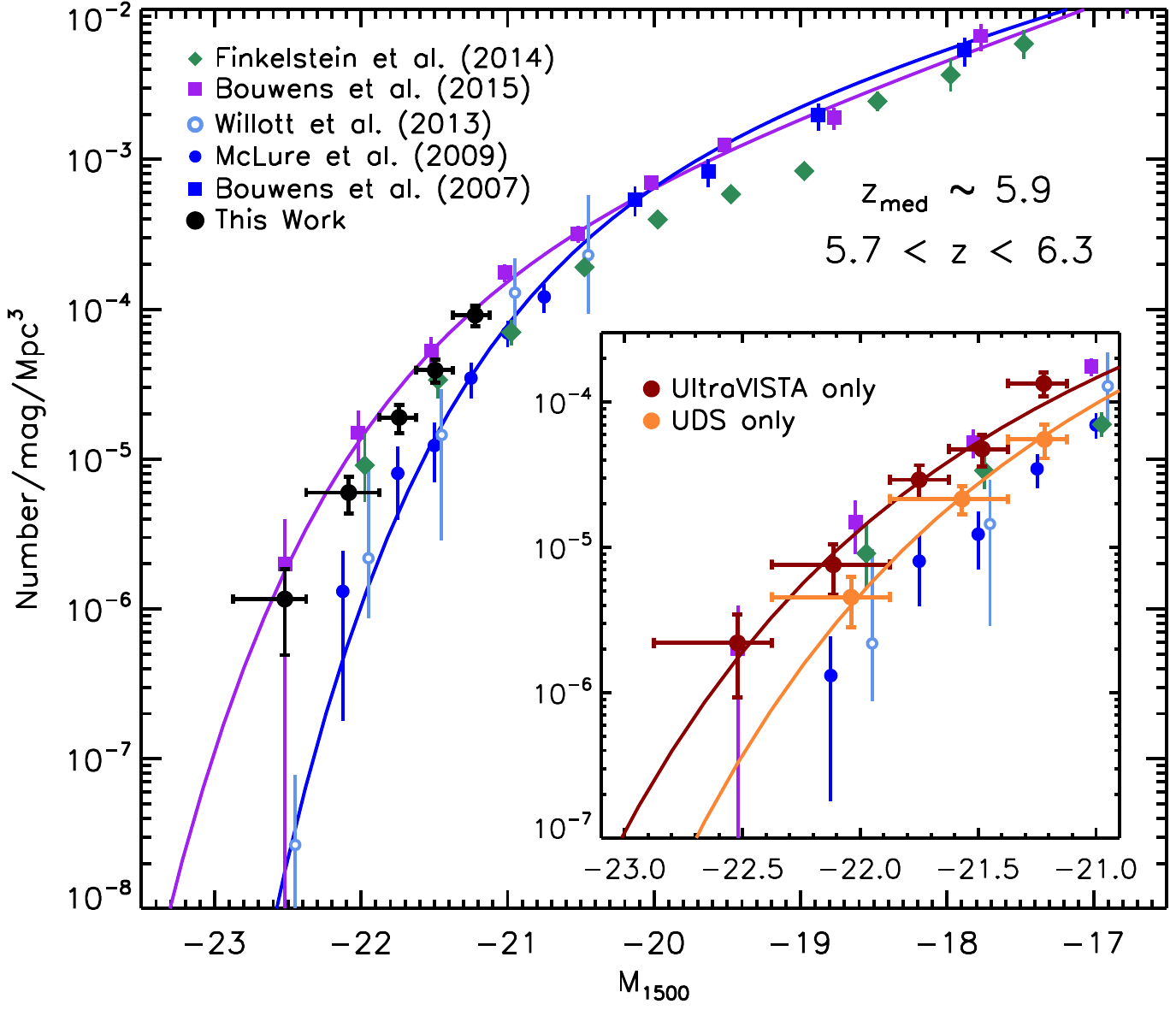} 

\caption{
The rest-frame UV LF at $z \simeq 6$, showing our results from the combined UltraVISTA/COSMOS and UDS/SXDS datasets as the black circles.
The redshift range is restricted to $5.7 < z < 6.3$ as in~\citet{McLure2009}, which results in a median redshift of $z_{\rm med} \sim 5.9$.
The inset plot shows our results from the UltraVISTA/COSMOS and UDS/SXDS fields separately as the red and orange points respectively, with the best fitting Schechter functions overplotted in identical colours.
Here we have not corrected the absolute magnitudes for dust extinction.
We have shifted the~\citet{Willott2013} points by $0.05\,$mag faint-ward for clarity.}
\label{fig:lf}
\end{figure*}

In Fig.~\ref{fig:lf} and Table~\ref{table:lf} we present our measured rest-frame UV LF at $z \simeq 6$.
The binned LF points were derived from 127 luminous LBGs with $M_{\rm UV} \le -21.125$ found within the combined UltraVISTA and UDS imaging, in the redshift range $5.7 < z_{\rm phot} < 6.3$.
The median redshift of the galaxies included in our LF determination is $z_{\rm med} = 5.9$.
Comparing to previous determinations of the $z \simeq 6$ LF from a compilation of \emph{HST} imaging from~\citet{Bouwens2007, Bouwens2015}, the larger area available from the combined UltraVISTA/COSMOS and UDS/SXDS fields allows us to more accurately probe lower space densities of objects (down to $\sim 1 \times 10^{-7}$ /mag/Mpc$^3$).
Furthermore, the error bars on our brightest points show that we are able to probe the number densities of the brightest galaxies more accurately than the previous determinations using ground-based imaging surveys from~\citet{Willott2013} and~\citet{McLure2009}.

The use of two independent fields in the present analysis also allows us to probe the cosmic variance and potential large scale structure effects in the number counts of bright objects.
~\citet{Willott2013} pointed out an over-density of $z \simeq 6$ galaxies in the COSMOS/UltraVISTA (CFHTLS D2) field, an observation that we are able to confirm using $\times 10$ the number of LBGs.
The inset plot in Fig.~\ref{fig:lf} shows our results at $z \simeq 6$ determined from the two fields separately.
There is a clear excess of galaxies in the COSMOS/UltraVISTA survey as compared to the UDS/SXDS field, which is present over the full magnitude range probed.
The discrepancy is most noticeable in the faintest bin at $M_{\rm UV} = -21.5$ where the number counts differ by a factor of $\gtrsim 2$.
The discrepancy between the fields is also evident in the observed $M_{\rm UV}$ histograms in Fig.~\ref{fig:megafigure}.
We note that our faintest bin is our most uncertain, however if lensing by foreground structures (as discussed in Section~\ref{sect:clusters}) was a factor in the increased number counts in UltraVISTA/COSMOS then we would expect the largest difference at the faint end of our sample, due to the rapidly increasing number counts of objects faint ward of the limiting magnitude.

\subsection{Comparison to previous work}

Previous determinations of the $ z \simeq 6$ rest-frame UV LF from~\citet{McLure2009},~\citet{Willott2013},~\citet{Bouwens2007,Bouwens2015} and~\citet{Finkelstein2014} are shown in Fig.~\ref{fig:lf}, with the best-fitting Schechter functions from~\citet{McLure2009} and~\citet{Bouwens2015} also shown.
The derived data points faint-ward of $M_{\rm UV} = -20.5$ show good agreement (with the exception of~\citealp{Finkelstein2014} in the range $-20.0 \gtrsim M_{\rm UV} \gtrsim -19.0$), although there is tension between the Schechter function fits derived by~\citet{McLure2009} and~\citet{Bouwens2015}.
At $M_{\rm UV} \lesssim -21$ however,  large discrepancies between the determination of~\citet{Bouwens2015} and the ground-based results of~\citet{McLure2009} and~\citet{Willott2013} become evident.
Using our wide-area \emph{and} deep dataset we find a $z \simeq 6$ LF that lies approximately midway between the previous determinations.

\subsubsection{\citet{McLure2009}}

Using the DR1 of the UKIDSS UDS near-infrared data in the UDS/SXDS field (Section~\ref{sect:mclure}),~\citet{McLure2009} determined the $z \simeq 5$ and $6$ LF following an analogous methodology to this work.
Comparing our determination of the LF from the combined UltraVISTA/COSMOS and UDS/SXDS datasets to the results of~\cite{McLure2009} however, we find a significantly higher number density of bright LBGs.
The difference is further highlighted because the Schechter function fit of~\citet{McLure2009} under-shoots the brightest two binned points from their work.
Considering our derived LF from the two fields separately (inset of Fig.~\ref{fig:lf}) sheds light on the discrepancy, as the results from this work exclusively in the UDS/SXDS field are in-fact in fair agreement (within $\simeq 1\sigma$) with the data points of~\citet{McLure2009}.
Overall there is still a lower number of objects found by~\citet{McLure2009}, which is likely due to the shallower near-infrared photometry available, resulting in a more conservative selection procedure being employed to ensure the removal of low-redshift interlopers or dwarf stars without secure near-infrared colours.

\subsubsection{\citet{Willott2013}}

The results of~\citet{Willott2013}, which were derived from 40 galaxies at $z \simeq 6$ found in the four CFHTLS fields (as described in Section~\ref{sect:willott}), are in good agreement with those of~\citet{McLure2009} and therefore fall below our determination.
The four independent fields analysed should make the~\citet{Willott2013} result more robust to the potentially large cosmic variance in the number counts of bright LBGs as found in this study.
Furthermore, one of the CFHTLS fields analysed by~\citet{Willott2013} overlaps with the UltraVISTA/COSMOS field, which we find to be over-dense compared to the UDS/SXDS, a result also corroborated by~\citet{Willott2013} who found 15 galaxies in UltraVISTA/COSMOS as compared to $\sim 8$ in each of the three other fields.
We note that two of the four fields used only had shallow $J$-band data available (D3 and D4), and hence the number counts here are the most uncertain (8 and 9 objects respectively).
Taking these uncertainties into account, it remains possible that either the UltraVISTA/COSMOS field is over-dense \emph{or} the true global number densities of these galaxies was not adequately probed by CFHTLS due to the inhomogeneous datasets (e.g. the four CFHTLS fields should have had $\sim 11.5$ galaxies in each).
Potential biases in the~\citet{Willott2013} selection were highlighted in the comparison between the~\citet{Willott2013} sample and that derived from this work, and in our rest-frame UV slope measurements (see Section~\ref{sect:sample}).

In particular, the very brightest datapoint derived by~\citet{Willott2013} at $M_{\rm UV} = -22.5$ disagrees strongly with our derived number density of galaxies.
However, as discussed in~\citet{Bowler2014},~\citet{Willott2013} do not directly measure such a low space density, rather they infer $\phi(M)$ using a maximum likelihood approach.
In~\citet{Bowler2014} we estimated the space density of galaxies in the~\citet{Willott2013} analysis at $M_{\rm UV} = -22.5$ using the brightest two objects presented (WHM5 and WHM29, neither of which is in the COSMOS/UltraVISTA field).
The result ($\phi \simeq 2.0 \pm 1.4 \times 10^{-7}$ /mag/Mpc$^3$), is still in tension with our brightest point, however the deeper $z'$-band and near-infrared imaging utilised here would suggest that our results are more reliable.

\subsubsection{\citet{Bouwens2015}}

Using a combination of the HUDF and two parallel fields, the Early Release Strip (ERS) and the CANDELS survey fields, \citet{Bouwens2015} selected a large sample of LBGs at $z \simeq 6$.
The~\emph{HST} surveys used in the analysis covered a total of 0.2\degsq~on the sky.
The LF determination derived from the full set of fields available is included in Fig.~\ref{fig:lf} along with the best fitting Schechter function.
The data points and the fit lie above our determination from the UltraVISTA/COSMOS and UDS/SXDS fields, and furthermore the simple evolving LF model we use assuming the~\citet{Bouwens2015} LFs at $z = 5$ and $z = 6$ predicts approximately double the number of galaxies than we find in these fields (see Fig.~\ref{fig:megafigure}).
Due to the relatively small area of the fields used by~\citet{Bouwens2015}, the number of $z \simeq 6$ galaxies at $M_{\rm UV} \simeq -22.5$ is poorly constrained, however at $M_{\rm UV} = -22.0$ there is a clear tension with our results.

At the very bright end of the LF however, the small area probed by~\citet{Bouwens2015} results in the samples being vulnerable to strong cosmic variance, as is evident from the distribution of the number counts of bright objects across the fields used (figure 14 and table 8 of~\citealt{Bouwens2015}), which can vary by up to $50$ per cent.
Furthermore, the small number statistics result in a large Poisson error as demonstrated by the brightest point shown in Fig.~\ref{fig:lf}.
Hence, the results of~\citet{Bouwens2015} cannot be relied upon in this magnitude regime ($M_{\rm UV} \lesssim -21$).
The area, depth and homogeneity of the data-sets utilised in this work enables a significantly improved determination of the bright-end of the LF at $z \simeq 6$ than can be provided by the combination of current and future~\emph{HST} surveys.
Finally, for the~\citet{Bouwens2015} results to be correct at the bright end of the LF we would expect to find roughly double the number of LBGs, which at $z' \sim 25.0$ would be individually detected at a significance of $> 20\sigma$.

\subsubsection{\citet{Finkelstein2014}}

From a combination of the UDF and parallels, the two GOODS fields and parallel imaging taken as part of the~\emph{Hubble} Frontier Fields program,~\citet{Finkelstein2014} selected a sample of $z = $4--$8$ LBGs using a photometric redshift fitting methodology.
In total, the area included was $\simeq 300\,{\rm arcmin}^2$, and the derived LF points are shown in Fig.~\ref{fig:lf}.
The results of~\citet{Finkelstein2014} are in excellent agreement with our determination of the bright end of the LF, however they appear to diverge from the results of~\citet{Bouwens2007} and~\citet{Bouwens2015} at fainter magnitudes, and furthermore show a step at $M_{\rm UV} \simeq -19.0$.
Although the~\citet{Finkelstein2014} analysis used a sub-set of the larger area of imaging used by~\citet{Bouwens2015},  the results are in good agreement with the ground-based analysis presented here, and could indicate large over-densities in the additional fields incorporated by~\citet{Bouwens2015} or contamination of the~\citet{Bouwens2015} sample by brown dwarfs (see the discussion in~\citealt{Finkelstein2014}).

\subsection{Gravitational lensing by foreground galaxies}\label{gravitationalensing}

As in~\citet{Bowler2014}, to determine whether the bright galaxies we find are only present in our sample as a result of `moderate' gravitational lensing by foreground galaxies, we estimated the expected magnification due to galaxies along the line of sight to each $z \simeq 6$ object in our sample.
Note that strong gravitational lensing by galaxies directly along the line of sight is ruled out by our deep optical non-detections.
The full details of our approach are described in Appendix~\ref{appendixgrav}.
We find that the galaxies in our sample show magnifications in the range $\Delta m_{\rm AB} = 0.0$--$0.6\,{\rm mag}$, with a median magnification of $\Delta m_{\rm AB} \simeq 0.1$--$0.2$.
However, the sample does not show `excess' lensing compared to random positions in the UltraVISTA/COSMOS or UDS/SXDS imaging, and no correlation between the magnification and the absolute magnitude of the galaxy was identified.
As the median magnification is relatively small (less than the LF bins we use) and likely an upper limit as a consequence of the assumed Faber-Jackson relation in the calculation of the line-of-sight mass distribution, we do not correct for the lensing effect in our LF analysis.
Furthermore, previous LF determinations at high-redshift do not generally correct for the magnification, and recent work by~\citet{Mason2015} has shown that the effect is small in the magnitude range currently probed by observations, and that the corrected Schechter function parameters agree with the results assuming no lensing effect within the errors.
The accurate determination of the lensing magnification and the impact on the observed LF will be extremely important however for future, wider area surveys, such as those performed by Euclid.

\subsection{The potential effect of large scale structure}\label{sect:clusters}

As described in Section~\ref{sect:sample}, we find a difference in the number counts of galaxies between the two fields studied, with the UltraVISTA/COSMOS field containing $\simeq 1.8$ times the number of objects found in the UDS/SXDS field.
Cosmic variance struggles to account for the difference, and hence further investigation is warranted.
The UltraVISTA/COSMOS field has been known to harbour an unusual richness of structure as measured by clustering analyses, particularly at $z \simeq 1$~\citep{McCracken2007, Meneux2009,Skibba2014}.
The $z \simeq 6$ objects here show no evidence for `excess' lensing by foreground galaxies as compared to a random position on the sky, and the UltraVISTA/COSMOS fields shows a similar predicted magnification distribution to the UDS/SXDS field.
However, such a calculation does not include gravitational lensing by galaxy clusters.
A full analysis of the lensing cross-section for the two fields is beyond the scope of this work, however an estimate of the magnification by the total matter in clusters can be made using the X-ray observations available in the fields.
The UltraVISTA/COSMOS and UDS/SXDS fields have been observed to similar depths ($\simeq 2$--$3 \times 10^{-15}\,{\rm ergs}\,{\rm cm}^{-2}{\rm s}^{-1}$ at $0.5$--$2\,$keV) with \emph{XMM-Newton}, and the X-ray bright clusters have been identified by \citet{Finoguenov2007} and~\citet{Finoguenov2010} in each field respectively.
From the X-ray luminosity we calculated $M_{200}$\footnote{The mass enclosed within a sphere of radius $R_{200}$, which contains a density of $200\, \rho_{\rm critical}$ at that redshift.} using the correlation presented by~\citet{Rykoff2008}.
The magnification due to the total matter present in clusters was then calculated using the SIS approximation as described in Section~\ref{gravitationalensing}, summing the contribution from the clusters at the position of each $z \simeq 6$ galaxy in our sample.

The cluster X-ray luminosity function reveals a higher density of X-ray luminous clusters in the UltraVISTA/COSMOS field compared to the UDS/SXDS~\citep{Finoguenov2010}, and calculating the predicted magnification from the clusters using the simple method described shows that the additional lensing magnification due to the difference in number density of high mass clusters is of the order of $\sim 0.05\,$mag.
Although small, such a magnification could have a significant effect on the determination of the bright end of the LF due to the declining number counts, and could be the origin of the discrepancy we find close to the $5\sigma$ limit of our survey.
Correcting our derived LF points by $\sim 0.05$ mag faint-ward would not impact on our conclusions described below, and would further strengthen the derived evolution in $M^*$.

Conversely, narrow-band studies of the UDS/SXDS field have revealed large voids in the distribution of $z = 5.7$ galaxies~\citep{Ouchi2005}, with comoving sizes of the order $10$--$40$ Mpc which corresponds to $\sim 4-17$ arcminutes on the sky (see figure 2 of~\citealp{Ouchi2005}).
Hence it remains possible that the $z\simeq 6$ LF derived from the UDS/SXDS field is biased low as a result of these voids at $z = 5.7 \pm 0.1$.

The discovery of cosmic variance between degree-scale fields at $z \simeq 6$ in this work further highlights the necessity of using multiple large fields to robustly determine the number density of bright galaxies.
A single, or even a collection of CANDELS fields could be strongly influenced by this large scale structure as a result of their small size.
Given the potential foreground structure in the COSMOS/UltraVISTA field and/or lack of structure in the UDS/SXDS field discussed above, future work on additional fields will be required to shed light on the origin of the discrepancy between them.

\section{Form and evolution of the UV LF}\label{sect:evolution}

\subsection{The functional form of the  ${\bf z \simeq 6}$ LF}\label{sect:fitting}

\begin{figure}
\includegraphics[width = 0.48\textwidth]{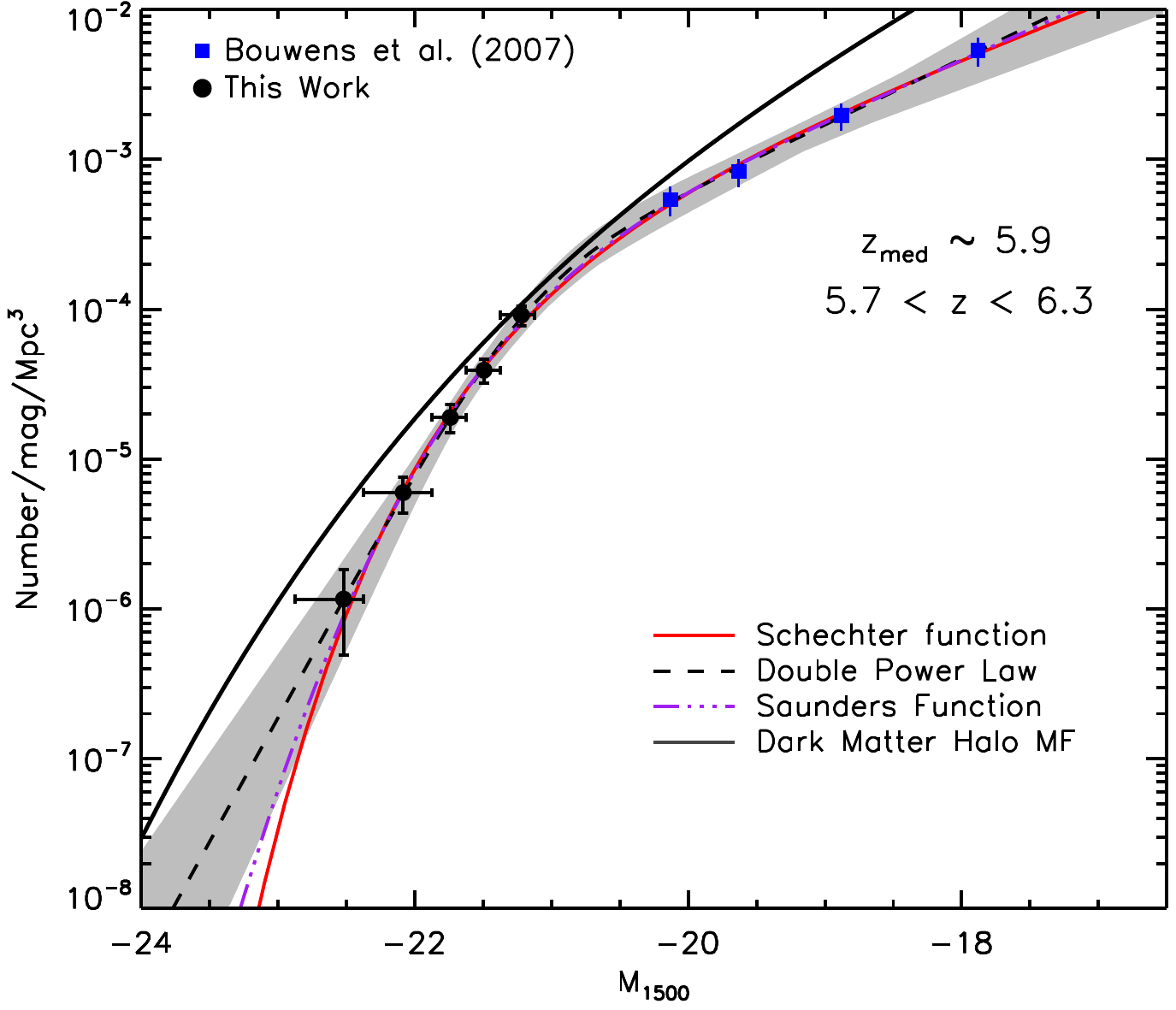}
\caption{The $z \simeq 6$ LF points from this work and~\citet{Bouwens2007}, with the best fitting Schechter (red solid line), double power law (black dashed line) and Saunders function (blue dot-dashed line) shown.
The grey shaded region shows the one-sigma confidence interval on the double power law fit.
The dark matter halo mass function, scaled as described in the text, is shown as the solid grey line.
}
\label{fig:functions}
\end{figure}

\begin{table*}
\caption{The best-fitting parameters derived from fitting the observed $z \simeq 6$ LF.
The results are shown for a Schechter function, double power-law and Saunders function, which are displayed in Fig.~\ref{fig:functions}.
The errors are the one sigma errors on that parameter, where the $\chi^2$ value has been minimised over all other parameters.
}
\begin{tabular}{l c c c c c c}
\hline
Function & $\phi^*$ & $M^*$ & $\alpha$ & $\beta$ & $\sigma$ \\
& /mag/Mpc$^3$ & /mag & &  \\
\hline

Schechter function & $5.7_{ -2.0}^{+  2.7}  \times 10^{-4}$ & $-20.77_{ -0.19}^{+  0.18}$ & $ -
1.88_{ -0.14}^{+  0.15}$ &  -- & -- \\[3pt]
Double power law & $1.9_{ -0.8}^{+  1.2}  \times 10^{-4}$ & $-21.20_{ -0.22}^{+  0.22}$ & $ -2.10_{
 -0.14}^{+  0.16}$ & $  -5.1_{  -0.6}^{+   0.5}$ & --\\[3pt]
Saunders function & $3.0_{ -2.3}^{+  6.1}  \times 10^{-4}$ & $-21.04_{ -1.12}^{+  0.91}$ & $ -2
.01_{ -0.17}^{+  0.21}$ &  -- &$   0.2_{  -0.1}^{+   0.1}$\\[3pt]

\hline
\end{tabular}
\label{table:fitting}
\end{table*}

The galaxy LF at high redshift is commonly fitted with a Schechter function, with the resulting best-fitting parameters used to determine the dominant form of the evolution(e.g.~\citealp{Bouwens2015, Bouwens2012,Bouwens2007, Finkelstein2014, McLure2013}).
A Schechter function (functional form detailed in Appendix A) tends to describe well the mass and luminosity functions at low redshift (e.g.~\citealp{Baldry2012, Loveday2012, Kelvin2014, Mortlock2014}).
Similarly at higher redshifts a Schechter function has also provided a good fit to the observations (e.g.~\citealp{McLure2013, Schenker2013, Bouwens2015, Finkelstein2014}), although qualitatively this is not surprising given the often reduced dynamic range and increased errors in LF determinations of $z > 4$ LBGs.
In contrast, at $z \simeq 7$~\citet{Bowler2014} showed that a double power-law provides a better description of the rest-frame UV LF, from measurements of the number of bright galaxies with $M_{\rm UV} \lesssim -21.5$.
Furthermore, theoretical models do not generally predict a Schechter function-type form without the addition of uncertain dust or feedback processes (e.g.~\citet{Cai2014, Dayal2014}, see Section~\ref{sect:sims}).
It is therefore important to consider alternatives to a Schechter function, if the data warrant such a conclusion, and to be aware that assuming a given functional form for the LF can potentially hide subtleties in any derived LF evolution~\citep{Jaacks2013}.

Indeed at lower redshift, a broken power-law or double power-law has been shown to provide a better description of the LF of galaxies in groups (e.g.~\citealp{Tempel2014,Tempel2009}), the far-infrared galaxy luminosity functions (e.g.~\citealp{Soifer1987}) and the LFs of quasars (e.g.~\citealp{McGreer2013}).
A shallower decline at the bright end of the LF than expected from a Schechter function has also been found in the NUV LF from the Wiggles survey~\citep{Jurek2013} and in the H$\alpha$ LF~\citep{Gunawardhana2013}.
From a theoretical standpoint,~\citep{Salim2012} has shown that scatter in the mass-to-light ratio of galaxies will naturally lead to a shallower function when measuring the luminosity function.
If the mass function is well described by a Schechter function, then one would expect any luminosity function measurement that directly traces the galaxy mass (e.g. the rest-frame optical) would also follow such a form~\citep{Bernhard2014}.
However, when the luminosity of the galaxies in question is measured from wavelength regimes that are dominated by recent star formation, and hence trace the SFR rather than the mass directly, then the shape of the observed LF will be convolved with the mass-to-SFR relation of the galaxies~\citep{Salim2012}.
The scatter in this relation tends to flatten the slope of the LF, resulting in a shallower function.
\citet{Salim2012} showed that a Saunders functional form (detailed in Appendix~\ref{sect:equations}, derived originally by~\citealp{Saunders1990} to model the $60\umu {\rm m}$ LF), where the bright-end of the LF declines as a log-normal, provides an improved fit to a LF that follows the SFR of the galaxies.

In addition, there are several observations effects and biases that can also lead to an apparent deviation from a Schechter function form.
Flux boosting of galaxies close to the limiting depth of a survey can cause inaccuracies in the derived LF unless properly accounted for with simulations, and gravitational lensing can strongly affect the observed LF~\citep{Wyithe2011} at the very bright-end.
The number counts of bright galaxies can also be affected by quasar contamination (e.g.~\citealp{BianFuyan2013}), although the number densities of quasars appear too low to strongly influence the $z > 5$ galaxy LF~\citep{Bowler2014}.
Furthermore we consider and rule out a strong gravitational lensing effect on our derived LF, and the effect of photometric scatter is accounted for by our completeness simulations.
Finally, cosmic variance and small number statistics can influence the derived LF and best-fitting functional parameters (e.g.~\citealp{Trenti2008, Eardley2015}). 
By analysing two independent, degree-scale, fields, we are able to directly measure the effect of cosmic variance on the bright-end of the LF using a larger sample of bright galaxies than previously obtained at $z \simeq 6$.

The bright end of the LF is sensitive to astrophysical effects such as dust obscuration and feedback mechanisms (we discuss theoretical predictions in Section~\ref{sect:sims}), and it is observations of the number density of rare and bright galaxies that place the tightest constraints on the position of the knee in the LF.
The typical errors on the determination of the bright end of the LF at high redshift, and more severely the systematic difference between different studies, make a secure determination of the form of the LF challenging (see Fig.~\ref{fig:theory} or the compilation of results by~\citealp{Bouwens2015}).
We therefore investigate the functional form of the rest-frame UV LF using a sub-set of the observed LF points at $z \simeq 6$, using the determination from this work at $M_{\rm UV} \lesssim -21.5$ due to the superior depth and/or area of the combined UltraVISTA/COSMOS and UDS/SXDS fields when compared to previous studies.
For the faint end of the LF we used the results derived by~\citet{Bouwens2007} given the slight discrepancy we find with the~\citet{Bouwens2015} results at $z \simeq 6$.
We fit these determinations of the LF with a Schechter, DPL and Saunders functional form and show the resulting fits in Fig.~\ref{fig:functions}.
The double power-law provides a slightly better fit to the data, even when corrected for the additional parameter available in the fitting, although a Schechter function also provides a good fit to the data.
As can be seen from the best-fitting parameters shown in Table~\ref{table:fitting}, the assumed functional form changes the derived characteristic magnitude $M^*$.
Although there is clearly a change in slope of the LF at brighter magnitudes, the exact position of the `break' is not clear from the current data and therefore depends strongly on the function assumed.
The Saunders function also provides a good fit to the data, however the function poorly constrains the characteristic magnitude $M^*$.
The uncertainty is a result of the parameter $\sigma$, which provides additional freedom in the shape of the bright end of the function.
Hence we only present the results of fitting with the Schechter function and DPL in the next section.

In~\citet{Bowler2014} we found good agreement between the observed rest-frame UV LF of galaxies at $z \simeq 7$ and the shape of the HMF when scaled using a constant mass to light ratio.
Evolving the HMF according to the~\citet{Reed2007} model using the online tool `HMFcalc'~\citep{Murray2013}, and using the same scaling as at $z \simeq 7$, we find the curve shown in Fig.~\ref{fig:functions}. 
Again there is a good agreement between the simple LF predicted from the HMF, especially considering that the only evolution incorporated is due to dark matter halo build-up.
There is a clear deficit of galaxies at the faint end, as would be expected from models of supernova feedback which rapidly quench the star-formation in low-mass galaxies.
In contrast to the results at $z \simeq 7$ from~\citet{Bowler2014}, the bright end of the rest-frame UV LF at $z\simeq 6$ also shows a deficit of objects compared to the underlying halo distribution.
Regardless of the exact scaling of the HMF into luminosity space, the comparison indicates that the bright-end slope of the LF is now steeper than the HMF at $z\simeq 6$.
Although tentative, this steepening could indicate that we are now observing the build-up of dust or the onset of AGN feedback (or some other mass quenching mechanism) in the brightest galaxies at $z \simeq 6$ as discussed further in Section~\ref{sect:sims}.

\begin{figure*}
\includegraphics[width = 0.65\textwidth]{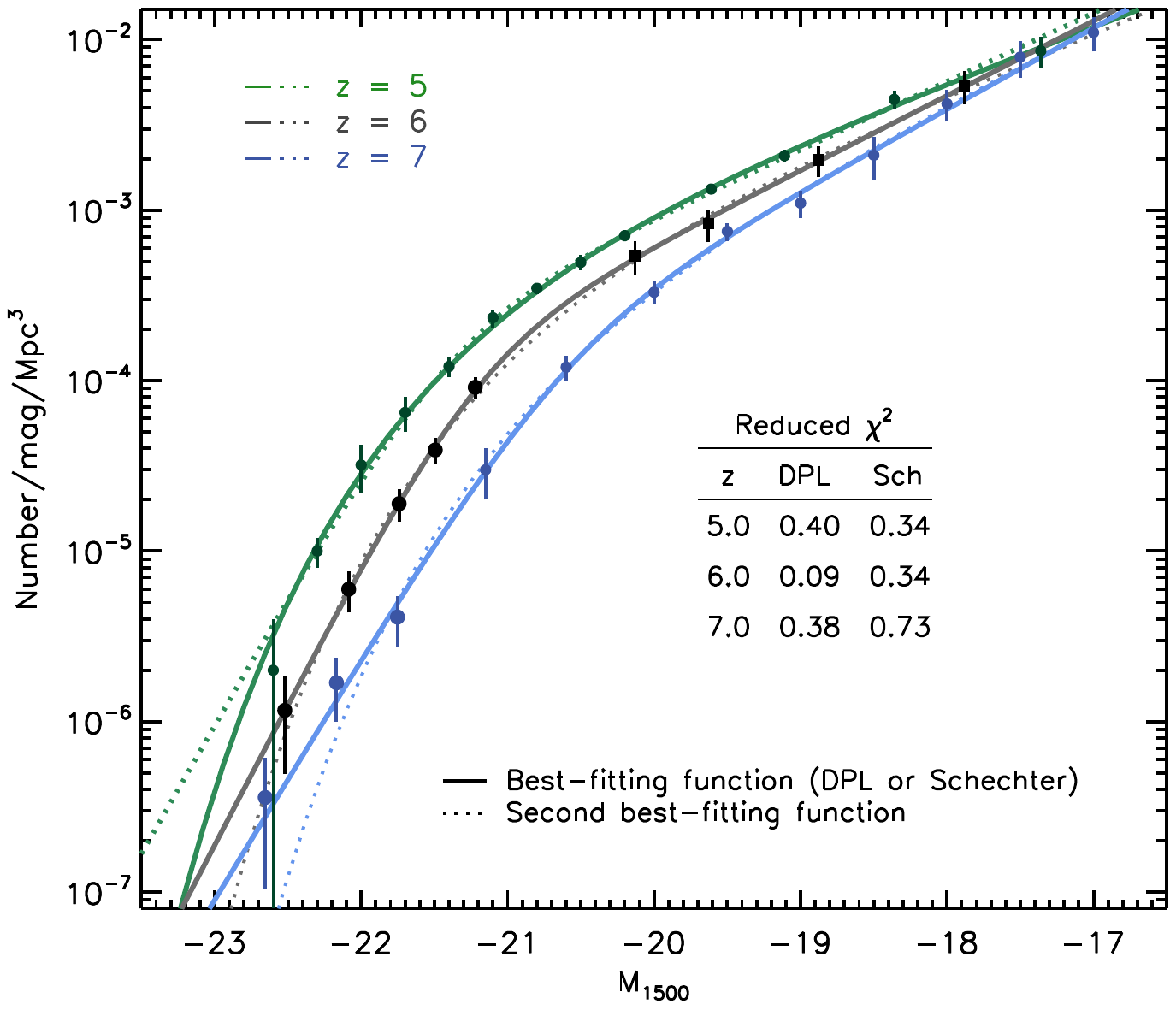}
\caption{The best-fitting DPL and Schechter function fits to a selection of observations of the rest-frame UV LF at $z = 5, 6$ and $7$ (green, black and blue lines/points respectively).
For each redshift, the best-fitting function is shown as the solid line and the second best fitting function is shown as a dotted line.
The inset table shows the reduced $\chi^2$ for each fit.
At $z \simeq 5$ the fitted points are from~\citet{vanderBurg2010},~\citet{Bouwens2015} and~\citet{Finkelstein2014}, and at $z \simeq 6$ the LF determination from this work is combined with the results from~\citet{Bouwens2007}.
Finally, at $z \simeq 7$ we fit to the LF determined by~\citet{Bowler2014} and~\citet{McLure2013}.
The best fitting parameters are presented in Table~\ref{table:fitting567}.
}
\label{fig:fitting}

\end{figure*}

\subsection{Evolution of the LF from ${\bf z \simeq 5}${\bf --}${\bf 7}$}

Observations of the rest-frame UV galaxy LF between $z\simeq 7$ and $z \simeq 5$ such as those shown in Fig.~\ref{fig:fitting} reveal a strong evolution in the number densities of galaxies at high redshift.
This work at $z\simeq 6$ and the analogous work at $z \simeq 7$ presented in~\citet{Bowler2014} allow the form and evolution of the bright end of the LF to be tightly constrained, and potential evolution of the functional form to be investigated.
As can be seen in Fig.~\ref{fig:fitting}, the observed LFs from $z \simeq 5$ to $z \simeq 7$ show little evolution faintward of $M_{\rm UV} \simeq -19.0$, however brighter than $M_{\rm UV} \simeq -20.0$ there is clear evolution in the number densities of galaxies, with bright LBGs ($M_{\rm UV} \simeq -21.5$) at $z \simeq 7$ being an order of magnitude less numerous than $z \simeq 5$ galaxies of the same luminosity.
Determining the exact evolution of the LFs however is not straightforward, as although the general agreement between different studies is good, there are systematic differences between the results that are larger than the errors estimated by each individual study.
Furthermore, the methodology used to derive the Schechter function parameters (i.e. $\chi^2$ versus maximum likelihood fitting) can also introduce systematic differences between studies based on small samples.
Hence, the exact parameterisation of the LF derived from each analysis can disagree; for example at $z \simeq 5$~\citet{vanderBurg2010} find a best-fitting characteristic magnitude $M^* = -20.93^{+0.10}_{-0.11}$ from an analysis of the CFHTLS data using a colour-colour selection, whereas the fit to the~\citet{McLure2009} results gives $M^* = -20.73 \pm 0.11$.
We therefore only fit to a subset of the available rest-frame UV LF points from different studies as motivated below, although we display a large compilation of studies in Fig.~\ref{fig:theory}.

To attempt to quantify the evolution of the bright end of the LF, we fit DPL and Schechter functions to a subset of derived rest-frame UV LF points from $z \simeq 5$--$7$.
A simple $\chi^2$ minimisation method was used, and the errors were found as the value of the parameter which gives a $\Delta \chi^2 = 1.0$, minimised over all other parameters.
We combine our results with those of~\citet{Bouwens2007} at $ z \simeq 6$, again excluding the~\cite{Bouwens2015} points due to the uncertainties in the bright-end of the LF where the sample is sensitive to cosmic variance (see Section~\ref{sect:cv}).
At $z \simeq 5$ we choose to fit to the~\citet{vanderBurg2010} results at the bright-end, excluding the results of~\citet{McLure2009} as they were based on a single field, and hence, as we have found in this study at $z \simeq 6$, could be vulnerable to cosmic variance.
The results of~\citet{Bouwens2015} and~\citet{Finkelstein2014}, although excluded in the fitting process at $z \simeq 6$ and $z\simeq 7$, agree well at $z\simeq 5$ at faint magnitudes and hence faint-wards of $M_{\rm UV} = -20.0$ we use the $ z\simeq 5$ points from~\citet{Bouwens2015}.
At $z \simeq 7$ we use the~\citet{McLure2013} determination of the LF, which follows a similar methodology to this work.
~\citet{Bouwens2015} showed that the total magnitudes of the brightest galaxies found by~\citet{McLure2013} were underestimated by assuming a point-source correction when using small apertures, as these objects are marginally extended.
We therefore boost the magnitudes of the points at $M_{\rm UV} = -21.0$ and $M_{\rm UV} = -20.5$ from the~\citet{McLure2013} analysis by 0.15 and 0.1 mag respectively when fitting, to account for the underestimation of the magnitudes here.
Uncertainties in the~\citet{Bouwens2015} analysis at the bright-end of the $z \simeq 7$ LF (which we exclude from the fitting process) are evident in Fig.~\ref{fig:theory}, where the implied number density of galaxies at $M_{\rm UV} = -21.86$ is comparable to that of $z \simeq 5$ galaxies at the same luminosity.

Fig.~\ref{fig:fitting} shows the result of fitting a DPL and Schechter function to the described subset of the observed LF points at $z = 5$--$7$, with the best-fitting function parameters presented in Table~\ref{table:fitting567}.
The reduced $\chi^2$ values for the fits are shown in Fig.~\ref{fig:fitting}, where it is evident that the current error bars available on the observed rest-frame UV LF result in an `over-fitting' of the data with the multi-variable fit provided by the Schechter or DPL functions, resulting in $\chi_{\rm red}^2 < 1$.

Reassuringly, our fits recover the steep faint-end slopes found in previous studies~\citep{Bouwens2012, McLure2013, Schenker2013}, showing that our measurement of $\alpha$ is not being strongly influenced by any tension in the fitting process.
For the DPL fit, the recovered bright-end slope values are relatively uncertain at $z = 6$ and $z = 7$, however there is tentative evidence for a steepening of $\beta_{}$ from $ z \simeq 7$ to $ z\simeq 6$.
At $z = 5$, the errors on $\beta_{}$ are much smaller, however as we have only fitted to the~\citet{vanderBurg2010} data, the derived value and uncertainty does not include the systematic error between the~\citet{vanderBurg2010} and~\citet{McLure2009} results (which can be seen in Fig.~\ref{fig:theory}).
Fitting the two studies separately we find $ \beta_{} = -4.8_{-0.5}^{+0.4}$ and $\beta_{} = -4.4^{+0.3}_{-0.3}$ for the~\citet{vanderBurg2010} and~\citet{McLure2009} results respectively, and hence it remains possible that a further steepening of the bright-end of the LF continues to $ z = 5$ and this is not excluded by the data.
Furthermore, at $z \simeq 5$, the Schechter function formally becomes the best-fitting function, demonstrating the steepening of the bright-end slope that is observed in the data.

The effects of cosmic variance on the our derived LF at $z\simeq 6$ can be clearly seen in Fig.~\ref{fig:lf}, where the LF was derived in the UltraVISTA/COSMOS and UDS/SXDS fields separately.
To quantify the differences we fit Schechter and DPL functions to the separate determinations and present the results in Table~\ref{table:lf}.
We find that both the best-fitting $M^*$ and $\phi^*$ differ between the individual degree scale fields, although they are consistent within the errors.
A deviation in these parameters with environments is to be expected theoretically~\citep{Trenti2008}, and has been observed at lower redshift (e.g.~\citealp{McNaughtRoberts2014, Eardley2015}).
The combined LF we present here therefore represents a closer approximation to the underlying bright-end of the LF at $z \simeq 6$ than that obtainable in an individual degree scale field, although additional sight-lines and wider area imaging will be necessary to constrain the LF and impact of cosmic variance further.

\subsubsection{Evolution in $M^*$ from $z = 5$--$7$?}

\begin{figure}
\includegraphics[width = 0.48\textwidth, trim = 0.7cm 0 0 0 ]{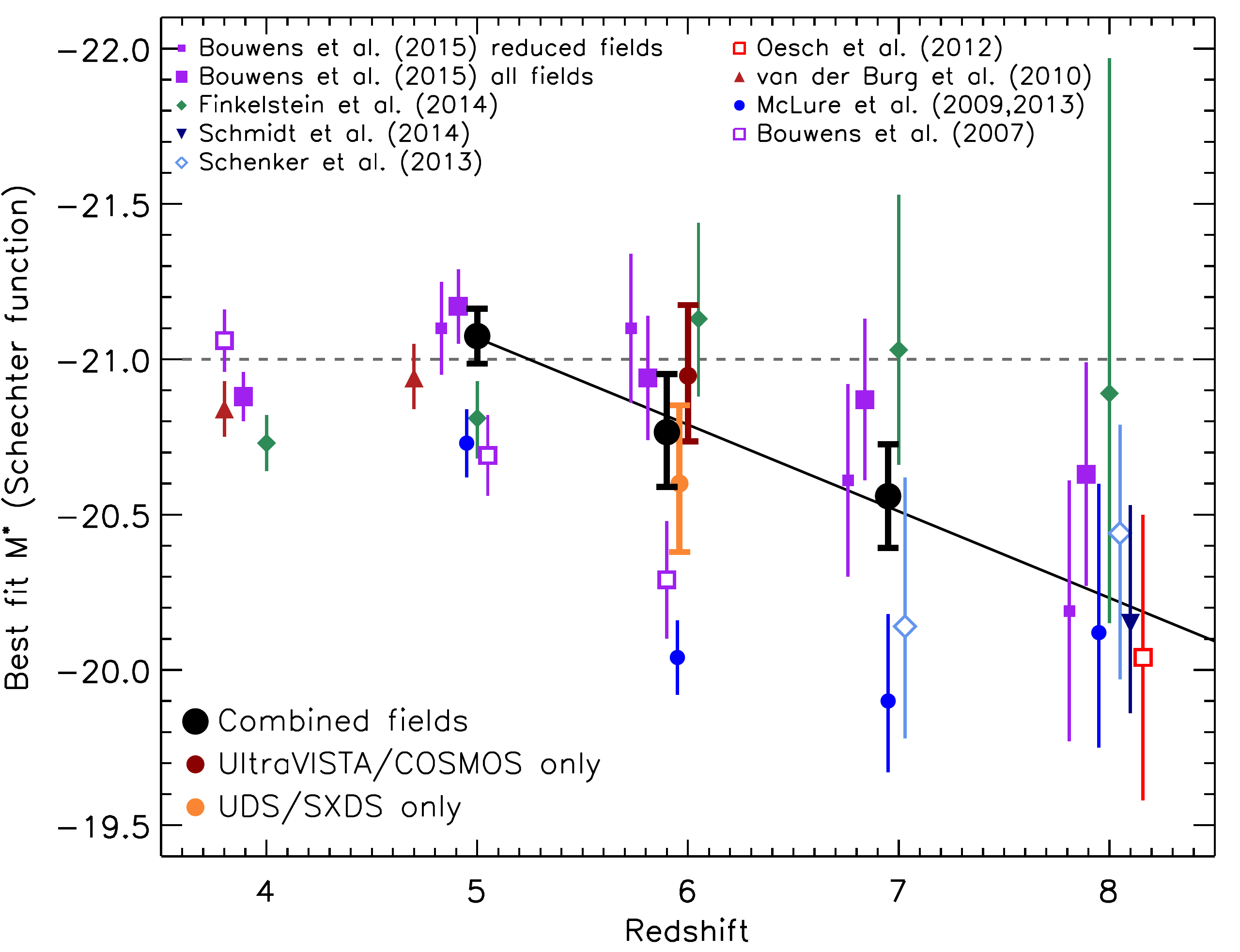}
\caption{The evolution in the characteristic magnitude derived from fitting a Schechter function to the $z \simeq 5$--$7$ data shown in Fig.~\ref{fig:fitting} are shown as the black circles.
The results derived from the UltraVISTA/COSMOS and UDS/SXDS fields separately at $z \simeq 6$ are shown in dark red and orange respectively.
The results of primarily~\emph{HST} surveys from~\citet{Finkelstein2014},~\citet{Bouwens2015,Bouwens2007},~\citet{Schmidt2014},~\citet{McLure2013},~\citet{Schenker2013} and~\citet{Oesch2012b} are shown, with additional results from wider-area ground-based imaging from~\citet{vanderBurg2010} and~\citet{McLure2009}.
The constant characteristic magnitude of $M^* = -21.0$ proposed by~\citet{Bouwens2015} and~\citet{Finkelstein2014} is shown as a horizontal dashed line, and the solid line shows a simple linear fit to our $M^*$ results from $z \simeq 5$--$7$.
For clarity, in some cases the plotted redshift of a point has been shifted by $\Delta z = 0.05$--$0.1$.
}
\label{fig:muv_ev}
\end{figure}

The evolution in the characteristic magnitude from $z~\simeq~4~-~8$ is displayed in Fig.~\ref{fig:muv_ev}, 
including the results derived from~\emph{HST} surveys~\citep{Finkelstein2014,Bouwens2015, Schmidt2014, McLure2013, Schenker2013,Oesch2012b, Bouwens2007} and ground-based analyses~\citep{vanderBurg2010, McLure2009}.
The results from this work were taken from the fitting of Schechter and DPL functions to the data shown in Fig.~\ref{fig:fitting}, and are displayed in Table~\ref{table:fitting567}.
While the error bars at $z \simeq 6$ are large, at $z \simeq 5$ and $z \simeq 7$ the data we fit to more completely fills the available magnitude space and hence $M^*$ is more securely defined.
The functional form assumed changes the derived values, as can be seen from comparing the DPL and Schechter function results, with the DPL fit tending to produce brighter characteristic magnitudes.
The DPL results cannot therefore be directly compared with the results from other studies that exclusively derive $M^*$ assuming a Schechter function, however as a DPL provides a better fit to the observed LF at $z\simeq 6$ and $7$ we include these results for comparison in Table~\ref{table:fitting567}.
Fig.~\ref{fig:muv_ev} illustrates the uncertainty in deriving the bright end of the LF from small fields such as those provided by the CANDELS survey, as the implied evolution from~\citet{Bouwens2015} changes substantially depending on whether the full or a reduced set of the CANDELS fields are included.
Furthermore, the errors on $M^*$ derived by~\citet{Finkelstein2014} are large at $z >6$ where the small area of the imaging used results in weak constraints on the break luminosity.
The effect of cosmic variance in degree-scale fields is also evident from the faint inferred $M^*$ found in the potentially under-dense UDS/SXDS field by~\citet{McLure2009} at $z \simeq 6$, and by the difference in the $M^*$ derived from the two separate fields in our analysis.

In contrast to the work by~\citet{Bouwens2015} and~\citet{Finkelstein2014} we find an evolution in $M^*$ between $z \simeq 5$ to $z \simeq 7$ of $\Delta M^* \simeq 0.4$--$0.5$ mag.
While our $M^*$ results agree with the determinations from~\citet{Bouwens2015} and~\citet{Finkelstein2014} within the errors, this is predominantly due to the large errors on $M^*$ derived by these studies, a consequence of the increased uncertainty in the number counts of bright galaxies when relatively small-area surveys are used.
Both~\citet{Bouwens2015} and~\citet{Finkelstein2014} found little evolution in the characteristic magnitude over the same redshift range, with both studies suggesting that an approximately constant $M^* \simeq -21.0$ provides a good fit at $z \leq 7$.
Instead, as shown in Fig~\ref{fig:muv_ev}, our results rule out a constant $M^*$ between $z \simeq 5$ and $7$.
The results of fitting the $z\simeq 7$ rest-frame UV LF, including the results of~\citet{Bowler2014}, with both a DPL and a Schechter function show a best-fitting $M^* > -21.0$; a result which, when combined with the faint characteristic magnitude found at $z \simeq 8$~\citep{Bouwens2015, Schmidt2014, McLure2013, Schenker2013, Oesch2012b} suggest a smooth brightening of $M^*$ from $z \simeq 8$ to $z \simeq 5$.
Although the error bars on $M^*$ are relatively large at $z \simeq 6$, they follow a smooth decline to the observed $M^* \simeq -20.5$ observed at $z \simeq 7$.
Such an evolution, primarily in the characteristic magnitude of galaxies, is qualitatively to be expected from the hierarchical coalescence and growth of the underlying HMF.
We caution however, that condensing any evolution in the LF to a single parameter is very uncertain and may be missing subtleties in the form of the evolution as illustrated by the potential change in functional form from $z \simeq 7$ to $z \simeq 5$ hinted at in Fig.~\ref{fig:fitting}.
For example we also find an evolution in $\phi^*$, however it is weaker than that obtained by~\citet{Bouwens2015} and~\citet{Finkelstein2014}.
Instead a detailed comparison of the full observed LF with the predictions of theoretical models is necessary to provide a more complete view of the evolution, and we compare the observed UV LFs at $z \simeq 5$, $6$ and $7$ to a compilation of semi-analytic and hydrodynamical models in Section~\ref{sect:sims}.
The evolution we observe over $z = 5$--$7$ is occurring in only $400\,$Myr and hence represents apparently rapid evolution in the characteristic magnitude of LBGs in the first billion years of cosmic time (however such evolution is arguably expected given the rapid evolution in the underlying HMF).
Improved constraints on the rest-frame UV LF around the apparent break magnitude from reconciling the various~\emph{HST}-based determinations, combined with future constraints on the form of the extreme bright-end of the LF from wider area imaging (e.g. VISTA VIDEO;~\citealp{Jarvis2013}), will reduce the current errors on the determination of the form and evolution of the LF.

\begin{table}
\caption{The best-fitting DPL and Schechter-function parameters derived from fitting the selection of observations of the rest-frame UV LF at $z \simeq 5$, $6$ and $7$ as described in the text and displayed in Fig.~\ref{fig:fitting}.
The results from the combined fields, the UltraVISTA/COSMOS field alone and the UDS/SXDS field alone are displayed in the upper, middle and lower parts of the table respectively.
For each field combination, the DPL results are shown above the Schechter function fits, and are identifiable by the presence of the bright-end slope ($\beta$) value.
Column 1 gives the approximate redshift, with the characteristic number density and absolute magnitude shown in Columns 2 and 3.
The faint and bright-end slope (for the DPL) are displayed in Columns 4 and 5.
}
\begin{tabular}{c c c c c}
\hline
$z$ & $\phi^*$ & $M^*$ & $\alpha$ & $\beta$ \\
& /mag/Mpc$^3$ & /mag &   \\
\hline
$  5.0$ & $2.5_{ -0.4}^{+  0.6}  \times 10^{-4}$ & $-21.40_{ -0.12}^{+  0.13}$ & $ -2.
00_{ -0.05}^{+  0.05}$ & $  -4.8_{  -0.4}^{+   0.3}$\\[3pt]
$  6.0$ & $1.9_{ -0.8}^{+  1.2}  \times 10^{-4}$ & $-21.20_{ -0.22}^{+  0.22}$ & $ -2.
10_{ -0.14}^{+  0.16}$ & $  -5.1_{  -0.6}^{+   0.5}$\\[3pt]
$  7.0$ & $2.2_{ -0.9}^{+  1.7}  \times 10^{-4}$ & $-20.61_{ -0.26}^{+  0.31}$ & $ -2.
19_{ -0.10}^{+  0.12}$ & $  -4.6_{  -0.5}^{+   0.4}$\\[3pt]

\hline
$  5.0$ & $6.4_{ -0.9}^{+  1.1}  \times 10^{-4}$ & $-21.07_{ -0.09}^{+  0.09}$ & $ -1.
81_{ -0.05}^{+  0.06}$ &  -- \\[3pt]
$  6.0$ & $5.7_{ -2.0}^{+  2.7}  \times 10^{-4}$ & $-20.77_{ -0.19}^{+  0.18}$ & $ -1.
88_{ -0.14}^{+  0.15}$ &  -- \\[3pt]
$  7.0$ & $3.7_{ -1.1}^{+  1.5}  \times 10^{-4}$ & $-20.56_{ -0.17}^{+  0.17}$ & $ -2.
09_{ -0.09}^{+  0.10}$ &  -- \\[3pt]
 
\hline
\multicolumn{5}{c}{UltraVISTA/COSMOS field only}\\
\hline
$  6.0$ & $1.6_{ -0.7}^{+  1.1}  \times 10^{-4}$ & $-21.35_{ -0.23}^{+  0.25}$ & $ -2.
08_{ -0.14}^{+  0.15}$ & $  -5.1_{  -0.8}^{+   0.6}$\\[3pt]
$  6.0$ & $4.8_{ -1.9}^{+  2.7}  \times 10^{-4}$ & $-20.95_{ -0.23}^{+  0.21}$ & $ -1.
88_{ -0.15}^{+  0.16}$ &  -- \\[3pt]

\hline
\multicolumn{5}{c}{UDS/SXDS field only}\\
\hline
$  6.0$ & $2.7_{ -1.3}^{+  3.1}  \times 10^{-4}$ & $-20.88_{ -0.30}^{+  0.45}$ & $ -2.
08_{ -0.17}^{+  0.20}$ & $  -4.8_{  -0.8}^{+   0.7}$\\[3pt]
$  6.0$ & $6.3_{ -2.8}^{+  4.0}  \times 10^{-4}$ & $-20.60_{ -0.25}^{+  0.22}$ & $ -1.
92_{ -0.18}^{+  0.19}$ &  -- \\[3pt]

\hline

\end{tabular}
\label{table:fitting567}
\end{table}

\subsection{Comparison to theory}\label{sect:sims}

\begin{figure}
\includegraphics[width = 0.45\textwidth, trim = 0 0 0 1cm]{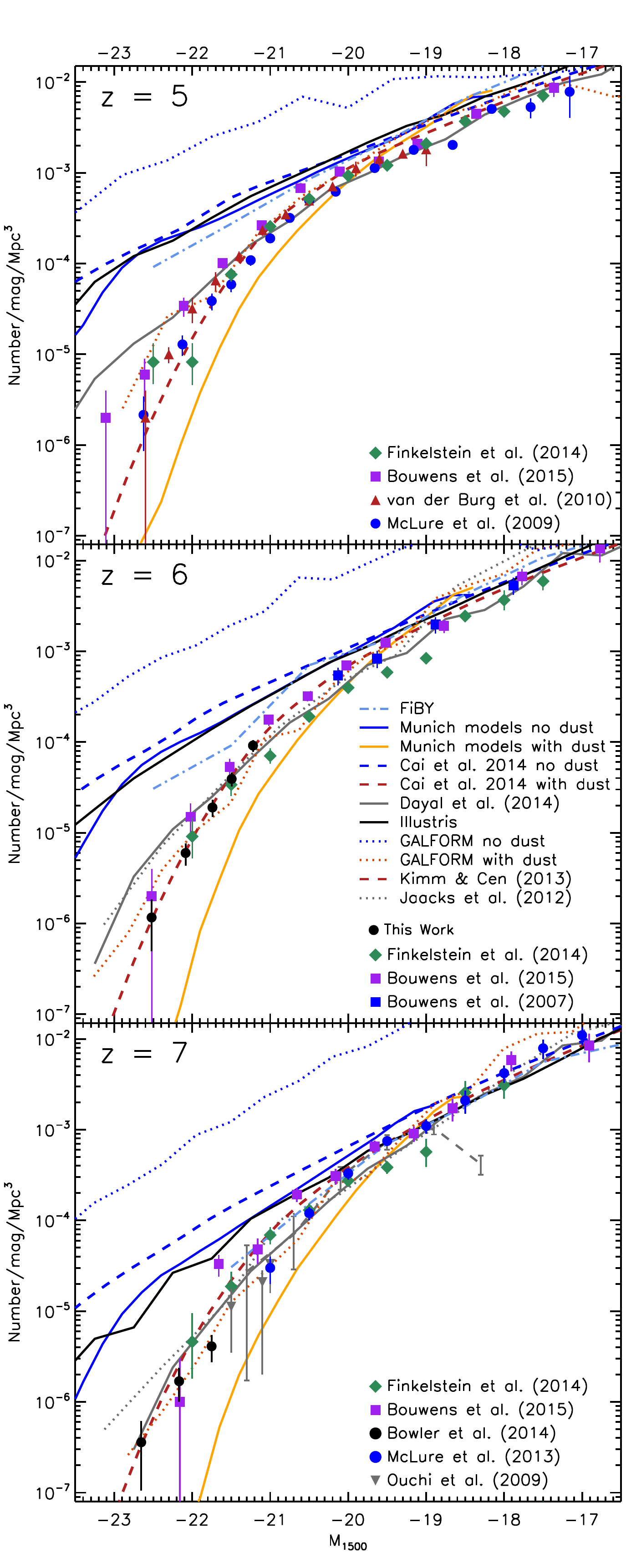}
\caption{
A comparison of the latest
observational data on the rest-frame UV galaxy LF
at $z \simeq 5$, 6 and 7 (including the new results
on the bright end presented here and in Bowler et al. 2014)
with the predictions of several of the latest semi-analytic
and hydrodynamical models of galaxy formation. The sources of the data points
are indicated in each panel, with the various model references provided in the central
($z \simeq 6$) panel. The implications of this comparison are discussed in the
text (Section~\ref{sect:sims}) but in general it can be seen that most of the models
struggle to reproduce the observations over the redshift range $z \simeq 5 - 7$
when faced with the large dynamic range now made possible by
the combined ground-based and~\emph{HST} dataset.
See the text for full references to the models displayed here.
}
\label{fig:theory}
\end{figure}

In Fig.~\ref{fig:theory} we present a comparison of the latest observational data on the rest-frame UV galaxy LF
at $z \simeq 5$, 6 and 7 (including the new results on the bright end presented here and in~\citealt{Bowler2014}) with the predictions of several of the latest semi-analytic and hydrodynamical models of galaxy formation.
This comparison is not completely fair, as some of the models have been (to some extent) tuned to explicitly match existing high redshift data (generally at the faint end of UV LF, e.g.~\citealt{Dayal2014}), while others have not been tuned at all (e.g. the First Billion Years (FiBY) simulations;~\citealp{Paardekooper2013}; Khochfar et al. in prep.). Moreover some models include the effects of dust obscuration (e.g. the SPH simulations of~\citealp{Jaacks2013} and the new Munich models of~\citealp{Henriques2014, Clay2015}) while others have yet to implement any form of dust obscuration at these redshifts (e.g. the Illustris simulation predictions from~\citealp{Genel2014}, the FiBY simulations from~\citealp{Paardekooper2013} and the semi-analytic results from~\citealp{Dayal2014}). Instructively, the predictions of the Munich models~\citep{Clay2015}, GALFORM~\citep{GonzalezPerez2013} and the~\citet{Cai2014} model were made available to us both with and without dust obscuration. Finally, it should be noted that several of the models do not cover large enough cosmological volumes for very useful comparison with the very bright end as derived from the degree-scale ground-based surveys (e.g.~\citealt{Kimm2013, Cen2014}).

Despite these complications, some useful conclusions can still be drawn from this figure. First, it is clear that while most models do a reasonable
job of reproducing the fainter end of the LF, there is a general problem of over-predicting the bright end (with the sole exception of the revised Munich models;~\citealp{Henriques2014, Clay2015}), even though the actual data produced by the work presented here indicate a shallower bright-end slope than would be inferred from a Schechter function
fit to the fainter data.
Second, with the possible exception of~\citet{Dayal2014} (although this dust free model seems to start to struggle at $ z \simeq 5$),
those models which do provide a satisfactory fit to the LF over this large dynamic range include substantial dust obscuration. In particular,
the model which apparently performs `best' in this comparison is the~\citet{Cai2014} model after application of dust obscuration, but it can be seen that the impact of this dust obscuration is enormous, equivalent to either an average depression of UV luminosity by $A_{1500} \simeq 2$\,mag at a number density of
$10^{-5}\,{\rm mag^{-1} Mpc^{-3}}$, or a depression in observed number density by $\simeq 2$ orders of magnitude at $M_{1500} \simeq -22.5$.

Thus, while much attention has been focussed on the faint end of the high-redshift galaxy LF in recent years (quite reasonably, especially given the important implications for reionization;~\citealp{Robertson2013}) it is clear that the full shape of the LF, extended to the brightest magnitudes through large-area ground-based surveys, has the potential to differentiate between alternative models of early galaxy formation and evolution. Moreover, while it currently remains unclear whether the shape of the bright end of the LF at $ z \simeq 5 - 7$ is really driven by evolution in dust properties
or by mass quenching (e.g.~\citealp{Peng2010}), or early AGN feedback
(or indeed by some other as yet poorly understood mechanism for regulating
star formation), forthcoming observations have the potential to clarify and quite possibly resolve these issues. For example, pointed ALMA
follow-up of bright UV-selected galaxies can address the prevalence of dust in such objects, while improved measurements of the stellar mass function at these early times (e.g. through improved deconfusion of deep
{\it Spitzer} IRAC data, and ultimately with {\it JWST} observations) will provide another important reference point for comparison with theoretical predictions.
At the same time, UltraVISTA DR3 (expected July 2015) should be deep enough to enable the work presented here at $z \simeq 6 - 7$ to be extended out to $z \simeq 8$
(with potentially useful constraints also at $z \simeq 9$), while wider-area surveys (e.g. with VISTA VIDEO at near-infrared wavelengths, and Subaru/Hyper-SuprimeCam
at red optical wavelengths) culminating in the {\it Euclid} Deep Survey~\citep{Laureijs2011} should
remove any remaining ambiguity over the shape of the bright end of the galaxy UV LF in the first billion years of cosmic history.

\section{Conclusion}\label{sect:conclusions}

We have selected a sample of star-forming galaxies at $z \simeq 6$ from the COSMOS/UltraVISTA and UDS/SXDS fields, which in total provide an area of $1.65\,$deg$^2$ of deep multiwavelength imaging in the optical/near-infrared.
The galaxies were selected using a full photometric redshift analysis, which allows the removal of low-redshift dusty galaxies and cool galactic brown dwarf stars.
The main findings of our work are as follows.

\begin{itemize}

\item Using a simple thin-disk galaxy model we find that the expected number of brown dwarf stars in each field greatly exceeds the number of LBGs at the very bright-end ($m_{\rm AB} < 25$), however brown dwarfs can be cleanly removed using fitting of stellar templates to the multiwavelength optical/near-infrared photometry.

\item We measure the rest-frame UV slope of the galaxies in our sample, finding that the derived values follow the colour-magnitude relation found at $z \simeq 5$ by~\citet{Rogers2014}, showing a mean $\beta_{\rm UV} = -1.8 \pm 0.1$ at $M_{\rm UV} < -22.0$, in contrast to the redder slopes found by~\citet{Willott2013}.
 
\item The number density of $z \simeq 6$ galaxies we find a factor of $\sim 1.8$ more galaxies in the UltraVISTA/COSMOS field than in the UDS/SXDS, a deviation that just exceeds that predicted from cosmic variance between fields of this size.
We consider the effect of gravitational lensing of our objects by galaxies close to the line of sight, finding no evidence that the objects in our sample have preferential boosting over random positions in the field.

\item We calculate the rest-frame UV galaxy LF from our sample, using the restricted redshift range $5.7 < z < 6.3$ to compare directly with the work of~\citet{McLure2009} and to ensure minimal contamination by brown dwarfs.
Our determination of the LF lies midway between previous determinations from~\emph{HST} surveys by~\citet{Bouwens2015} and the ground-based analysis by~\citet{McLure2009}.
In particular, the recent determination of the UV LF from~\citet{Bouwens2015} over-predicts the expected number of LBGs in the fields we analyse by approximately a factor of two.
We find a good agreement with the results of~\citet{Finkelstein2014} in the magnitude range where our results overlap.

\item By comparing the LF derived from the COSMOS/UltraVISTA and UDS/SXDS fields separately, we conclude that part of the discrepancy between the results of~\citet{Bouwens2015} and~\citet{McLure2009} at the bright end of the LF is a result of the UDS/SXDS (analysed by~\citealp{McLure2009}) appearing under-dense at $z \simeq 6$.
Our results show that cosmic variance on scales of $1\,$deg$^2$ can be significant and therefore determining the bright-end of the LF from relatively small area datasets such as the $200\,$ arcmin$^2$ CANDELS fields could be highly uncertain.

\item We fit Schechter and double power-law functions to the observed $z \simeq 6$ rest-frame UV LF, showing that a DPL is marginally preferred, although an exponential decline also provides an acceptable description of the current data.
The fits show that the bright-end slope of the LF appears to steepen from $z\simeq 7$ to $z \simeq 5$ suggesting we may be observing the onset of feedback (e.g. from AGN or some other form of mass quenching, e.g.~\citealp{Peng2010}) or the build-up of dust in the brightest LBGs~\citep{Rogers2014}.

\item In contrast to~\citet{Bouwens2015} and~\citet{Finkelstein2014}, we find clear evidence for a brightening of the characteristic magnitude of $\Delta M^* \sim 0.4$--$0.5$ between $z \simeq 7$ and $5$.
Our results show that the evolution can still be well described as predominately luminosity evolution, as expected if the star-formation of the galaxies follows the hierarchical build-up of the underlying dark matter haloes.
We caution however that there still exist unexplained systematic errors between LF determinations at $z \simeq 5$--$7$ that can impact the results of functional fitting, and the future analysis of wider area imaging along independent sight-lines is required to further quantify these systematics and similarly the effect of cosmic variance at the bright-end.

\item Finally, comparison of a collection of the latest semi-analytical and hydrodynamical models of galaxy formation to the observed rest-frame UV galaxy LF at $z \simeq 5$, $6$ and $7$ reveals that most models tend to over-predict the number density of bright galaxies and substantial attenuation is required ($A_{1500} \simeq 1.5$--$2.0$) to bring the models into agreement with the data. 

\end{itemize}

\section*{Acknowledgements}
RAAB and JSD acknowledge the support of the European Research Council via
the award of an Advanced Grant to JSD. 
JSD also acknowledges the contribution of the EC FP7 SPACE project ASTRODEEP (Ref.No: 312725).
RJM acknowledges the support of the European Research Council via the award of a Consolidator Grant (PI McLure).
JPUF and BM acknowledge the support of the European Research Council via the award of a Consolidator Grant (PI Fynbo).
We thank Zhen-Yi Cai, Cen Renyue, Taysun Kimm, Sadegh Khochfar, Shy Genel, Mark Vogelsberger, Bruno Henriques, Scott Clay, Simon White, Jason Jaacks, Violeta Gonzalez-Perez and Pratika Dayal for generously supplying their theoretical results, in some cases in advance of publication.
We thank the anonymous referee for comments that improved this paper.

This work is based on data products from observations made with ESO Telescopes at the La Silla Paranal Observatory as part of programme ID 179.A-2005, using data products produced by TERAPIX and the Cambridge Astronomy Survey Unit on behalf of the UltraVISTA consortium, and as part of the VIDEO survey, under programme ID 179.A-2006 (PI: Jarvis).
 This study was based in part on observations obtained with MegaPrime/MegaCam, 
a joint project of CFHT and CEA/DAPNIA, at the Canada-France-Hawaii Telescope (CFHT) which is operated by the National Research Council (NRC) of Canada, the Institut National des Science de l'Univers of the Centre National de la Recherche Scientifique (CNRS) of France, and the University of Hawaii. This work is based in part on data products produced at TERAPIX and the Canadian Astronomy Data Centre as part of the Canada-France-Hawaii Telescope Legacy Survey, a collaborative project of NRC and CNRS. 
This research has benefitted from the SpeX Prism Spectral Libraries, maintained by Adam Burgasser at \url{http://pono.ucsd.edu/~adam/browndwarfs/spexprism}.

\label{lastpage}

\bibliographystyle{mn2e}
 \bibliography{../library_abbrv}

\begin{thebibliography}{114}
\expandafter\ifx\csname natexlab\endcsname\relax\def\natexlab#1{#1}\fi

\bibitem[{Baldry {et~al}\mbox{.}(2012)Baldry, Driver, Loveday, Taylor, Kelvin,
  Liske, Norberg, Robotham, Brough, Hopkins, Bamford, Peacock, Bland-Hawthorn,
  Conselice, Croom, Jones, Parkinson, Popescu, Prescott, Sharp, \&
  Tuffs}]{Baldry2012}
Baldry I.~K. {et~al.}, 2012, MNRAS, 421, 621

\bibitem[{Bernardi {et~al}\mbox{.}(2003)Bernardi, Sheth, Annis, Burles,
  Eisenstein, Finkbeiner, Hogg, Lupton, Schlegel, SubbaRao, Bahcall, Blakeslee,
  Brinkmann, Castander, Connolly, Csabai, Doi, Fukugita, Frieman, Heckman,
  Hennessy, Ivezi\'{c}, Knapp, Lamb, McKay, Munn, Nichol, Okamura, Schneider,
  Thakar, \& York}]{Bernardi2003}
Bernardi M. {et~al.}, 2003, AJ, 125, 1849

\bibitem[{Bernhard {et~al}\mbox{.}(2014)Bernhard, Bethermin, Sargent, Buat,
  Mullaney, Pannella, Heinis, \& Daddi}]{Bernhard2014}
Bernhard E., Bethermin M., Sargent M., Buat V., Mullaney J.~R., Pannella M.,
  Heinis S., Daddi E., 2014, MNRAS, 442, 509

\bibitem[{Bertin \& Arnouts(1996)}]{Bertin1996}
Bertin E., Arnouts S., 1996, A\&AS, 117, 393

\bibitem[{Bian {et~al}\mbox{.}(2013)Bian, Fan, Jiang, McGreer, Dey, Green,
  Maiolino, Walter, Lee, \& Dav\'{e}}]{BianFuyan2013}
Bian F. {et~al.}, 2013, ApJ, 774, 28

\bibitem[{Bouwens {et~al}\mbox{.}(2007)Bouwens, Illingworth, Franx, \&
  Ford}]{Bouwens2007}
Bouwens R.~J., Illingworth G.~D., Franx M., Ford H., 2007, ApJ, 670, 928

\bibitem[{Bouwens {et~al}\mbox{.}(2008)Bouwens, Illingworth, Franx, \&
  Ford}]{Bouwens2008}
Bouwens R.~J., Illingworth G.~D., Franx M., Ford H., 2008, ApJ, 686, 230

\bibitem[{Bouwens {et~al}\mbox{.}(2011)Bouwens, Illingworth, Labbe, Oesch,
  Trenti, Carollo, van Dokkum, Franx, Stiavelli, Gonz\'{a}lez, Magee, \&
  Bradley}]{Bouwens2011}
Bouwens R.~J. {et~al.}, 2011, Nat, 469, 504

\bibitem[{Bouwens {et~al}\mbox{.}(2014)Bouwens, Illingworth, Oesch, Labb\'{e},
  van Dokkum, Trenti, Franx, Smit, Gonzalez, \& Magee}]{Bouwens2014beta}
Bouwens R.~J. {et~al.}, 2014, ApJ, 793, 115

\bibitem[{Bouwens {et~al}\mbox{.}(2015)Bouwens, Illingworth, Oesch, Trenti,
  Labb\'{e}, Bradley, Carollo, van Dokkum, Gonzalez, Holwerda, Franx, Spitler,
  Smit, \& Magee}]{Bouwens2015}
Bouwens R.~J. {et~al.}, 2015, ApJ, 803, 34

\bibitem[{Bouwens {et~al}\mbox{.}(2012)Bouwens, Illingworth, Oesch, Trenti,
  Labb\'{e}, Franx, Stiavelli, Carollo, van Dokkum, \& Magee}]{Bouwens2012}
Bouwens R.~J. {et~al.}, 2012, ApJ, 752, L5

\bibitem[{Bouwens {et~al}\mbox{.}(2010)Bouwens, Illingworth, Oesch, Trenti,
  Stiavelli, Carollo, Franx, van Dokkum, Labb\'{e}, \& Magee}]{Bouwens2010}
Bouwens R.~J. {et~al.}, 2010, ApJ, 708, L69

\bibitem[{Bower {et~al}\mbox{.}(2012)Bower, Benson, \& Crain}]{Bower2012}
Bower R.~G., Benson A.~J., Crain R.~A., 2012, MNRAS, 422, 2816

\bibitem[{Bowler {et~al}\mbox{.}(2012)Bowler, Dunlop, McLure, McCracken,
  Milvang-Jensen, Furusawa, Fynbo, {Le F\`{e}vre}, Holt, Ideue, Ihara, Rogers,
  \& Taniguchi}]{Bowler2012}
Bowler R. A.~A. {et~al.}, 2012, MNRAS, 426, 2772

\bibitem[{Bowler {et~al}\mbox{.}(2014)Bowler, Dunlop, McLure, Rogers,
  McCracken, Milvang-Jensen, Furusawa, Fynbo, Taniguchi, Afonso, Bremer, \&
  Fevre}]{Bowler2014}
Bowler R. A.~A. {et~al.}, 2014, MNRAS, 440, 2810

\bibitem[{Bruzual \& Charlot(2003)}]{Bruzual2003}
Bruzual G., Charlot S., 2003, MNRAS, 344, 1000

\bibitem[{Caballero {et~al}\mbox{.}(2008)Caballero, Burgasser, \&
  Klement}]{Caballero2008}
Caballero J.~A., Burgasser A.~J., Klement R., 2008, A\&A, 488, 181

\bibitem[{Cai {et~al}\mbox{.}(2014)Cai, Lapi, Bressan, {De Zotti}, Negrello, \&
  Danese}]{Cai2014}
Cai Z.-Y., Lapi A., Bressan A., {De Zotti} G., Negrello M., Danese L., 2014,
  ApJ, 785, 65

\bibitem[{Calzetti {et~al}\mbox{.}(2000)Calzetti, Armus, Bohlin, Kinney,
  Koornneef, \& Storchi-Bergmann}]{Calzetti2000}
Calzetti D., Armus L., Bohlin R.~C., Kinney A.~L., Koornneef J.,
  Storchi-Bergmann T., 2000, ApJ, 533, 682

\bibitem[{Castellano {et~al}\mbox{.}(2010{\natexlab{a}})Castellano, Fontana,
  Boutsia, Grazian, Pentericci, Bouwens, Dickinson, Giavalisco, Santini,
  Cristiani, Fiore, Gallozzi, Giallongo, Maiolino, Mannucci, Menci, Moorwood,
  Nonino, Paris, Renzini, Rosati, Salimbeni, Testa, \&
  Vanzella}]{Castellano2010a}
Castellano M. {et~al.}, 2010{\natexlab{a}}, A\&A, 511, A20

\bibitem[{Castellano {et~al}\mbox{.}(2010{\natexlab{b}})Castellano, Fontana,
  Paris, Grazian, Pentericci, Boutsia, Santini, Testa, Dickinson, Giavalisco,
  Bouwens, Cuby, Mannucci, Cl\'{e}ment, Cristiani, Fiore, Gallozzi, Giallongo,
  Maiolino, Menci, Moorwood, Nonino, Renzini, Rosati, Salimbeni, \&
  Vanzella}]{Castellano2010b}
Castellano M. {et~al.}, 2010{\natexlab{b}}, A\&A, 524, A28

\bibitem[{Cen \& Kimm(2014)}]{Cen2014}
Cen R., Kimm T., 2014, ApJ, 782, 32

\bibitem[{Chen {et~al}\mbox{.}(2001)Chen, Stoughton, Smith, Uomoto, Pier,
  Yanny, Ivezi\'{c}, York, Anderson, Annis, Brinkmann, Csabai, Fukugita,
  Hindsley, Lupton, \& Munn}]{Chen2001}
Chen B. {et~al.}, 2001, ApJ, 553, 184

\bibitem[{Clay {et~al}\mbox{.}(2015)Clay, Thomas, Wilkins, \&
  Henriques}]{Clay2015}
Clay S., Thomas P., Wilkins S., Henriques B., 2015, preprint (arXiv:1504.03321)

\bibitem[{Curtis-Lake {et~al}\mbox{.}(2014)Curtis-Lake, McLure, Dunlop, Rogers,
  Targett, Dekel, Ellis, Faber, Ferguson, Grogin, Huang, Kocevski, Koekemoer,
  Lai, \& Robertson}]{CurtisLake2014}
Curtis-Lake E. {et~al.}, 2014, preprint (arXiv:1409.1832)

\bibitem[{Curtis-Lake {et~al}\mbox{.}(2013)Curtis-Lake, McLure, Dunlop,
  Schenker, Rogers, Targett, Cirasuolo, Almaini, Ashby, Bradshaw, Finkelstein,
  Dickinson, Ellis, Faber, Fazio, Ferguson, Fontana, Grogin, Hartley, Kocevski,
  Koekemoer, Lai, Robertson, Vanzella, \& Willner}]{CurtisLake2013}
Curtis-Lake E. {et~al.}, 2013, MNRAS, 429, 302

\bibitem[{Dayal {et~al}\mbox{.}(2014)Dayal, Ferrara, Dunlop, \&
  Pacucci}]{Dayal2014}
Dayal P., Ferrara A., Dunlop J.~S., Pacucci F., 2014, MNRAS, 445, 2545

\bibitem[{Dunlop {et~al}\mbox{.}(2013)Dunlop, Rogers, McLure, Ellis, Robertson,
  Koekemoer, Dayal, Curtis-Lake, Wild, Charlot, Bowler, Schenker, Ouchi, Ono,
  Cirasuolo, Furlanetto, Stark, Targett, \& Schneider}]{Dunlop2013}
Dunlop J.~S. {et~al.}, 2013, MNRAS, 432, 3520

\bibitem[{Eardley {et~al}\mbox{.}(2015)Eardley, Peacock, McNaught-Roberts,
  Heymans, Norberg, Alpaslan, Baldry, Bland-Hawthorn, Brough, Cluver, Driver,
  Farrow, Liske, Loveday, \& Robotham}]{Eardley2015}
Eardley E. {et~al.}, 2015, MNRAS, 448, 3665

\bibitem[{Ellis {et~al}\mbox{.}(2013)Ellis, Mclure, Dunlop, Robertson, Ono,
  Schenker, Bowler, Ouchi, Rogers, Curtis-Lake, Schneider, Charlot, Stark,
  Furlanetto, Cirasuolo, \& Koekemoer}]{Ellis2013}
Ellis R.~S. {et~al.}, 2013, ApJ, 763, L2

\bibitem[{Findlay {et~al}\mbox{.}(2012)Findlay, Sutherland, Venemans,
  Reyl\'{e}, Robin, Bonfield, Bruce, \& Jarvis}]{Findlay2012}
Findlay J.~R., Sutherland W.~J., Venemans B.~P., Reyl\'{e} C., Robin A.~C.,
  Bonfield D.~G., Bruce V.~A., Jarvis M.~J., 2012, MNRAS, 419, 3354

\bibitem[{Finkelstein {et~al}\mbox{.}(2014)Finkelstein, {Ryan, Russell E.},
  Papovich, Dickinson, Song, Somerville, Ferguson, Salmon, Giavalisco,
  Koekemoer, Ashby, Behroozi, Castellano, Dunlop, Faber, Fazio, Fontana,
  Grogin, Hathi, Jaacks, Kocevski, Livermore, McLure, Merlin, Mobasher, Newman,
  Rafelski, Tilvi, \& Willner}]{Finkelstein2014}
Finkelstein S.~L. {et~al.}, 2014, preprint (arXiv:1410.5439)

\bibitem[{Finoguenov {et~al}\mbox{.}(2007)Finoguenov, Guzzo, Hasinger,
  Scoville, Aussel, Bohringer, Brusa, Capak, Cappelluti, Comastri, Giodini,
  Griffiths, Impey, Koekemoer, Kneib, Leauthaud, {Le Fevre}, Lilly, Mainieri,
  Massey, McCracken, Mobasher, Murayama, Peacock, Sakelliou, Schinnerer,
  Silverman, Smol\v{c}i\'{c}, Taniguchi, Tasca, Taylor, Trump, \&
  Zamorani}]{Finoguenov2007}
Finoguenov A. {et~al.}, 2007, ApJS, 172, 182

\bibitem[{Finoguenov {et~al}\mbox{.}(2010)Finoguenov, Watson, Tanaka, Simpson,
  Cirasuolo, Dunlop, Peacock, Farrah, Akiyama, Ueda, Smol\v{c}i\'{c}, Stewart,
  Rawlings, van Breukelen, Almaini, Clewley, Bonfield, Jarvis, Barr, Foucaud,
  McLure, Sekiguchi, \& Egami}]{Finoguenov2010}
Finoguenov A. {et~al.}, 2010, MNRAS, 403, 2063

\bibitem[{Frei \& Gunn(1994)}]{Frei1994}
Frei Z., Gunn J.~E., 1994, AJ, 108, 1476

\bibitem[{Furusawa {et~al}\mbox{.}(2008)Furusawa, Kosugi, Akiyama, Takata,
  Sekiguchi, Tanaka, Iwata, Kajisawa, Yasuda, Doi, Ouchi, Simpson, Shimasaku,
  Yamada, Furusawa, Morokuma, Ishida, Aoki, Fuse, Imanishi, Iye, Karoji,
  Kobayashi, Kodama, Komiyama, Maeda, Miyazaki, Mizumoto, Nakata, Noumaru,
  Ogasawara, Okamura, Saito, Sasaki, Ueda, \& Yoshida}]{Furusawa2008}
Furusawa H. {et~al.}, 2008, ApJS, 176, 1

\bibitem[{Genel {et~al}\mbox{.}(2014)Genel, Vogelsberger, Springel, Sijacki,
  Nelson, Snyder, Rodriguez-Gomez, Torrey, \& Hernquist}]{Genel2014}
Genel S. {et~al.}, 2014, MNRAS, 445, 175

\bibitem[{Gonzalez-Perez {et~al}\mbox{.}(2013)Gonzalez-Perez, Lacey, Baugh,
  Frenk, \& Wilkins}]{GonzalezPerez2013}
Gonzalez-Perez V., Lacey C.~G., Baugh C.~M., Frenk C.~S., Wilkins S.~M., 2013,
  MNRAS, 429, 1609

\bibitem[{Grogin {et~al}\mbox{.}(2011)Grogin, Kocevski, Faber, Ferguson,
  Koekemoer, Riess, Acquaviva, Alexander, Almaini, Ashby, Barden, Bell,
  Bournaud, Brown, Caputi, Casertano, Cassata, Castellano, Challis, Chary,
  Cheung, Cirasuolo, Conselice, Cooray, Croton, Daddi, Dahlen, Dav\'{e},
  de~Mello, Dekel, Dickinson, Dolch, Donley, Dunlop, Dutton, Elbaz, Fazio,
  Filippenko, Finkelstein, Fontana, Gardner, Garnavich, Gawiser, Giavalisco,
  Grazian, Guo, Hathi, H\"{a}ussler, Hopkins, Huang, Huang, Jha, Kartaltepe,
  Kirshner, Koo, Lai, Lee, Li, Lotz, Lucas, Madau, McCarthy, McGrath, McIntosh,
  McLure, Mobasher, Moustakas, Mozena, Nandra, Newman, Niemi, Noeske, Papovich,
  Pentericci, Pope, Primack, Rajan, Ravindranath, Reddy, Renzini, Rix, Robaina,
  Rodney, Rosario, Rosati, Salimbeni, Scarlata, Siana, Simard, Smidt,
  Somerville, Spinrad, Straughn, Strolger, Telford, Teplitz, Trump, van~der
  Wel, Villforth, Wechsler, Weiner, Wiklind, Wild, Wilson, Wuyts, Yan, \&
  Yun}]{Grogin2011}
Grogin N.~A. {et~al.}, 2011, ApJS, 197, 35

\bibitem[{Guhathakurta {et~al}\mbox{.}(1990)Guhathakurta, Tyson, \&
  Majewski}]{Guhathakurta1990}
Guhathakurta P., Tyson J.~A., Majewski S.~R., 1990, ApJ, 357, L9

\bibitem[{Gunawardhana {et~al}\mbox{.}(2013)Gunawardhana, Hopkins,
  Bland-Hawthorn, Brough, Sharp, Loveday, Taylor, Jones, Lara-Lopez, Bauer,
  Colless, Owers, Baldry, Lopez-Sanchez, Foster, Bamford, Brown, Driver,
  Drinkwater, Liske, Meyer, Norberg, Robotham, Ching, Cluver, Croom, Kelvin,
  Prescott, Steele, Thomas, \& Wang}]{Gunawardhana2013}
Gunawardhana M. L.~P. {et~al.}, 2013, MNRAS, 433, 2764

\bibitem[{Henriques {et~al}\mbox{.}(2014)Henriques, White, Thomas, Angulo, Guo,
  Lemson, Springel, \& Overzier}]{Henriques2014}
Henriques B., White S., Thomas P., Angulo R., Guo Q., Lemson G., Springel V.,
  Overzier R., 2014, preprint (arXiv:1410.0365)

\bibitem[{Holwerda {et~al}\mbox{.}(2014)Holwerda, Trenti, Clarkson, Sahu,
  Bradley, Stiavelli, Pirzkal, {De Marchi}, Andersen, Bouwens, \&
  Ryan}]{Holwerda2014}
Holwerda B.~W. {et~al.}, 2014, ApJ, 788, 77

\bibitem[{Ilbert {et~al}\mbox{.}(2009)Ilbert, Capak, Salvato, Aussel,
  McCracken, Sanders, Scoville, Kartaltepe, Arnouts, Floc'h, Mobasher,
  Taniguchi, Lamareille, Leauthaud, Sasaki, Thompson, Zamojski, Zamorani,
  Bardelli, Bolzonella, Bongiorno, Brusa, Caputi, Carollo, Contini, Cook,
  Coppa, Cucciati, de~la Torre, de~Ravel, Franzetti, Garilli, Hasinger, Iovino,
  Kampczyk, Kneib, Knobel, Kovac, {Le Borgne}, {Le Brun}, F\`{e}vre, Lilly,
  Looper, Maier, Mainieri, Mellier, Mignoli, Murayama, Pell\`{o}, Peng,
  P\'{e}rez-Montero, Renzini, Ricciardelli, Schiminovich, Scodeggio, Shioya,
  Silverman, Surace, Tanaka, Tasca, Tresse, Vergani, \& Zucca}]{Ilbert2009}
Ilbert O. {et~al.}, 2009, ApJ, 690, 1236

\bibitem[{Ilbert {et~al}\mbox{.}(2013)Ilbert, McCracken, {Le F\`{e}vre}, Capak,
  Dunlop, Karim, Renzini, Caputi, Boissier, Arnouts, Aussel, Comparat, Guo,
  Hudelot, Kartaltepe, Kneib, Krogager, {Le Floc’h}, Lilly, Mellier,
  Milvang-Jensen, Moutard, Onodera, Richard, Salvato, Sanders, Scoville,
  Silverman, Taniguchi, Tasca, Thomas, Toft, Tresse, Vergani, Wolk, \&
  Zirm}]{Ilbert2013}
Ilbert O. {et~al.}, 2013, A\&A, 556, A55

\bibitem[{Ilbert {et~al}\mbox{.}(2008)Ilbert, Salvato, Capak, {Le Floc'h},
  Aussel, McCracken, Arnouts, Mobasher, Sanders, Scoville, \&
  Taniguchi}]{Ilbert2008}
Ilbert O. {et~al.}, 2008, ASP Conf. Ser., 399, 169

\bibitem[{Jaacks {et~al}\mbox{.}(2013)Jaacks, Thompson, \&
  Nagamine}]{Jaacks2013}
Jaacks J., Thompson R., Nagamine K., 2013, ApJ, 766, 94

\bibitem[{Jarvis {et~al}\mbox{.}(2013)Jarvis, Bonfield, Bruce, Geach, McAlpine,
  McLure, Gonzalez-Solares, Irwin, Lewis, Yoldas, Andreon, Cross, Emerson,
  Dalton, Dunlop, Hodgkin, Le, Karouzos, Meisenheimer, Oliver, Rawlings,
  Simpson, Smail, Smith, Sullivan, Sutherland, White, \& Zwart}]{Jarvis2013}
Jarvis M.~J. {et~al.}, 2013, MNRAS, 428, 1281

\bibitem[{Jurek {et~al}\mbox{.}(2013)Jurek, Drinkwater, Pimbblet, Glazebrook,
  Blake, Brough, Colless, Contreras, Couch, Croom, Croton, {M. Davis}, Forster,
  Gilbank, Gladders, Jelliffe, Li, Madore, Martin, Poole, Pracy, Sharp,
  Wisnioski, Woods, Wyder, \& Yee}]{Jurek2013}
Jurek R.~J. {et~al.}, 2013, MNRAS, 434, 257

\bibitem[{Kelvin {et~al}\mbox{.}(2014)Kelvin, Driver, Robotham, Graham,
  Phillipps, Agius, Alpaslan, Baldry, Bamford, Bland-Hawthorn, Brough, Brown,
  Colless, Conselice, Hopkins, Liske, Loveday, Norberg, Pimbblet, Popescu,
  Prescott, Taylor, \& Tuffs}]{Kelvin2014}
Kelvin L.~S. {et~al.}, 2014, MNRAS, 439, 1245

\bibitem[{Kimm \& Cen(2013)}]{Kimm2013}
Kimm T., Cen R., 2013, ApJ, 776, 35

\bibitem[{Koekemoer {et~al}\mbox{.}(2011)Koekemoer, Faber, Ferguson, Grogin,
  Kocevski, Koo, Lai, Lotz, Lucas, McGrath, Ogaz, Rajan, Riess, Rodney,
  Strolger, Casertano, Castellano, Dahlen, Dickinson, Dolch, Fontana,
  Giavalisco, Grazian, Guo, Hathi, Huang, van~der Wel, Yan, Acquaviva,
  Alexander, Almaini, Ashby, Barden, Bell, Bournaud, Brown, Caputi, Cassata,
  Challis, Chary, Cheung, Cirasuolo, Conselice, Cooray, Croton, Daddi,
  Dav\'{e}, de~Mello, de~Ravel, Dekel, Donley, Dunlop, Dutton, Elbaz, Fazio,
  Filippenko, Finkelstein, Frazer, Gardner, Garnavich, Gawiser, Gruetzbauch,
  Hartley, H\"{a}ussler, Herrington, Hopkins, Huang, Jha, Johnson, Kartaltepe,
  Khostovan, Kirshner, Lani, Lee, Li, Madau, McCarthy, McIntosh, McLure,
  McPartland, Mobasher, Moreira, Mortlock, Moustakas, Mozena, Nandra, Newman,
  Nielsen, Niemi, Noeske, Papovich, Pentericci, Pope, Primack, Ravindranath,
  Reddy, Renzini, Rix, Robaina, Rosario, Rosati, Salimbeni, Scarlata, Siana,
  Simard, Smidt, Snyder, Somerville, Spinrad, Straughn, Telford, Teplitz,
  Trump, Vargas, Villforth, Wagner, Wandro, Wechsler, Weiner, Wiklind, Wild,
  Wilson, Wuyts, \& Yun}]{Koekemoer2011}
Koekemoer A.~M. {et~al.}, 2011, ApJS, 197, 36

\bibitem[{Laureijs {et~al}\mbox{.}(2011)Laureijs, Gondoin, Duvet, {Saavedra
  Criado}, Hoar, Amiaux, Augu\`{e}res, Cole, Cropper, Ealet, Ferruit, {Escudero
  Sanz}, Jahnke, Kohley, Maciaszek, Mellier, Oosterbroek, Pasian, Sauvage,
  Scaramella, Sirianni, \& Valenziano}]{Laureijs2011}
Laureijs R. {et~al.}, 2011, preprint (arXiv:1110.3193)

\bibitem[{Lawrence {et~al}\mbox{.}(2007)Lawrence, Warren, Almaini, Edge,
  Hambly, Jameson, Lucas, Casali, Adamson, Dye, Emerson, Foucaud, Hewett,
  Hirst, Hodgkin, Irwin, Lodieu, McMahon, Simpson, Smail, Mortlock, \&
  Folger}]{Lawrence2007}
Lawrence A. {et~al.}, 2007, MNRAS, 379, 1599

\bibitem[{Loveday {et~al}\mbox{.}(2012)Loveday, Norberg, Baldry, Driver,
  Hopkins, Peacock, Bamford, Liske, Bland-Hawthorn, Brough, Brown, Cameron,
  Conselice, Croom, Frenk, Gunawardhana, Hill, Jones, Kelvin, Kuijken, Nichol,
  Parkinson, Phillipps, Pimbblet, Popescu, Prescott, Robotham, Sharp,
  Sutherland, Taylor, Thomas, Tuffs, van Kampen, \& Wijesinghe}]{Loveday2012}
Loveday J. {et~al.}, 2012, MNRAS, 420, 1239

\bibitem[{Madau(1995)}]{Madau1995}
Madau P., 1995, ApJ, 441, 18

\bibitem[{Madau \& Dickinson(2014)}]{Madau2014}
Madau P., Dickinson M., 2014, ARA\&A, 52, 415

\bibitem[{Mason {et~al}\mbox{.}(2015)Mason, Treu, Schmidt, Collett, Trenti,
  Marshall, Barone-Nugent, Bradley, Stiavelli, \& Wyithe}]{Mason2015}
Mason C.~A. {et~al.}, 2015, preprint (arXiv:1502.03795)

\bibitem[{McCracken {et~al}\mbox{.}(2013)McCracken, Milvang-Jensen, Dunlop,
  Franx, Fynbo, {Le Fevre}, Holt, Caputi, Goranova, Buitrago, Emerson,
  Freudling, Herent, \& Hudelot}]{McCracken2013}
McCracken H.~J. {et~al.}, 2013, The Messenger, 154, 29

\bibitem[{McCracken {et~al}\mbox{.}(2012)McCracken, Milvang-Jensen, Dunlop,
  Franx, Fynbo, {Le F\`{e}vre}, Holt, Caputi, Goranova, Buitrago, Emerson,
  Freudling, Hudelot, L\'{o}pez-Sanjuan, Magnard, Mellier, M\o~ller, Nilsson,
  Sutherland, Tasca, \& Zabl}]{McCracken2012}
McCracken H.~J. {et~al.}, 2012, A\&A, 544, A156

\bibitem[{McCracken {et~al}\mbox{.}(2007)McCracken, Peacock, Guzzo, Capak,
  Porciani, Scoville, Aussel, Finoguenov, James, Kitzbichler, Koekemoer,
  Leauthaud, {Le Fevre}, Massey, Mellier, Mobasher, Norberg, Rhodes, Sanders,
  Sasaki, Taniguchi, Thompson, White, \& El‐Zant}]{McCracken2007}
McCracken H.~J. {et~al.}, 2007, ApJS, 172, 314

\bibitem[{McGreer {et~al}\mbox{.}(2013)McGreer, Jiang, Fan, Richards, Strauss,
  Ross, White, Shen, Schneider, Myers, Brandt, DeGraf, Glikman, Ge, \&
  Streblyanska}]{McGreer2013}
McGreer I.~D. {et~al.}, 2013, ApJ, 768, 105

\bibitem[{McLure {et~al}\mbox{.}(2009)McLure, Cirasuolo, Dunlop, Foucaud, \&
  Almaini}]{McLure2009}
McLure R.~J., Cirasuolo M., Dunlop J.~S., Foucaud S., Almaini O., 2009, MNRAS,
  395, 2196

\bibitem[{McLure {et~al}\mbox{.}(2006)McLure, Cirasuolo, Dunlop, Sekiguchi,
  Almaini, Foucaud, Simpson, Watson, Hirst, Page, \& Smail}]{McLure2006}
McLure R.~J. {et~al.}, 2006, MNRAS, 372, 357

\bibitem[{McLure {et~al}\mbox{.}(2013)McLure, Dunlop, Bowler, Curtis-Lake,
  Schenker, Ellis, Robertson, Koekemoer, Rogers, Ono, Ouchi, Charlot, Wild,
  Stark, Furlanetto, Cirasuolo, \& Targett}]{McLure2013}
McLure R.~J. {et~al.}, 2013, MNRAS, 432, 2696

\bibitem[{McLure {et~al}\mbox{.}(2010)McLure, Dunlop, Cirasuolo, Koekemoer,
  Sabbi, Stark, Targett, \& Ellis}]{McLure2010}
McLure R.~J., Dunlop J.~S., Cirasuolo M., Koekemoer A.~M., Sabbi E., Stark
  D.~P., Targett T.~A., Ellis R.~S., 2010, MNRAS, 403, 960

\bibitem[{McNaught-Roberts {et~al}\mbox{.}(2014)McNaught-Roberts, Norberg,
  Baugh, Lacey, Loveday, Peacock, Baldry, Bland-Hawthorn, Brough, Driver,
  Robotham, \& Vazquez-Mata}]{McNaughtRoberts2014}
McNaught-Roberts T. {et~al.}, 2014, MNRAS, 445, 2125

\bibitem[{Meneux {et~al}\mbox{.}(2009)Meneux, Guzzo, de~la Torre, Porciani,
  Zamorani, Abbas, Bolzonella, Garilli, Iovino, Pozzetti, Zucca, Lilly, {Le
  F\`{e}vre}, Kneib, Carollo, Contini, Mainieri, Renzini, Scodeggio, Bardelli,
  Bongiorno, Caputi, Coppa, Cucciati, de~Ravel, Franzetti, Kampczyk, Knobel,
  Kova\v{c}, Lamareille, {Le Borgne}, {Le Brun}, Maier, Pell\`{o}, Peng, {Perez
  Montero}, Ricciardelli, Silverman, Tanaka, Tasca, Tresse, Vergani, Bottini,
  Cappi, Cimatti, Cassata, Fumana, Koekemoer, Leauthaud, Maccagni, Marinoni,
  McCracken, Memeo, Oesch, \& Scaramella}]{Meneux2009}
Meneux B. {et~al.}, 2009, A\&A, 505, 463

\bibitem[{Meurer {et~al}\mbox{.}(1999)Meurer, Heckman, \&
  Calzetti}]{Meurer1999}
Meurer G.~R., Heckman T.~M., Calzetti D., 1999, ApJ, 521, 64

\bibitem[{Mortlock {et~al}\mbox{.}(2014)Mortlock, Conselice, Hartley, Duncan,
  Lani, Ownsworth, Almaini, Wel, Huang, Ashby, Willner, Fontana, Dekel,
  Koekemoer, Ferguson, Faber, Grogin, \& Kocevski}]{Mortlock2014}
Mortlock A. {et~al.}, 2014, MNRAS, 447, 2

\bibitem[{Mu\~{n}oz \& Loeb(2008)}]{Munoz2008}
Mu\~{n}oz J.~A., Loeb A., 2008, MNRAS, 386, 2323

\bibitem[{Murray {et~al}\mbox{.}(2013)Murray, Power, \& Robotham}]{Murray2013}
Murray S., Power C., Robotham A., 2013, A\&C, 3, 23

\bibitem[{Oesch {et~al}\mbox{.}(2010)Oesch, Bouwens, Illingworth, Carollo,
  Franx, Labb\'{e}, Magee, Stiavelli, Trenti, \& van Dokkum}]{Oesch2010}
Oesch P.~A. {et~al.}, 2010, ApJ, 709, L16

\bibitem[{Oesch {et~al}\mbox{.}(2012)Oesch, Bouwens, Illingworth, Gonzalez,
  Trenti, van Dokkum, Franx, Labb\'{e}, Carollo, \& Magee}]{Oesch2012b}
Oesch P.~A. {et~al.}, 2012, ApJ, 759, 135

\bibitem[{Oesch {et~al}\mbox{.}(2013)Oesch, Bouwens, Illingworth, Labb\'{e},
  Franx, van Dokkum, Trenti, Stiavelli, Gonzalez, \& Magee}]{Oesch2013}
Oesch P.~A. {et~al.}, 2013, ApJ, 773, 75

\bibitem[{Oke(1974)}]{Oke1974}
Oke J.~B., 1974, ApJS, 27, 21

\bibitem[{Oke \& Gunn(1983)}]{Oke1983}
Oke J.~B., Gunn J.~E., 1983, ApJ, 266, 713

\bibitem[{Ono {et~al}\mbox{.}(2012)Ono, Ouchi, Mobasher, Dickinson, Penner,
  Shimasaku, Weiner, Kartaltepe, Nakajima, Nayyeri, Stern, Kashikawa, \&
  Spinrad}]{Ono2012}
Ono Y. {et~al.}, 2012, ApJ, 744, 83

\bibitem[{Ouchi {et~al}\mbox{.}(2009)Ouchi, Mobasher, Shimasaku, Ferguson,
  Fall, Ono, Kashikawa, Morokuma, Nakajima, Okamura, Dickinson, Giavalisco, \&
  Ohta}]{Ouchi2009}
Ouchi M. {et~al.}, 2009, ApJ, 706, 1136

\bibitem[{Ouchi {et~al}\mbox{.}(2005)Ouchi, Shimasaku, Akiyama, Sekiguchi,
  Furusawa, Okamura, Kashikawa, Iye, Kodama, Saito, Sasaki, Simpson, Takata,
  Yamada, Yamanoi, Yoshida, \& Yoshida}]{Ouchi2005}
Ouchi M. {et~al.}, 2005, ApJ, 620, L1

\bibitem[{Paardekooper {et~al}\mbox{.}(2013)Paardekooper, Khochfar, \&
  Dalla}]{Paardekooper2013}
Paardekooper J.-P., Khochfar S., Dalla C.~V., 2013, MNRAS: Letters, 429, L94

\bibitem[{Peng {et~al}\mbox{.}(2010)Peng, Lilly, Kova\v{c}, Bolzonella,
  Pozzetti, Renzini, Zamorani, Ilbert, Knobel, Iovino, Maier, Cucciati, Tasca,
  Carollo, Silverman, Kampczyk, de~Ravel, Sanders, Scoville, Contini, Mainieri,
  Scodeggio, Kneib, {Le F\`{e}vre}, Bardelli, Bongiorno, Caputi, Coppa, de~la
  Torre, Franzetti, Garilli, Lamareille, {Le Borgne}, {Le Brun}, Mignoli,
  Montero, Pello, Ricciardelli, Tanaka, Tresse, Vergani, Welikala, Zucca,
  Oesch, Abbas, Barnes, Bordoloi, Bottini, Cappi, Cassata, Cimatti, Fumana,
  Hasinger, Koekemoer, Leauthaud, Maccagni, Marinoni, McCracken, Memeo, Meneux,
  Nair, Porciani, Presotto, \& Scaramella}]{Peng2010}
Peng Y. {et~al.}, 2010, ApJ, 721, 193

\bibitem[{Pickles(1998)}]{Pickles1998}
Pickles A.~J., 1998, PASP, 110, 863

\bibitem[{Pirzkal {et~al}\mbox{.}(2009)Pirzkal, Burgasser, Malhotra, Holwerda,
  Sahu, Rhoads, Xu, Bochanski, Walsh, Windhorst, Hathi, \& Cohen}]{Pirzkal2009}
Pirzkal N. {et~al.}, 2009, ApJ, 695, 1591

\bibitem[{Press {et~al}\mbox{.}(1992)Press, Teukolsky, Vetterling, \&
  Flannery}]{Press1992}
Press W.~H., Teukolsky S.~A., Vetterling W.~T., Flannery B.~P., 1992, Numerical
  recipes in FORTRAN. The art of scientific computing, CUP

\bibitem[{Reed {et~al}\mbox{.}(2007)Reed, Bower, Frenk, Jenkins, \&
  Theuns}]{Reed2007}
Reed D.~S., Bower R., Frenk C.~S., Jenkins A., Theuns T., 2007, MNRAS, 374, 2

\bibitem[{Robertson {et~al}\mbox{.}(2013)Robertson, Furlanetto, Schneider,
  Charlot, Ellis, Stark, McLure, Dunlop, Koekemoer, Schenker, Ouchi, Ono,
  Curtis-Lake, Rogers, Bowler, \& Cirasuolo}]{Robertson2013}
Robertson B.~E. {et~al.}, 2013, ApJ, 768, 71

\bibitem[{Rogers {et~al}\mbox{.}(2013)Rogers, McLure, \& Dunlop}]{Rogers2013}
Rogers A.~B., McLure R.~J., Dunlop J.~S., 2013, MNRAS, 429, 2456

\bibitem[{Rogers {et~al}\mbox{.}(2014)Rogers, McLure, Dunlop, Bowler,
  Curtis-Lake, Dayal, Faber, Ferguson, Finkelstein, Grogin, Hathi, Kocevski,
  Koekemoer, \& Kurczynski}]{Rogers2014}
Rogers A.~B. {et~al.}, 2014, MNRAS, 440, 3714

\bibitem[{Ryan {et~al}\mbox{.}(2011)Ryan, Thorman, Yan, Fan, Yan, Mechtley,
  Hathi, Cohen, Windhorst, McCarthy, \& Wittman}]{Ryan2011}
Ryan R.~E. {et~al.}, 2011, ApJ, 739, 83

\bibitem[{Rykoff {et~al}\mbox{.}(2008)Rykoff, Evrard, McKay, Becker, Johnston,
  Koester, Nord, Rozo, Sheldon, Stanek, \& Wechsler}]{Rykoff2008}
Rykoff E.~S. {et~al.}, 2008, MNRAS: Letters, 387, L28

\bibitem[{Salim \& Lee(2012)}]{Salim2012}
Salim S., Lee J.~C., 2012, ApJ, 758, 134

\bibitem[{Saunders {et~al}\mbox{.}(1990)Saunders, Rowan-Robinson, Lawrence,
  Efstathiou, Kaiser, Ellis, \& Frenk}]{Saunders1990}
Saunders W., Rowan-Robinson M., Lawrence A., Efstathiou G., Kaiser N., Ellis
  R.~S., Frenk C.~S., 1990, MNRAS, 242, 318

\bibitem[{Schechter(1976)}]{Schechter1976}
Schechter P., 1976, ApJ, 203, 297

\bibitem[{Schenker {et~al}\mbox{.}(2013)Schenker, Robertson, Ellis, Ono,
  McLure, Dunlop, Koekemoer, Bowler, Ouchi, Curtis-Lake, Rogers, Schneider,
  Charlot, Stark, Furlanetto, \& Cirasuolo}]{Schenker2013}
Schenker M.~A. {et~al.}, 2013, ApJ, 768, 196

\bibitem[{Schmidt {et~al}\mbox{.}(2014)Schmidt, Treu, Trenti, Bradley, Kelly,
  Oesch, Holwerda, Shull, \& Stiavelli}]{Schmidt2014}
Schmidt K.~B. {et~al.}, 2014, ApJ, 786, 57

\bibitem[{Schmidt(1968)}]{Schmidt1968}
Schmidt M., 1968, ApJ, 151, 393

\bibitem[{Scoville {et~al}\mbox{.}(2007)Scoville, Abraham, Aussel, Barnes,
  Benson, Blain, Calzetti, Comastri, Capak, Carilli, Carlstrom, Carollo,
  Colbert, Daddi, Ellis, Elvis, Ewald, Fall, Franceschini, Giavalisco, Green,
  Griffiths, Guzzo, Hasinger, Impey, Kneib, Koda, Koekemoer, Lefevre, Lilly,
  Liu, McCracken, Massey, Mellier, Miyazaki, Mobasher, Mould, Norman,
  Refregier, Renzini, Rhodes, Rich, Sanders, Schiminovich, Schinnerer,
  Scodeggio, Sheth, Shopbell, Taniguchi, Tyson, Urry, {Van Waerbeke},
  Vettolani, White, \& Yan}]{Scoville2007a}
Scoville N. {et~al.}, 2007, ApJS, 172, 38

\bibitem[{Skibba {et~al}\mbox{.}(2014)Skibba, Smith, Coil, Moustakas, Aird,
  Blanton, Bray, Cool, Eisenstein, Mendez, Wong, \& Zhu}]{Skibba2014}
Skibba R.~A. {et~al.}, 2014, ApJ, 784, 128

\bibitem[{Soifer {et~al}\mbox{.}(1987)Soifer, Sanders, Madore, Neugebauer,
  Danielson, Elias, Lonsdale, \& Rice}]{Soifer1987}
Soifer B.~T., Sanders D.~B., Madore B.~F., Neugebauer G., Danielson G.~E.,
  Elias J.~H., Lonsdale C.~J., Rice W.~L., 1987, ApJ, 320, 238

\bibitem[{Stanway \& Davies(2014)}]{Stanway2014}
Stanway E.~R., Davies L. J.~M., 2014, MNRAS, 439, 2474

\bibitem[{Stark {et~al}\mbox{.}(2011)Stark, Ellis, \& Ouchi}]{Stark2011}
Stark D.~P., Ellis R.~S., Ouchi M., 2011, ApJ, 728, L2

\bibitem[{Steidel {et~al}\mbox{.}(1999)Steidel, Adelberger, Giavalisco,
  Dickinson, \& Pettini}]{Steidel1999}
Steidel C.~C., Adelberger K.~L., Giavalisco M., Dickinson M., Pettini M., 1999,
  ApJ, 519, 1

\bibitem[{Steidel \& Hamilton(1992)}]{Steidel1992}
Steidel C.~C., Hamilton D., 1992, AJ, 104, 941

\bibitem[{Takeuchi {et~al}\mbox{.}(2012)Takeuchi, Yuan, Ikeyama, Murata, \&
  Inoue}]{Takeuchi2012}
Takeuchi T.~T., Yuan F.-T., Ikeyama A., Murata K.~L., Inoue A.~K., 2012, ApJ,
  755, 144

\bibitem[{Tempel {et~al}\mbox{.}(2009)Tempel, Einasto, Einasto, Saar, \&
  Tago}]{Tempel2009}
Tempel E., Einasto J., Einasto M., Saar E., Tago E., 2009, A\&A, 495, 37

\bibitem[{Tempel {et~al}\mbox{.}(2014)Tempel, Tamm, Gramann, Tuvikene,
  Liivam\"{a}gi, Suhhonenko, Kipper, Einasto, \& Saar}]{Tempel2014}
Tempel E. {et~al.}, 2014, A\&A, 566, A1

\bibitem[{Trenti {et~al}\mbox{.}(2011)Trenti, Bradley, Stiavelli, Oesch, Treu,
  Bouwens, Shull, MacKenty, Carollo, \& Illingworth}]{Trenti2011}
Trenti M. {et~al.}, 2011, ApJ, 727, L39

\bibitem[{Trenti \& Stiavelli(2008)}]{Trenti2008}
Trenti M., Stiavelli M., 2008, ApJ, 676, 767

\bibitem[{van~der Burg {et~al}\mbox{.}(2010)van~der Burg, Hildebrandt, \&
  Erben}]{vanderBurg2010}
van~der Burg R. F.~J., Hildebrandt H., Erben T., 2010, A\&A, 523, A74

\bibitem[{Willott {et~al}\mbox{.}(2015)Willott, Carilli, Wagg, \&
  Wang}]{Willott2015}
Willott C.~J., Carilli C.~L., Wagg J., Wang R., 2015, preprint
  (arXiv:1504.05875)

\bibitem[{Willott {et~al}\mbox{.}(2009)Willott, Delorme, Reyl\'{e}, Albert,
  Bergeron, Crampton, Delfosse, Forveille, Hutchings, McLure, Omont, \&
  Schade}]{Willott2009}
Willott C.~J. {et~al.}, 2009, AJ, 137, 3541

\bibitem[{Willott {et~al}\mbox{.}(2013)Willott, McLure, Hibon, Bielby,
  McCracken, Kneib, Ilbert, Bonfield, Bruce, \& Jarvis}]{Willott2013}
Willott C.~J. {et~al.}, 2013, AJ, 145, 4

\bibitem[{Wyithe {et~al}\mbox{.}(2011)Wyithe, Yan, Windhorst, \&
  Mao}]{Wyithe2011}
Wyithe J. S.~B., Yan H., Windhorst R.~A., Mao S., 2011, Nat, 469, 181

\end{thebibliography}

\appendix

\section{Schechter, DPL and Saunders functional forms}\label{sect:equations}

For reference, the Schechter function parameterisation of the LF in magnitudes is: 
\begin{equation}
\phi(M) = 0.4 \,{\rm ln} 10 \,\phi^* [10^{-0.4(M - M^*)}]^{(1+\alpha)} e^{(-10^{0.4(M-M^*)})}
\end{equation}
 where $M^*$ and $\phi^*$ are the characteristic magnitude and number density respectively, and $\alpha$ denotes the faint-end slope.  The double power law (DPL) function parameterisation of the LF is: 
 \begin{equation}
 \phi(M) = \frac{\phi^*}{10^{0.4(\alpha + 1)(M - M^*)} + 10^{0.4(\beta + 1)(M - M^*)}}
\end{equation} 
Here an additional parameter $\beta_{}$ determines the slope of the bright-end of the LF, as opposed to the assumed exponential decline in the Schechter function.
Finally, the Saunders function (which has been highlighted as appropriate choice to fit the LF by~\citealt{Salim2012}) is parameterised as:
\begin{equation}
  \phi (M) = 0.4 \,{\rm ln}10\, \phi^*[10^{- \Delta M'}]^{(\alpha + 1)}\, {\rm exp}\left(-\frac{{\rm log}^2( 1 + 10^{- \Delta M'})}{2\sigma^2}\right)
\end{equation}
with $\Delta M' = 0.4 (M - M^*)$.

\section{Potential Brown Dwarfs}\label{sect:possible stars}

In this appendix we present the high-redshift galaxy candidates that were excluded from the final sample of $z\simeq 6$ objects based on a good stellar fit to the photometry.

\begin{table*}
\caption{
High-redshift galaxy candidates that were excluded based on a good fit to a stellar template (defined as $\chi^{2}_{\star} < 10.0$).
Objects have been ordered by their best-fitting galaxy photometric redshift and separated by field (UltraVISTA/COSMOS DR2 in the upper part of the table, followed by objects in the UDS/SXDS).
Note that the brightest object (S1) formally has a somewhat poor stellar fit, however this object is clearly stellar from inspection of the SED fit.
The large $\chi^{2}_{\star}$ here is a result of the limited template set available and high-S/N photometry. 
}

\begin{tabular}{ll l c c c c c c c rc r l}

\hline
\multicolumn{13}{c}{UltraVISTA/COSMOS field}\\
\hline
ID & R.A. & Dec. & $i$ & $z'$ & $Y$ & $J$ & $H$ & $K$ & Star & $\chi_{\star}^2$ & $z_{\rm phot}$ & $\chi_{\rm gal}^2$  \\
 & (J2000)& (J2000)& & & & & & & type & & &  \\
\hline
S1 & 10:00:10.71 & +02:06:37.1 & $  27.2_{-  0.4}^{+  0.6} $ & $  23.8_{-  0.1}^{+  0.1} $ & $  22.6_{-  0.1}^{+  0.1} $ & $  21.6_{-  0.1}^{+  0.1} $ & $  21.0_{-  0.1}^{+  0.1} $ & $  20.7_{-  0.1}^{+  0.1} $ & L8 & $      27.5$ & $       6.4$ & $      11.1$ \\
S2 & 10:00:23.90 & +01:59:05.6 & $  27.1_{-  0.4}^{+  0.7} $ & $  25.1_{-  0.1}^{+  0.1} $ & $  24.3_{-  0.1}^{+  0.1} $ & $  23.9_{-  0.1}^{+  0.1} $ & $  23.7_{-  0.1}^{+  0.1} $ & $  23.6_{-  0.1}^{+  0.1} $ & M8 & $       3.9$ & $       6.3$ & $       2.9$ \\
S3 & 10:00:09.93 & +02:22:07.2 & $ >  27.0 $ & $  25.5_{-  0.1}^{+  0.1} $ & $  25.1_{-  0.2}^{+  0.3} $ & $  24.5_{-  0.3}^{+  0.3} $ & $  25.1_{-  0.4}^{+  0.5} $ & $ >  25.5 $ & M4 & $       6.7$ & $       6.3$ & $       4.8$ \\
S4 & 10:02:14.87 & +02:11:04.9 & $  26.8_{-  0.3}^{+  0.4} $ & $  24.9_{-  0.1}^{+  0.1} $ & $  24.2_{-  0.1}^{+  0.1} $ & $  23.9_{-  0.1}^{+  0.1} $ & $  24.0_{-  0.1}^{+  0.1} $ & $  23.7_{-  0.1}^{+  0.1} $ & M7 & $       9.0$ & $       6.2$ & $       4.9$ \\
S5 & 09:58:56.08 & +02:35:08.1 & $ >  27.3 $ & $  24.9_{-  0.1}^{+  0.1} $ & $  24.1_{-  0.1}^{+  0.1} $ & $  23.2_{-  0.1}^{+  0.1} $ & $  22.7_{-  0.1}^{+  0.1} $ & $  22.3_{-  0.1}^{+  0.1} $ & L8 & $       4.6$ & $       6.2$ & $       9.6$ \\
S6 & 09:58:42.41 & +02:26:06.7 & $  26.9_{-  0.3}^{+  0.5} $ & $  24.8_{-  0.1}^{+  0.1} $ & $  24.3_{-  0.1}^{+  0.1} $ & $  23.9_{-  0.1}^{+  0.1} $ & $  24.0_{-  0.1}^{+  0.2} $ & $  23.9_{-  0.1}^{+  0.1} $ & M7 & $       7.0$ & $       6.2$ & $       5.4$ \\
S7 & 10:02:07.38 & +02:25:44.1 & $ >  27.4 $ & $  25.6_{-  0.1}^{+  0.1} $ & $  25.3_{-  0.2}^{+  0.3} $ & $  24.7_{-  0.2}^{+  0.2} $ & $  24.6_{-  0.2}^{+  0.3} $ & $  25.0_{-  0.2}^{+  0.3} $ & M7 & $       3.0$ & $       6.2$ & $       4.1$ \\
S8 & 10:02:02.78 & +02:24:00.0 & $  27.1_{-  0.4}^{+  0.7} $ & $  25.0_{-  0.1}^{+  0.1} $ & $  24.5_{-  0.1}^{+  0.2} $ & $  24.4_{-  0.1}^{+  0.1} $ & $  24.1_{-  0.1}^{+  0.2} $ & $  24.4_{-  0.2}^{+  0.2} $ & M7 & $       6.5$ & $       6.2$ & $       3.5$ \\
S9 & 10:00:53.64 & +02:09:45.5 & $ >  27.9 $ & $  25.7_{-  0.1}^{+  0.1} $ & $  25.3_{-  0.2}^{+  0.2} $ & $  24.8_{-  0.2}^{+  0.2} $ & $  25.3_{-  0.3}^{+  0.5} $ & $  25.2_{-  0.3}^{+  0.4} $ & M7 & $       8.6$ & $       6.2$ & $       3.7$ \\
S10 & 10:00:45.18 & +02:31:40.3 & $ >  27.3 $ & $  25.6_{-  0.1}^{+  0.1} $ & $  25.1_{-  0.1}^{+  0.1} $ & $  25.1_{-  0.3}^{+  0.4} $ & $  25.1_{-  0.4}^{+  0.6} $ & $  25.1_{-  0.2}^{+  0.3} $ & M7 & $       9.7$ & $       6.2$ & $       0.1$ \\
S11 & 09:58:45.50 & +02:23:24.7 & $ >  27.6 $ & $  25.4_{-  0.1}^{+  0.2} $ & $  25.2_{-  0.2}^{+  0.3} $ & $  24.5_{-  0.2}^{+  0.2} $ & $  24.8_{-  0.3}^{+  0.3} $ & $  25.1_{-  0.3}^{+  0.4} $ & M7 & $       9.1$ & $       6.2$ & $       5.7$ \\
S12 & 09:58:45.02 & +02:29:04.3 & $ >  26.2 $ & $  25.7_{-  0.1}^{+  0.2} $ & $ >  25.4 $ & $ >  25.6 $ & $ >  25.2 $ & $ >  25.0 $ & M4 & $       5.6$ & $       5.9$ & $       1.2$ \\
S13 & 10:01:55.84 & +02:37:51.9 & $ >  27.7 $ & $  25.7_{-  0.1}^{+  0.1} $ & $  25.7_{-  0.3}^{+  0.4} $ & $  24.9_{-  0.2}^{+  0.3} $ & $  24.7_{-  0.2}^{+  0.2} $ & $  24.6_{-  0.2}^{+  0.2} $ & M7 & $       6.5$ & $       5.7$ & $       1.9$ \\
S14 & 09:59:29.41 & +01:46:40.3 & $  27.4_{-  0.4}^{+  0.7} $ & $  25.4_{-  0.1}^{+  0.1} $ & $  25.1_{-  0.2}^{+  0.3} $ & $  25.1_{-  0.3}^{+  0.4} $ & $  25.1_{-  0.3}^{+  0.5} $ & $  25.1_{-  0.4}^{+  0.5} $ & M5 & $       9.4$ & $       5.7$ & $       1.1$ \\
S15 & 10:02:00.39 & +02:23:52.3 & $ >  27.1 $ & $  25.6_{-  0.1}^{+  0.1} $ & $  25.8_{-  0.4}^{+  0.6} $ & $  25.5_{-  0.3}^{+  0.4} $ & $ >  25.7 $ & $ >  25.5 $ & M4 & $       8.2$ & $       5.7$ & $       0.5$ \\
S16 & 09:58:53.79 & +02:22:37.1 & $ >  26.8 $ & $  25.2_{-  0.1}^{+  0.1} $ & $  24.9_{-  0.2}^{+  0.2} $ & $  24.6_{-  0.2}^{+  0.2} $ & $  24.3_{-  0.2}^{+  0.2} $ & $  24.2_{-  0.1}^{+  0.1} $ & M5 & $       5.9$ & $       5.7$ & $       0.5$ \\
S17 & 09:59:15.77 & +01:51:06.2 & $ >  27.5 $ & $  25.7_{-  0.1}^{+  0.1} $ & $  25.9_{-  0.3}^{+  0.4} $ & $  25.3_{-  0.2}^{+  0.3} $ & $  24.9_{-  0.3}^{+  0.3} $ & $  25.0_{-  0.3}^{+  0.3} $ & M5 & $       9.9$ & $       5.7$ & $       2.0$ \\
S18 & 09:59:07.45 & +02:35:03.2 & $  27.0_{-  0.3}^{+  0.4} $ & $  25.3_{-  0.1}^{+  0.1} $ & $  25.0_{-  0.1}^{+  0.2} $ & $  24.9_{-  0.2}^{+  0.2} $ & $  24.3_{-  0.2}^{+  0.2} $ & $  24.8_{-  0.2}^{+  0.2} $ & M5 & $       9.0$ & $       5.7$ & $       6.2$ \\
S19 & 10:00:07.62 & +02:21:25.0 & $  26.7_{-  0.3}^{+  0.5} $ & $  25.4_{-  0.1}^{+  0.1} $ & $  25.4_{-  0.3}^{+  0.4} $ & $ >  25.2 $ & $ >  25.1 $ & $ >  25.3 $ & M4 & $       8.9$ & $       5.7$ & $       0.9$ \\
S20 & 10:02:18.39 & +02:20:25.9 & $  26.9_{-  0.3}^{+  0.4} $ & $  25.2_{-  0.1}^{+  0.1} $ & $  24.8_{-  0.1}^{+  0.2} $ & $  24.6_{-  0.1}^{+  0.2} $ & $  24.4_{-  0.2}^{+  0.2} $ & $  24.6_{-  0.2}^{+  0.2} $ & M5 & $       3.4$ & $       5.7$ & $       6.0$ \\
S21 & 10:01:42.49 & +02:38:24.0 & $  26.5_{-  0.2}^{+  0.3} $ & $  24.8_{-  0.1}^{+  0.1} $ & $  24.6_{-  0.2}^{+  0.2} $ & $  24.3_{-  0.1}^{+  0.1} $ & $  24.3_{-  0.2}^{+  0.2} $ & $  24.4_{-  0.2}^{+  0.2} $ & M5 & $       7.6$ & $       5.6$ & $       4.0$ \\
S22 & 10:01:47.42 & +02:06:18.9 & $  27.0_{-  0.4}^{+  0.6} $ & $  25.5_{-  0.1}^{+  0.1} $ & $  25.6_{-  0.2}^{+  0.2} $ & $  25.2_{-  0.3}^{+  0.3} $ & $  25.4_{-  0.4}^{+  0.6} $ & $  25.4_{-  0.3}^{+  0.4} $ & M4 & $       8.3$ & $       5.6$ & $       1.0$ \\
S23 & 10:01:54.85 & +02:33:33.4 & $  27.3_{-  0.4}^{+  0.7} $ & $  25.5_{-  0.1}^{+  0.1} $ & $  25.2_{-  0.2}^{+  0.3} $ & $  25.1_{-  0.3}^{+  0.4} $ & $  25.0_{-  0.3}^{+  0.4} $ & $  24.6_{-  0.2}^{+  0.3} $ & M5 & $       6.4$ & $       5.6$ & $       0.7$ \\
S24 & 09:58:44.72 & +02:25:37.2 & $  27.1_{-  0.3}^{+  0.5} $ & $  25.6_{-  0.1}^{+  0.2} $ & $  25.5_{-  0.3}^{+  0.3} $ & $  25.4_{-  0.3}^{+  0.4} $ & $ >  25.5 $ & $  25.4_{-  0.3}^{+  0.4} $ & M4 & $       5.5$ & $       5.6$ & $       0.6$ \\
S25 & 10:00:38.46 & +01:56:22.3 & $  26.6_{-  0.2}^{+  0.2} $ & $  25.2_{-  0.1}^{+  0.1} $ & $  24.8_{-  0.1}^{+  0.1} $ & $  24.5_{-  0.1}^{+  0.2} $ & $  24.5_{-  0.2}^{+  0.2} $ & $  24.8_{-  0.2}^{+  0.2} $ & M5 & $       4.9$ & $       5.6$ & $       9.9$ \\
S26 & 09:58:41.39 & +02:23:03.7 & $  26.9_{-  0.2}^{+  0.3} $ & $  25.4_{-  0.1}^{+  0.1} $ & $  25.2_{-  0.2}^{+  0.3} $ & $  25.5_{-  0.4}^{+  0.7} $ & $  25.1_{-  0.3}^{+  0.4} $ & $  25.2_{-  0.3}^{+  0.4} $ & M4 & $       7.7$ & $       5.5$ & $       1.6$ \\
S27 & 10:01:35.94 & +02:23:03.3 & $  27.0_{-  0.3}^{+  0.4} $ & $  25.5_{-  0.1}^{+  0.1} $ & $  25.2_{-  0.3}^{+  0.3} $ & $  24.5_{-  0.2}^{+  0.2} $ & $  24.9_{-  0.3}^{+  0.4} $ & $ >  25.1 $ & M5 & $       4.8$ & $       5.5$ & $       7.7$ \\
S28 & 10:02:15.54 & +02:36:46.9 & $  26.8_{-  0.2}^{+  0.3} $ & $  25.7_{-  0.1}^{+  0.1} $ & $  25.4_{-  0.2}^{+  0.2} $ & $  25.8_{-  0.4}^{+  0.6} $ & $  25.1_{-  0.4}^{+  0.7} $ & $ >  25.8 $ & M4 & $       7.5$ & $       5.5$ & $       5.2$ \\
S29 & 10:02:25.51 & +02:33:32.1 & $  26.7_{-  0.2}^{+  0.3} $ & $  25.4_{-  0.1}^{+  0.1} $ & $  24.9_{-  0.2}^{+  0.3} $ & $  24.5_{-  0.2}^{+  0.3} $ & $  24.7_{-  0.2}^{+  0.3} $ & $ >  25.3 $ & M5 & $       7.3$ & $       5.5$ & $      10.0$ \\
S30 & 10:02:21.87 & +02:24:01.0 & $  26.7_{-  0.3}^{+  0.4} $ & $  25.3_{-  0.1}^{+  0.1} $ & $  25.2_{-  0.3}^{+  0.4} $ & $  25.4_{-  0.4}^{+  0.7} $ & $  25.1_{-  0.4}^{+  0.5} $ & $  24.7_{-  0.2}^{+  0.3} $ & M4 & $       7.7$ & $       5.5$ & $       1.8$ \\
S31 & 10:00:08.43 & +02:20:49.1 & $  26.8_{-  0.3}^{+  0.4} $ & $  25.3_{-  0.1}^{+  0.1} $ & $  25.6_{-  0.3}^{+  0.5} $ & $  25.2_{-  0.3}^{+  0.4} $ & $  25.0_{-  0.4}^{+  0.6} $ & $  25.1_{-  0.3}^{+  0.4} $ & M4 & $       7.5$ & $       5.5$ & $       0.9$ \\
S32 & 10:02:05.41 & +02:14:46.4 & $  26.9_{-  0.3}^{+  0.5} $ & $  25.5_{-  0.1}^{+  0.1} $ & $  25.5_{-  0.2}^{+  0.3} $ & $  25.0_{-  0.2}^{+  0.2} $ & $  25.3_{-  0.4}^{+  0.6} $ & $  25.1_{-  0.3}^{+  0.3} $ & M4 & $       5.6$ & $       5.5$ & $       2.1$ \\
S33 & 10:01:40.07 & +02:11:51.1 & $  27.0_{-  0.3}^{+  0.3} $ & $  25.7_{-  0.1}^{+  0.1} $ & $  25.6_{-  0.3}^{+  0.4} $ & $  25.5_{-  0.4}^{+  0.5} $ & $ >  25.4 $ & $ >  25.6 $ & M4 & $       6.7$ & $       5.5$ & $       1.0$ \\
S34 & 10:02:22.64 & +01:51:57.1 & $  26.7_{-  0.3}^{+  0.4} $ & $  25.5_{-  0.2}^{+  0.2} $ & $ >  25.3 $ & $ >  25.5 $ & $ >  25.1 $ & $ >  24.9 $ & M4 & $       4.8$ & $       5.5$ & $       1.5$ \\
S35 & 10:00:24.21 & +02:39:07.2 & $  26.8_{-  0.3}^{+  0.4} $ & $  25.6_{-  0.1}^{+  0.1} $ & $  25.5_{-  0.2}^{+  0.3} $ & $  25.0_{-  0.2}^{+  0.2} $ & $ >  25.6 $ & $ >  25.6 $ & M4 & $       6.5$ & $       5.5$ & $       4.9$ \\
S36 & 10:01:33.84 & +02:03:30.9 & $  26.7_{-  0.3}^{+  0.5} $ & $  25.7_{-  0.1}^{+  0.1} $ & $ >  25.8 $ & $ >  25.5 $ & $ >  25.3 $ & $ >  24.7 $ & M4 & $       7.8$ & $       5.5$ & $       2.2$ \\
S37 & 09:59:08.92 & +01:50:50.7 & $  24.7_{-  0.1}^{+  0.1} $ & $  23.5_{-  0.1}^{+  0.1} $ & $  23.1_{-  0.1}^{+  0.1} $ & $  22.8_{-  0.1}^{+  0.1} $ & $  22.9_{-  0.1}^{+  0.1} $ & $  22.9_{-  0.1}^{+  0.1} $ & M5 & $       9.9$ & $       5.5$ & $      10.3$ \\
\hline
\multicolumn{13}{c}{UDS/SXDS field}\\
\hline
ID & R.A. & Dec. & $i$ & $z'$ & $Y$ & $J$ & $H$ & $K$ & Star & $\chi_{\star}^2$ & $z_{\rm phot}$ & $\chi_{\rm gal}^2$  \\
 & (J2000) & (J2000) & & & & & & & type & & & \\
 \hline
S38 & 02:18:01.62 & -04:52:22.3 & $  27.4_{-  0.3}^{+  0.4} $ & $  25.4_{-  0.1}^{+  0.1} $ & $  24.6_{-  0.2}^{+  0.2} $ & $  24.1_{-  0.1}^{+  0.1} $ & $  24.2_{-  0.1}^{+  0.1} $ & $  24.0_{-  0.1}^{+  0.1} $ & M8 & $       6.8$ & $       6.4$ & $       6.1$ \\
S39 & 02:16:13.17 & -04:51:40.5 & $  27.4_{-  0.3}^{+  0.5} $ & $  25.4_{-  0.1}^{+  0.1} $ & $  24.8_{-  0.2}^{+  0.2} $ & $  24.1_{-  0.1}^{+  0.1} $ & $  24.1_{-  0.1}^{+  0.1} $ & $  23.9_{-  0.1}^{+  0.1} $ & M8 & $       5.4$ & $       6.3$ & $       9.3$ \\
S40 & 02:19:36.16 & -05:03:16.2 & $ >  27.8 $ & $  25.5_{-  0.1}^{+  0.1} $ & $  24.9_{-  0.2}^{+  0.2} $ & $  23.8_{-  0.1}^{+  0.1} $ & $  23.7_{-  0.1}^{+  0.1} $ & $  23.2_{-  0.1}^{+  0.1} $ & L4 & $       7.9$ & $       6.3$ & $      11.0$ \\
S41 & 02:16:50.79 & -05:28:44.5 & $  26.8_{-  0.2}^{+  0.3} $ & $  25.6_{-  0.1}^{+  0.1} $ & $  25.1_{-  0.3}^{+  0.4} $ & $  25.3_{-  0.1}^{+  0.2} $ & $  25.6_{-  0.3}^{+  0.5} $ & $  25.5_{-  0.3}^{+  0.4} $ & M4 & $       9.7$ & $       5.6$ & $       3.5$ \\
S42 & 02:17:40.34 & -04:40:06.4 & $  26.5_{-  0.1}^{+  0.1} $ & $  25.1_{-  0.1}^{+  0.1} $ & $  25.0_{-  0.2}^{+  0.3} $ & $  24.6_{-  0.1}^{+  0.1} $ & $  24.5_{-  0.2}^{+  0.2} $ & $  24.4_{-  0.1}^{+  0.1} $ & M5 & $       8.2$ & $       5.6$ & $       1.3$ \\
S43 & 02:18:17.95 & -05:25:47.9 & $  26.8_{-  0.2}^{+  0.2} $ & $  25.7_{-  0.1}^{+  0.2} $ & $ >  25.6 $ & $  25.6_{-  0.3}^{+  0.3} $ & $ >  25.6 $ & $  26.1_{-  0.4}^{+  0.6} $ & M4 & $       7.1$ & $       5.5$ & $       0.6$ \\
S44 & 02:17:06.25 & -04:49:21.6 & $  26.7_{-  0.2}^{+  0.2} $ & $  25.6_{-  0.1}^{+  0.1} $ & $  25.6_{-  0.3}^{+  0.5} $ & $  25.7_{-  0.3}^{+  0.3} $ & $ >  25.5 $ & $  25.8_{-  0.4}^{+  0.6} $ & M4 & $       9.5$ & $       5.5$ & $       0.3$ \\
S45 & 02:16:38.53 & -05:02:08.2 & $  26.8_{-  0.2}^{+  0.2} $ & $  25.7_{-  0.1}^{+  0.1} $ & $  25.4_{-  0.2}^{+  0.3} $ & $  25.7_{-  0.3}^{+  0.4} $ & $  25.4_{-  0.3}^{+  0.4} $ & $  25.6_{-  0.3}^{+  0.5} $ & M4 & $       7.1$ & $       5.5$ & $       0.9$ \\
S46 & 02:19:09.36 & -04:40:00.4 & $  26.3_{-  0.1}^{+  0.1} $ & $  25.2_{-  0.1}^{+  0.1} $ & $  25.5_{-  0.3}^{+  0.5} $ & $  25.0_{-  0.2}^{+  0.2} $ & $  24.9_{-  0.2}^{+  0.3} $ & $  25.2_{-  0.2}^{+  0.3} $ & M4 & $       8.1$ & $       5.5$ & $       2.2$ \\
S47 & 02:17:32.33 & -04:39:36.4 & $  26.9_{-  0.2}^{+  0.2} $ & $  25.7_{-  0.1}^{+  0.1} $ & $  25.7_{-  0.3}^{+  0.5} $ & $  25.7_{-  0.3}^{+  0.4} $ & $  24.9_{-  0.2}^{+  0.3} $ & $  25.4_{-  0.3}^{+  0.4} $ & M4 & $       8.6$ & $       5.5$ & $       3.4$ \\
\hline

\end{tabular}
\label{table:stars}
\end{table*}

\section{Gravitational lensing by foreground galaxies}\label{appendixgrav}

Following the approach presented in~\citet{McLure2009} and~\citet{Bowler2014}, we estimated the magnification, $\mu$, from foreground galaxies at a separation, $\theta$, from our high-redshift galaxy as:
\begin{equation}
\mu = \frac{\theta}{\theta - \theta_{\rm E}}
\end{equation}
using the singular isothermal sphere (SIS) approximation to describe the dark matter halo of each foreground galaxy.
Here $\theta_{\rm E}$ denotes the Einstein radius which depends on the velocity dispersion, $\sigma_{V}$, of the dark matter halo as:
\begin{equation}
\theta_{\rm E} = \frac{4\pi(\sigma_{V} / c)^2 \, D_{LS}}{D_{S}}
\end{equation}
in the SIS model, where $D_{LS}$ denotes the luminosity distance from the lens object the the source and $D_{S}$ denotes the luminosity distance to the source.

In each field we created a $K$-band (mass selected) catalogue using the MAG\_AUTO from {\sc SExtractor} as an estimate for the total magnitude of the foreground galaxies, which then allowed an estimate of the velocity dispersion of the dark matter halo from the $i$-band absolute magnitude using the Faber-Jackson relation from~\citet{Bernardi2003}.
The photometric redshifts of the foreground $K$-band selected objects were calculated using the {\sc le Phare} code using 3-arcsec diameter circular aperture photometry following~\citet{Ilbert2009, Ilbert2013}.
The~\citet{Ilbert2009} SED template set was used in the SED fitting process rather than the high and low-redshift model subsets utilised in our sample selection, to better represent the range of galaxies found at $z < 4$.
We select galaxies based on an acceptable galaxy solution and a superior galaxy fit over that from stars (using the PICKLES library of stellar templates;~\citealp{Pickles1998}).

\begin{figure}

\includegraphics[width = 0.5\textwidth]{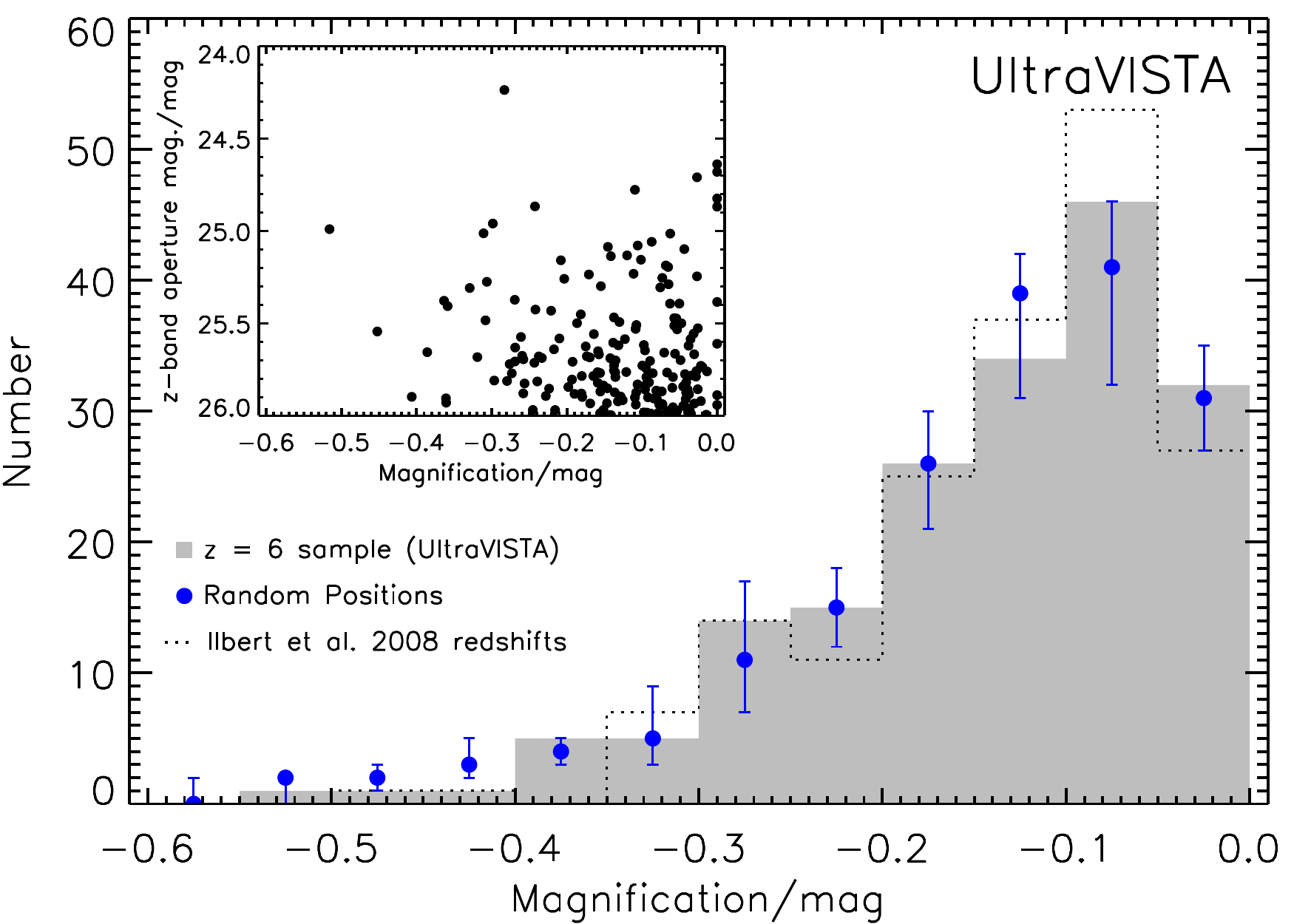}
\includegraphics[width = 0.5\textwidth]{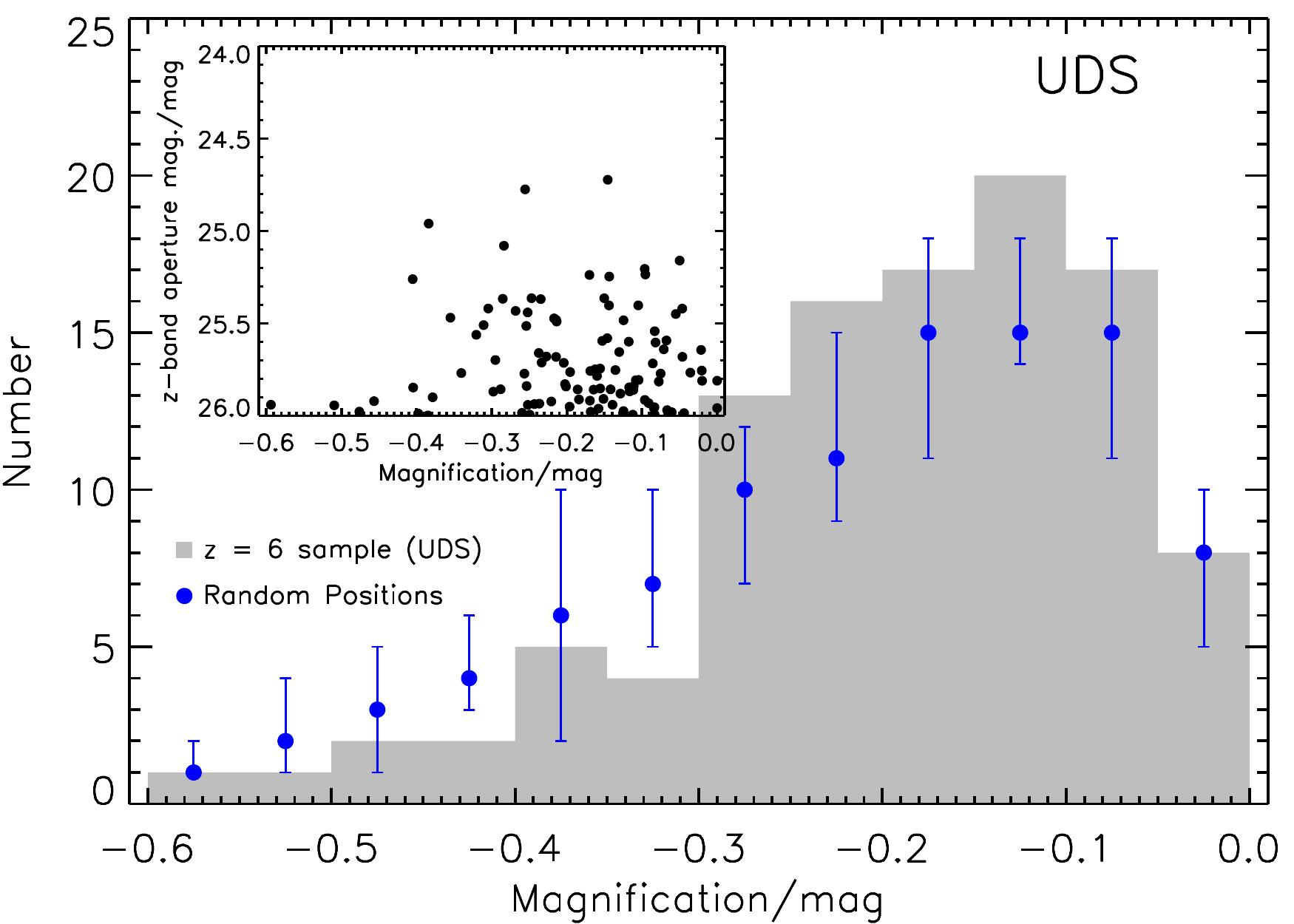}

\caption{The magnification distribution of the full $z \simeq 6$ sample due to gravitational lensing by the mass associated with foreground galaxies close to the line of sight.
The observed magnification distribution is shown in grey, for the UltraVISTA/COSMOS and UDS/SXDS samples in the upper and lower plots respectively.
The predicted magnification distribution for random positions in the field is shown as the blue points, where we plot the median and 68-percentiles.
The inset plot shows the gravitational magnification plotted against the $z'$-band magnitude for the sample. }
\label{fig:lensing}

\end{figure}

The resulting magnification distributions for our samples of objects in the UltraVISTA/COSMOS and UDS/SXDS fields are shown in Fig.~\ref{fig:lensing}, where we also plot the magnification against the $z'$-band magnitude of the galaxy.
The magnification sums the contribution from all foreground galaxies closer than 10-arcsec to the high-redshift galaxy.
We find that the majority of galaxies have some magnification of the order of $\sim 0.1\,$mag (median values are $0.11$ in the UltraVISTA/COSMOS field and 0.16 in the UDS/SXDS field), with several objects showing magnifications as large as $\sim 0.6\,$mag.
If we use the~\citet{Ilbert2008} photometric redshifts in the UltraVISTA/COSMOS field we find a similar shape of magnification distribution as shown in Fig.~\ref{fig:lensing} with an identical median magnification.
The inset plot of magnification against $z'$-band magnitude shows no evidence for the brightest objects having the largest magnification, which, if true,  could influence the derived shape of the LF.

The magnification of the objects we derive could impact the measured LF, however we must determine if this magnification is unusual given that all astronomical imaging surveys show foreground objects close to the line of sight of the background high-redshift galaxies.
Hence, we calculated the expected magnification for random positions in the field, using the full $K$-band catalogues used to determine the magnification of our sample.
A minimum separation of 1-arcsec was applied when calculating the lensing at a given position, to exclude very high magnification of objects directly along the line of sight.
The resulting distribution of magnification values, for a randomly drawn sample of 159 or 107 objects for the UltraVISTA/COSMOS and UDS/SXDS fields respectively, is shown in Fig.~\ref{fig:lensing}.
The simulated samples of objects were run 1000 times for each field, and the median and 68-percentiles were calculated from the derived magnification distributions.
The random distributions are very similar to those observed in our samples, showing median values of 0.11 and 0.18, indicating that modest gravitational lensing of our objects is not unusual for high-redshift sources in the field.
We therefore do not correct the absolute magnitudes of our objects for this magnification when determining the LF.

We find a slight difference in the observed and predicted magnification distributions between the UltraVISTA/COSMOS and UDS/SXDS fields.
The $K$-band data in the UDS/SXDS field is deeper than that in UltraVISTA/COSMOS ($m_{\rm AB} = 24.6$ as compared to $m_{\rm AB} = 24.2$), which would imply a higher surface density of sources in the UDS/SXDS field and hence a higher derived magnification.
Inspection of the distribution of the number of sources within a 10-arcsec radius of the sample of high-redshift galaxies shows that this is indeed the case, and the magnification distributions can be brought into closer agreement if we cut the UDS/SXDS catalogue at the same depth as the UltraVISTA/COSMOS data, which results in a shift in the peak of the magnification distribution faint-wards by $\sim 0.05\,$mag.
Any residual difference is likely due to the slightly different redshift distribution between the two fields, a result of large-scale structure in the fields or the different relative depths in the multiwavelength images which can subtly bias the photometric redshifts.

\end{document}